\pdfoutput=1
\documentclass[preprint2]{aastex}

\shorttitle{The Spitzer SDSS GALEX Spectroscopic Survey}
\shortauthors{O'Dowd et al. 2011}

\begin{document}

\title{SSGSS: The Spitzer SDSS GALEX Spectroscopic Survey}

\author{Matthew J. O'Dowd$^{*, 1,2}$, David Schiminovich$^{3}$, Benjamin D. Johnson$^{4}$, Marie A. Treyer$^{5}$, Christopher D. Martin$^{5}$, Ted K. Wyder$^{5}$, St\'ephane Charlot$^{6}$, Timothy M. Heckman$^{7}$, Lucimara P. Martins$^{8}$, Mark Seibert$^{9}$, J. M. van der Hulst$^{10}$}

\email{$^*$ matt@astro.columbia.edu}
\affil{\small $^{1}$ Lehman College, City University of New York, 250 Bedford Park Blvd. West, Bronx, NY 10468, USA}
\affil{\small $^{2}$ American Museum of Natural History, Central Park West at 79th Street, New York, NY 10024, USA}
\affil{\small $^{3}$ Astronomy Department, Columbia University, New York, NY 10027, USA}
\affil{\small $^{4}$ Institute of Astronomy, Madingley Road, Cambridge CB3 0HA, UK}
\affil{\small $^{5}$ California Institute of Technology, MC 405-47, 1200 East California Boulevard, Pasadena, CA 91125, USA}
\affil{\small $^{6}$ Institut d'Astrophysique de Paris, UMR 7095, 98 bis Bvd Arago, 75014, Paris, France}
\affil{\small $^{7}$ Department of Physics and Astronomy, Johns Hopkins University, Baltimore, MD 21218, USA}
\affil{\small $^{8}$ NAT - Universidade Cruzeiro do Sul, Rua Galvao Bueno, 868 S\~ao Paulo, SP, Brazil)}
\affil{\small $^{9}$ Observatories of the Carnegie Institution of Washington, 813 Santa Barbara Street, Pasadena, CA 91101, USA}
\affil{\small $^{10}$ Kapteyn Astronomical Institute, University of Groningen, the Netherlands}


\begin{abstract}
The Spitzer SDSS GALEX Spectroscopic Survey (SSGSS) provides a new sample
of 101 star-forming galaxies at $z<0.2$ with unprecedented multi-wavelength coverage. 
New mid- to far-infrared spectroscopy from the Spitzer Space Telescope is
added to a rich suite of previous imaging and spectroscopy, including ROSAT,
GALEX, SDSS, 2MASS, and {\it Spitzer}/SWIRE. Sample selection ensures an
even coverage of the full range of normal galaxy properties, spanning
two orders of magnitude in stellar mass, colour, and dust attenuation. 
In this paper we present the SSGSS data set, describe the science drivers, and detail the
sample selection, observations, data reduction, and quality
assessment. Also in this paper, we compare the shape of the thermal continuum and the
degree of silicate absorption of these typical, star-forming galaxies
to those of starburst galaxies. We investigate the link between
star formation rate, infrared luminosity, and total PAH luminosity,
with a view to calibrating the latter for SED models in
photometric samples and at high redshift.
Lastly, we take advantage of the 5-40~\micron\ spectroscopic and
far infrared photometric coverage of this sample to perform
detailed fitting of the \citet{Draineetal07} dust models, and investigate
the link between dust mass and star formation history and AGN properties.

\end{abstract}


\section{Introduction}

Spectroscopy in the mid- to far-infrared is emerging as a
critical tool for understanding the physical properties of galaxies, 
and is the missing link in the new era of very large 
multi-wavelength spectroscopic galaxy surveys. This
spectral region is dominated by emission from interstellar gas and
dust, largely powered by active star formation. Observation of this
reprocessed starlight enables measurement of embedded star formation
that is inaccessible to ultraviolet (UV) or optical diagnostics. 
Spectroscopy in the infrared (IR) also directly probes the makeup and
physical state of the interstellar medium (ISM) through the shape of the
thermal continuum, aromatic molecular bands and a wealth of emission lines.

The improved understanding of star formation on a galactic scale
granted by IR spectroscopy is
increasingly important as we strive to understand the evolution of
cosmic star formation, and in particular its decline since 
z$\sim$1. While IR spectroscopy has enabled substantial progress in
mapping the most intensely star-forming (SF) galaxies in the local
universe, these galaxies are not the principal stellar factories of the current
epoch, nor are they the closest analogs to the dominant sources of 
star-formation at its cosmic peak. 

At z$\sim$1, luminous infrared as well as ultra-luminous infrared galaxies
(LIRGS and ULIRGS) appear to dominate star-forming activity \citep{LeFloch05}, in stark contrast to the
local universe where lower-luminosity 'normal' galaxies contribute the
bulk of current star formation. However a number of independent
studies \citep{Melbourne05, Bell05, Noeske07, Zheng07}
show that these z$\sim$1 LIRGS and ULIRGS
do not display the signs of violent interactions observed in their 
local counterparts. In fact, it appears that over much of cosmic
history the bulk of stars were formed in galaxies dynamically similar
to normal, disk-dominated galaxies at low redshift. 

Thus, low-redshift disk galaxies are important laboratories of
cosmic star formation---both in their own right and as dynamical
analogs of higher redshift LIRGS and ULIRGS. The Sloan Digital Sky
Survey (SDSS) \citep{York00} has enabled dramatic strides in our
understanding of this galaxy population \citep{Bell03,
Brinchmann04, Kauffmann03, Kauffmann04, Blanton03a, Balogh04}, and
GALEX observations of the SDSS sample has provided a powerful additional
lever arm for understanding---and selecting for---the star formation properties of
these galaxies (e.g. \citealt{Heckman05, Yi05}.

Mid-IR (MIR) to far-IR (FIR) spectroscopy is now the key missing ingredient. 
Its addition to the GALEX and SDSS data allows a thorough accounting
of star formation, opens up a wealth of new diagnostics, and grants the
potential to self-consistently treat stellar populations and dust
absorption and emission in models (e.g. Da Cunha, Charlot, \& Elbaz 2008). 
A sample of star-forming galaxies spanning a representative range of
physical properties with comprehensive multiwavelength coverage from
the far-UV (FUV) to the FIR is necessary.

The {\it Spitzer-SDSS-GALEX Spectroscopy Survey} (SSGSS) provides such
a sample. Selecting from the low-extinction Lockman Hole within the
Sloan+GALEX footprint, FUV and optical diagnostics are used to define a
broad, representative sample of 100 normal, star-forming galaxies between
redshifts 0 and 0.2, boasting rich, multi-wavelength coverage. Deep
{\it Spitzer} low-resolution 
spectroscopy has been obtained for the entire sample, and
high-resolution spectroscopy for the brightest 33 galaxies.

In this paper we describe the SSGSS dataset. In
Section~\ref{motivation} we outline the scientific motivation and
goals of the survey, in Section~\ref{sample} we describe the sample
selection and properties, in Section~\ref{observations} we detail the
observations, in Section~\ref{reduction} we describe the data
reduction and quality assessment, in Section~\ref{data} we present the
spectra and compare to the IR spectra of starburst galaxies, 
and in Section~\ref{products} we describe the SSGSS data products.

\section{Scientific Motivation}
\label{motivation}

The primary goal of SSGSS is, broadly, the detailed characterization
of the IR spectra of normal, star-forming galaxies spanning a
comprehensive range of physical properties.
Through comparison with multi-wavelength data from the FUV to
the FIR, SSGSS seeks to disentangle the physical processes responsible
for the wide range of emission features observed in galaxies' IR
spectra, and so to calibrate these features as diagnostics of galaxies'
physical states.

\subsection{The Importance of MIR Spectroscopy}

Interstellar dust absorbs from 50\% to as much as 90\% of the UV and
optical starlight in star-forming galaxies
(e.g., Calzetti, Kinney, \& Storchi-Bergmann 1994; Wang \& Heckman 1996; Buat et al. 1999, 
2002; Sullivan et al. 2000; Bell \& Kennicutt 2001; Calzetti 
2001; Goldader et al. 2002), and the re-radiated IR photons
consitute half of the bolometric luminosity in the local universe.
Broad-band measures of the IR Spectral Energy Distribution (SED)
of galaxies have taken us a long way towards understanding 
the link between this emission and star formation; first with the
{\it Infrared Astronomical Satellite}
(IRAS; e.g., \citealt{Helou86, RR89, Devereux91, Sauvage92, Buat96,
  Walterbos96, Kennicutt98, Kewley02}), and 
subsequently with the {\it Spitzer Space Telescope}, with deep imaging surveys tracing obscured
star formation in LIRGs, ULIRGs and “normal” disk systems out to
redshift 3 and beyond (e.g. Perez-Gonzalez et al. 2005). 

However several processes complicate the
interpretation of broad-band measurements in the IR:
the thermal IR continuum depends on both the temperature distribution and
composition of the dust; aromatic bands in the MIR are linked to 
star formation, dust composition, and AGN activity, but are poorly
understood; atomic and molecular lines span a wide range of
ionizations and hence arise from a wide variety of environments; AGN and
stellar emission compete with thermal dust emission in the continuum;
and all of these may be affected by extinction (although more weakly
than in optical and UV bands.)
To begin to deconvolve the effects of these processes, and hence to unlock the diagnostic 
power of the IR spectrum of galaxies, high quality spectroscopy is necessary.

The {\it Spitzer} Infrared Spectrograph (IRS) has enabled great strides in this regard, 
especially for very luminous IR sources. 
We now have detailed MIR spectroscopic libraries 
of the most actively star-forming galaxies, both in the local universe
(eg. \citealt{Brandl, Imanishi10}), and, increasingly, at higher redshifts
(eg. \citealt{Yan07, Murphy09, Dasyra, Desai, HernanCaballero09}).
The Spitzer Infrared Nearby Galaxies
Survey (SINGS; \citealt{Kennicutt}) has been especially important
in linking the primary components of the MIR spectrum with physical properties,
including: linking the effects of metallicity, ionizing field,
star-formation, and H$_2$ intensity with
Polycyclic Aromatic Hydrocarbon (PAH) features 
\citep{Smith07, Roussel07}, physical modeling and characterization of dust content
\citep{Draineetal07, MunozMateos09b}, and diagnosing abundances and
the relative contributions of star formation versus AGN activity
\citep{Dale09, Moustakas10}.
With its local sample, one of the strengths of SINGS is its capacity
for spatial resolution within its galaxies. This has enabled, for example,
the mapping of the radial distributions of dust, gas, stars, and their evolution \citep{MunozMateos09a, MunozMateos09b,
  MunozMateos11} and the use of H$\alpha$ attenuation as a tracer of obscured star-formation \citep{Prescott07}.

These results have provided a powerful insight into the inner workings of
galaxies, however it is challenging to draw general conclusions on the
global properties of the SINGS sample due to its focus on the nuclear
regions of these extended  sources. In addition,
SINGS is typical of other IRS samples, in that it is
dominated by IR luminous galaxies.
To characterize the IR spectra of a
representative population of star-forming galaxies, we need  increased
depth, a greater redshift range, and more careful selection criteria.

\subsection{MIR Spectral Diagnostics}

SSGSS spectroscopy spans the 5--40~\micron\ range, covering 
a wealth of diagnostic features. These include: prominent emission bands from PAH
molecules, a continuum fueled by reprocessed as well as direct
starlight and by AGN, and abundant emission lines revealing a very
wide range of ionization states. Each of these features promises
significant diagnostic potential.

\subsubsection{PAH Bands}

PAH molecules produce a convoluted spectrum of broad features and
complexes that dominate the MIR below 20~\micron. These arise 
from multiple vibrational modes of aromatic carbon lattices with a
wide range of grain sizes and ionization fractions (see \citet{Tielens} for a review). 

The SSGSS spectra span the full complement of major PAH bands from
6.2~\micron\ to 17~\micron, and
include features arising from carbon-carbon stretching, and
carbon-hydrogen in-plane and out-of-plane bending modes.
The strengths of these PAH bands, relative to both the continuum and
to each other, have the potential to serve as powerful diagnostics of
the environments of these molecules. Examples of physical properties
probed by PAH spectra include:

{\it Star formation history:} 
the overall luminosity radiated in the PAH bands
is directly linked to the Star Formation Rate (SFR; e.g. \citealt{Teplitz07, Brandl}), and may be an important proxy at
higher redshifts where PAH features fall in MIR bands.
As the PAH spectrum 
varies depending on star formation history and other properties, its diagnostic potential
depends on a suite of templates measured at low-redshift and
calibrated to a range of properties more robustly measurable at high
redshift.

{\it PAH grain size:} grain size distribution affects the relative strength of
short-to-long wavelenth PAH bands. Larger PAH molecules tend to emit more
efficiently at longer wavelengths than do smaller grains 
\citep{Tielens, DraineLi, Schutte} and so PAH
ratios can be used as an indicator of grain size distribution. This in
turn is an indicator of both the growth and destruction of PAH
molecules, and so will gauge the likely governing parameters,
such as abundance, star formation, and AGN activity.

{\it PAH ionization:}
ionized PAH molecules radiate via carbon-carbon (CC) stretching modes
with significantly greater efficiency that do neutral PAH
molecules, while carbon-hydrogen (CH) modes do not show the same dependence
\citep{Tielens}. The ratios of bands arising from CC modes, such as
those centered at 6.2~\micron\ and 7.7~\micron, to CH bands,
such as the prominant 11.3~\micron\ feature, provide a sensitive
indicator of PAH ionization fraction, and so probe the FUV photon flux
and the temperature distribution of the PAH grains.

As indicated by the models of \citet{DraineLi}, a combined suite of
PAH ratios can be used to disentangle the effects of grain size
distribution and ionized fraction.

\subsubsection{The Thermal Continuum}

Combined with MIPS 70~\micron\ and 160~\micron\ flux, the SSGSS spectra enable the
most comprehensive characterization of dust temperature distributions
in normal galaxies to date. While the 60~\micron/100~\micron\ flux
ratio has been shown to be a good diagnostic of the general shape of
the IR spectrum \citep{Helou}, and hence of dust temperature,
the MIR continuum can exhibit a range of shapes beyond its slope. 
These variations are likely due to variations in the grain size and
composition of the warm dust component, as well as any AGN component. 
Coupled with the UV/Optical dust attenuation curves, spectral
shapes to 40~\micron\ grant insight into these grain properties in
galaxies with no significant AGN component.
Comparison of continuum shapes with optical diagnostics of star
formation and metallicity will be important in disentangling these
competing effects.

Furthermore, short wavelength spectra allow a better determination of the 
contribution of thermal and stellar emission in galaxies over the wavelength range from 
5~\micron\ to 10~\micron, where dust is expected to fully dominate the MIR spectrum. 

SSGSS's MIR spectra and FIR photometry allow for a detailed
understanding of dust emission in these galaxies, and so provide a
thorough account of reprocessed starlight. The multi-wavelength nature
of this sample is important here; optical diagnostics such as the
$H\alpha$ emission line and the 4000\AA\ break, as well as UV
photometry, give independent measures of star formation that may be
used to study the efficiency of reprocessing. This is important for
the use of IR photometry as a measure of star formation, and also to
help understand the fate of galaxies' ionizing radiation. GALEX data
is especially helpful here: the comparison of Lyman-$\alpha$ emission
in the FUV band to the MIR--FIR intensity and slope allows us to
investigate the fate of FUV photons, and hence escape fraction, as a
function of galaxy type (see for example \citealt{Hanish2010}). 

\subsubsection{Emission Lines}

\begin{figure*}[ht!]
\centering
\includegraphics*[width=15cm]{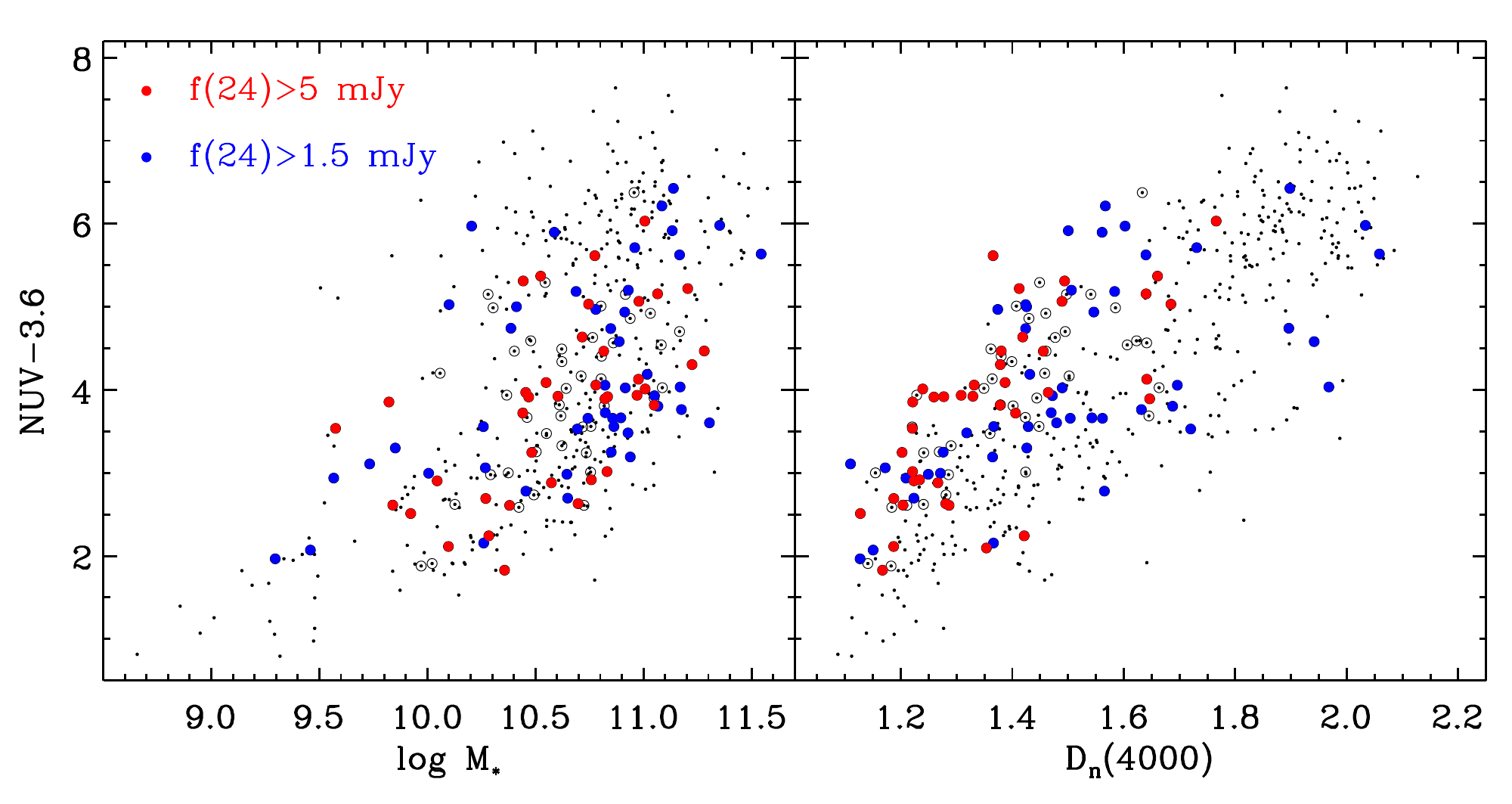}
\caption{NUV-3.6~\micron\ flux vs. log M$_*$ (left) and D$_n$(4000)
  (right) for the SSGSS sample (solid circles). The blue circles are
  galaxies with 1.5~mJy~$<$~F$_{24~\micron} < 5$~mJy and are observed with only
  low-res spectroscopy, while the red circles have F$_{24~\micron} > 1.5$~mJy
  and are observed with both low-res and hi-res modules. Unfilled
  circles are galaxies above the flux and surface brightness cuts
  but rejected due to high source density in those regions of
  parameter space. Black points are galaxies of the Lockman Hole
  sample \citep{Johnson06}, the parent sample of SSGSS. Stellar masses are from \citet{Brinchmann04}.
\label{NUV-3.6}}
\end{figure*}
 
MIR line ratios provide valuable diagnostics of HII
regions. Electron densities can be studied using 
S III]$_{18.7~\micron}$/[S III]$_{33.4~\micron}$ and
[Ne~III]$_{15.5~\micron}$/[Ne~III]$_{36.0~\micron}$, 
while ionization ratios can be studied using
[Ne~III]/[Ne~II], [S IV]/[S III] and [Ar III]/[Ar II], and elemental
abundances using various optical and MIR line species. 

Many high stellar mass galaxies (log~$M_* > 10$) in the local universe
and at z=1 possess composite optical emission-line spectra showing evidence of
excitation by both young stars and AGN. These signatures are
degenerate and are necessarily incomplete in characterizing reddened
AGN. MIR ionization indicators such as [Ne~III]/[Ne~II] are invaluable 
for decomposing star formation and AGN emission because of the minimal levels of
attenuation seen at these wavelengths \citep{Weedman06}. 
Comparison of these with other AGN indicators, such as
UV/optical line ratios \citep{Kewley,KauffmannAGN03}, and even PAH equivalent widths (EWs), allow us to
analyse the diagnostic power of all of these measures.
Combined with detailed MIR spectra, we can 
investigate the relative effects of AGN-sourced and
starburst-sourced hard radiation fields on the ISM.

The abundance and composition of raw materials probably has a significant effect on
the build-up and behaviour of dust grains. To better understand this effect, MIR 
abundance indicators combined with optical measures of gas-phase
metallicity \citep{Tremonti} can be compared with PAH bands, thermal dust continuum emission,
and with MIR silicate absorption.

The molecular hydrogen IR fluorescence spectrum provides an
indication of the physical conditions in photo-dissociation regions
around molecular clouds and, while linked to the FUV radiation field,
may also be produced by shocks in these regions.

\section{Sample Definition and Properties}
\label{sample}

The SSGSS sample consists of 100 star-forming galaxies observed with
{\it Spitzer} IRS in the Lockman Hole region, and is selected to have
ancillary multi-wavelength imaging and spectroscopy from the FUV to
the FIR, and to include galaxies with SFRs spanning
the full range observed in the normal, star-forming population, short
of the most extreme starbursting sources.


\subsection{The Lockman Hole}

The Lockman Hole, situated around $\alpha=10^h.7$, $\delta=50\deg$, 
is a $\sim$10 square degree field with one of the lowest column
densities of interstellar material in the Milky Way, with $N_{HI} <
7\times10^{19}$, approximately 5\% of the Galactic median.
This makes it ideal for low-surface
brightness surveys, and so it has been the subject of extensive 
multi-wavelength surveys. These include 
deep ROSAT imaging, deep GALEX (m$_{FUV,NUV}$~24.5 AB), SDSS imaging
(m$_{r}\sim22.2$ AB) and spectroscopy, Deep 2MASS  ($\times6$,
m$_K\sim$17.8 AB, \citealt{Beichman03}) and Spitzer/SWIRE \citep{Lonsdale03} IRAC (m $\sim$22.5, 20 for 3.6, 8.0~\micron) surveys,  
as well as future programs (e.g. UKIDSS deep survey over the full region).
The SDSS primary spectroscopic sample targets all galaxies with
$r<17.8$ and yields $\sim100$ galaxies deg$^{-1}$.

\subsection{Selection Criteria}

\begin{figure*}
\centering
\includegraphics*[width=15cm]{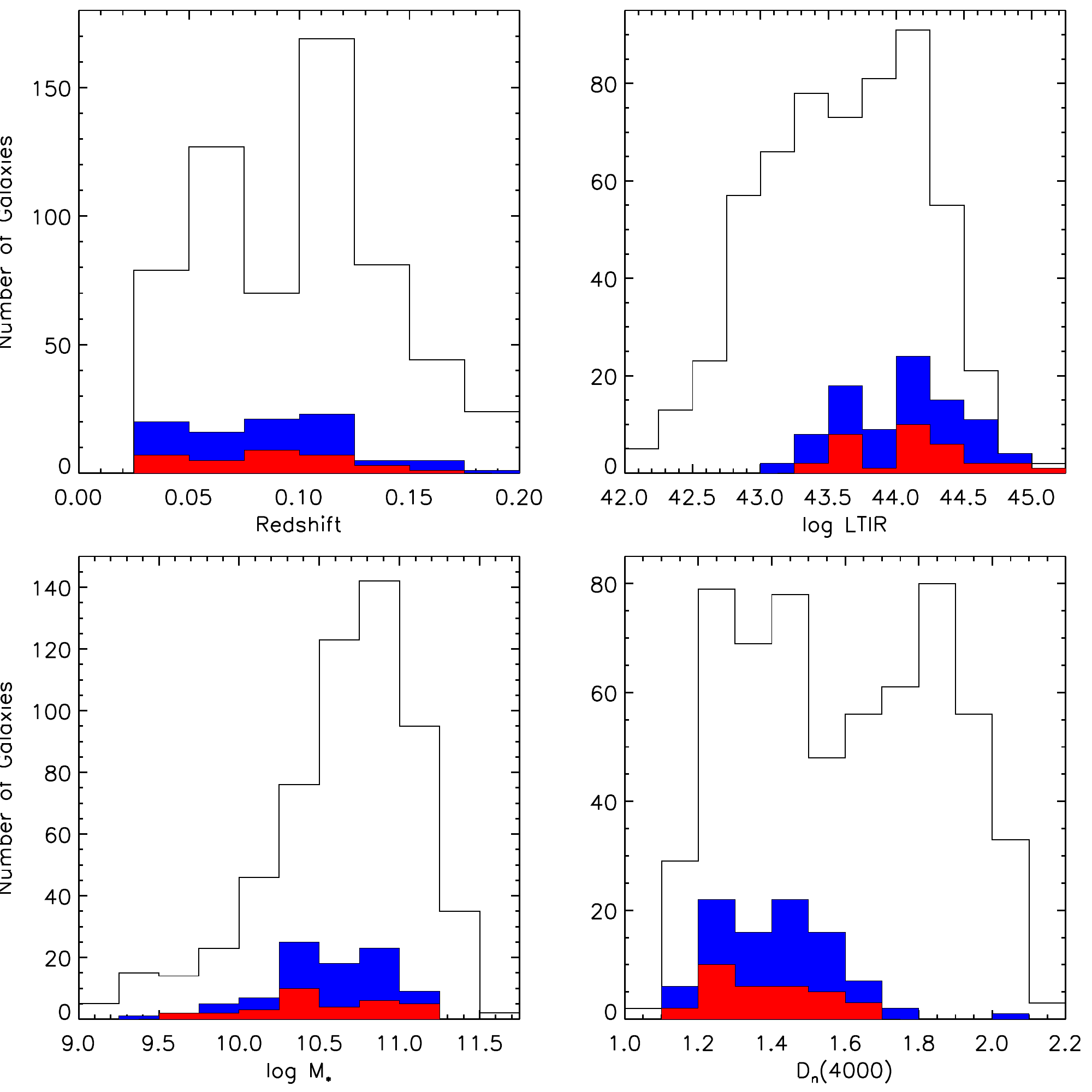}
\caption{Distribution of SDSS-measured redshift, $L_{TIR}$, stellar mass, and
  D$_n$(4000) for SSGSS galaxies. The blue histograms show the 101
  galaxies of the entire SSGSS sample, red histograms show the 33 galaxies of the
  SSGSS bright sample, and open histograms show the 
  galaxies of the Lockman Hole sample of \citet{Johnson06}.
\label{histograms}}
\end{figure*}

The Lockman Hole sample of \citet{Johnson06} forms the parent sample
for SSGSS, and consists of 916 galaxies at $z<0.2$
with GALEX-SDSS-Spitzer/SWIRE all-band coverage. The wide spectral
coverage of this sample---from FUV to FIR---allows a 
consistent treatment of dust absorption and the corresponding dust
emission, providing a total accounting of the star formation, stellar
mass, and dust attenuation for these galaxies. A detailed description
of these and other derived measurements and physical parameters can be
found in Section~\ref{cproducts}.
SSGSS targets a subset of this Lockman Hole sample for observation with
the {\it Spitzer} IRS. These galaxies are selected based on two
brightness criteria: 

\begin{itemize}

\item {\bf 5.8~\micron\ surface brightness:}
Galaxies fill the IRS Short-Low slit, and so at short-wavelengths we apply
a surface brightness limit of I$_{5.8\mu m} > 0.75$~MJy~sr$^{-1}$, based
on IRAC 5.8~\micron\  flux and half-light diameter (the median/mean
half-light diameter is 5\AA\ for the entire sample). 

\item {\bf 24~\micron\  flux:} At long wavelengths
most of the galaxies are unresolved, and so a flux limit of
F$_{24\mu m} > 1.5$~mJy is applied.

\end{itemize}

The above criteria yield 154 galaxies. We further restrict the sample
by  performing an even sampling of the planes of near-UV (NUV)-3.6~\micron\
vs. stellar mass (from \citealt{Brinchmann04}) and NUV-3.6~\micron\ vs. $D_n(4000)$; 
in the dense regions of these spaces, we select the galaxy with the
highest 24~\micron\ flux. This process
ensures that the sample spans a comprehensive range of physical
properties. Figure~\ref{NUV-3.6} shows the
distributions of NUV-3.6~\micron\ color versus stellar mass and
$D_n(4000)$ for the SSGSS galaxies compared to the Lockman Hole
sample, and illustrates that the two span a similar range
of galaxy properties. The Lockman Hole sample itself is representative of
the SDSS galaxy sample with GALEX FUV detections; that is to 
say, normal, star-forming galaxies, consisting of both early- and
late-type,s with $1.1 \lesssim D_n(4000) \lesssim 2.1$.

The sample observed with Spitzer consists of 101 galaxies. 
Due to a fault with the SL spectrum for SSGSS 40, 
the current SSGSS data release consists of 100 galaxies. 
We preserve theoriginal indexing for consistency.


Within the SSGSS sample, we designate the {\it bright sample} and the 
{\it faint sample}, with the bright sample consisting of the 33
galaxies with F$_{24\mu m} > 5$~mJy. The entire sample is the
subject of low-resolution 5~\micron--40~\micron\ IRS spectroscopy, while the
bright sample is also observed with 10-20~\micron\
high-resolution spectroscopy.

\subsection{Sample Properties}

This final SSGSS sample has a redshift range of $0.03 < z < 0.2$ with
median redshift 0.085, and a total infrared luminosity ($L_{TIR}$)
range of $3.7\times10^9 L_\odot < L_{TIR} < 3.2\times10^{11} L_\odot$,
with median $3.9\times10^{10} L_\odot$.
The bright sample has redshift range of $0.04 < z < 0.2$ with median
redshift 0.077, and an $L_{TIR}$ range of 
$5.1\times10^9 L_\odot < L_{TIR} < 1.9\times10^{10} L_\odot$ with
median $6.9\times10^{10} L_\odot$. 
The flux and surface brightness cuts do eliminate galaxies with very
low stellar masses ($2\times10^{9} M_\odot\le M \le 2\times10^{11} M_\odot$), metallicities 
($8.7 \le log(O/H) + 12 \le 9.2$), and extinctions ($0.4 < A_{H\alpha}
< 2.3$).

Table~\ref{sampletable} details the 101 galaxies in the original SSGSS sample,
and Figure~\ref{histograms} shows the distribution of
redshift, stellar mass, $L_{TIR}$, and $D_n(4000)$ of the sample.

\begin{deluxetable}{cccccccccccc}[ht!]
\tabletypesize{\tiny}
\tablewidth{0pt}
\tablecaption{
\label{sampletable}}
\tablehead{
\colhead{SSGSS} & \colhead{RA} & \colhead{DEC} & \colhead{z} &
\colhead{NUV} & \colhead{r} & \colhead{L$_{TIR}$ (L$_{\odot}$)} &
\colhead{M$_*$ (M$_{\odot}$)$^{a}$} & \colhead{D$_n(4000)^{b}$} &
\colhead{S/N$^{c}$} & \colhead{Ap.Cor.$^{d}$} & \colhead{Notes}
}
\startdata
 1 & 160.34398 & 58.89201 & 0.066 & 18.04 & 16.22 & 3.90E+10 & 1.06E+10 & 1.14 & 17.1  & -0.36& 1\\
  2 & 159.86748 & 58.79165 & 0.045 & 19.56 & 17.21 & 7.49E+09 & 3.67E+09 & 1.21 &  1.3 & -0.37 & 1\\
  3 & 162.41000 & 59.58426 & 0.117 & 22.42 & 17.17 & 4.21E+10 & 1.15E+11 & 1.57 & 10.6 & -0.33 & \\
  4 & 162.54131 & 59.50806 & 0.066 & 19.92 & 16.92 & 1.85E+10 & 1.74E+10 & 1.43 &  ... & -0.90 & \\
  5 & 162.36443 & 59.54812 & 0.217 & 20.94 & 17.71 & 2.64E+11 & 1.20E+11 & 1.28 & 10.8 & -0.35 & \\
  6 & 162.52991 & 59.54828 & 0.115 & 20.01 & 17.03 & 1.10E+11 & 8.50E+10 & 1.31 & 14.6 & -0.59 & \\
  7 & 161.78737 & 59.63707 & 0.090 & 20.35 & 17.17 & 2.76E+10 & 2.86E+10 & 1.42 &  7.4 & -0.81 & 2\\
  8 & 161.48123 & 59.15443 & 0.044 & 18.52 & 15.79 & 1.05E+10 & 2.88E+10 & 1.57 & 12.0 & -1.15 & \\
  9 & 161.59111 & 59.73368 & 0.121 & 20.79 & 17.23 & 5.33E+10 & 7.45E+10 & 1.42 & 12.3 & -0.56 & 2\\
 10 & 161.11412 & 59.74155 & 0.117 & 21.02 & 17.12 & 5.62E+10 & 8.59E+10 & 1.43 &  6.3 & -0.50 & \\
 11 & 161.71980 & 56.25187 & 0.047 & 20.14 & 16.44 & 5.85E+09 & 1.14E+10 & 1.46 &  5.9 & -0.25 & 1\\
 12 & 162.26756 & 56.22390 & 0.072 & 20.34 & 16.26 & 7.33E+10 & 1.20E+11 & 1.64 & 26.8 & -0.41 & 1\\
 13 & 163.00845 & 56.55043 & 0.124 & 20.96 & 17.46 & 4.31E+10 & 6.57E+10 & 1.49 & 15.5 & -0.51 & \\
 14 & 161.92709 & 56.31395 & 0.153 & 19.74 & 16.94 & 1.29E+11 & 7.32E+10 & 1.28 & 11.0 & -0.37 & 1\\
 15 & 161.75783 & 56.30670 & 0.153 & 21.13 & 17.56 & 1.93E+11 & 6.85E+10 & 1.41 & 33.9 & -0.30 & 1\\
 16 & 162.04231 & 56.38041 & 0.072 & 20.07 & 16.82 & 2.91E+10 & 2.87E+10 & 1.46 & 12.7 & -0.31 & 1\\
 17 & 161.76901 & 56.34029 & 0.047 & 20.65 & 16.00 & 7.43E+10 & 5.86E+10 & 1.37 & 33.9 & -0.23 & 1\\
 18 & 163.39658 & 56.74202 & 0.128 & 20.43 & 17.76 & 3.24E+11 & 4.27E+10 & 1.35 & 30.8 & -0.58 & \\
 19 & 163.44330 & 56.73859 & 0.076 & 19.07 & 16.41 & 3.31E+10 & 3.30E+10 & 1.27 &  4.1 & -1.01 & 2\\
 20 & 163.26968 & 56.55812 & 0.070 & 20.87 & 16.76 & 3.29E+10 & 3.57E+10 & 1.45 & 21.8 & -0.29 & 3\\
 21 & 163.19810 & 56.48840 & 0.077 & 19.77 & 16.28 & 3.17E+10 & 9.41E+10 & 1.64 &  5.1 & -0.88 & 2\\
 22 & 163.09050 & 56.50836 & 0.045 & 20.07 & 16.35 & 9.86E+09 & 1.89E+10 & 1.32 &  5.1 & -0.52 & 2, 3\\
 23 & 163.47926 & 56.44407 & 0.046 & 21.42 & 16.54 & 3.66E+09 & 1.60E+10 & 1.60 &  ... &  ...  & 3\\
 24 & 163.53931 & 56.82104 & 0.046 & 17.65 & 15.26 & 4.27E+10 & 2.50E+10 & 1.29 & 14.5 & -0.95 & 2, 3\\
 25 & 158.22482 & 58.10917 & 0.073 & 19.27 & 16.89 & 2.24E+10 & 1.16E+10 & 1.24 &  8.9 & -0.74 & 3\\
 26 & 159.04880 & 57.72258 & 0.093 & 19.65 & 16.46 & 2.89E+10 & 5.75E+10 & 1.50 &  9.7 & -0.57 & \\
 27 & 159.34668 & 57.52069 & 0.072 & 19.18 & 16.06 & 1.15E+11 & 4.04E+10 & 1.33 & 31.6 & -0.55 & 1\\
 28 & 158.91122 & 57.59536 & 0.103 & 21.54 & 17.15 & 8.89E+10 & 1.01E+11 & 1.77 & 16.5 & -0.76 & 1, 2\\
 29 & 158.91344 & 57.71219 & 0.113 & 19.88 & 16.50 & 4.22E+10 & 1.14E+11 & 1.47 &  6.7 & -0.85 & \\
 30 & 159.73558 & 57.26361 & 0.046 & 19.99 & 17.39 & 1.42E+10 & 3.84E+09 & 1.22 & 15.3 & -0.23 & 1\\
 31 & 159.63510 & 57.40035 & 0.047 & 18.70 & 15.41 & 7.99E+09 & 4.90E+10 & 1.72 & 10.4 & -1.26 & \\
 32 & 161.48724 & 57.45520 & 0.117 & 19.96 & 17.59 & 7.23E+10 & 2.47E+10 & 1.15 & 13.5 & -0.20 & 1\\
 33 & 160.29099 & 56.93161 & 0.185 & 20.76 & 17.13 & 1.92E+11 & 1.68E+11 & 1.38 & 19.1 & -0.34 & 1\\
 34 & 160.30701 & 57.08246 & 0.046 & 20.35 & 17.05 & 1.01E+10 & 6.62E+09 & 1.22 &  7.5 & -0.21 & 1\\
 35 & 160.31664 & 56.89586 & 0.185 & 20.33 & 17.69 & 3.19E+11 & 1.06E+11 & 1.24 &  ... &  ...  & 1\\
 36 & 159.98523 & 57.40522 & 0.072 & 19.53 & 16.55 & 2.76E+10 & 3.74E+10 & 1.36 &  9.9 & -0.73 & \\
 37 & 160.10915 & 57.43977 & 0.047 & 19.86 & 15.61 & 1.40E+10 & 5.65E+10 & 1.68 & 11.9 & -1.01 & \\
 38 & 160.20963 & 57.39475 & 0.118 & 20.07 & 16.92 & 9.40E+10 & 6.77E+10 & 1.28 & 17.4 & -0.60 & \\
 39 & 159.38356 & 57.38491 & 0.074 & 19.81 & 17.35 & 1.57E+10 & 1.04E+10 & 1.27 &  6.5 & -0.30 & 1\\
 40 & 159.79688 & 57.27135 & 0.102 & 20.55 & 17.00 & 4.61E+10 & 1.09E+11 & 1.39 &  ... &  ...  & \\
 41 & 158.99098 & 57.41671 & 0.102 & 21.03 & 17.28 & 2.46E+10 & 5.96E+10 & 1.48 &  6.5 & -0.49 & \\
 42 & 158.97563 & 58.31007 & 0.155 & 19.80 & 17.31 & 1.18E+11 & 9.13E+10 & 1.36 & 10.2 & -0.58 & \\
 43 & 158.85551 & 58.28600 & 0.114 & 20.14 & 17.22 & 4.76E+10 & 4.82E+10 & 1.35 & 10.5 & -0.60 & \\
 44 & 159.74423 & 58.37608 & 0.068 & 20.00 & 16.58 & 5.44E+10 & 3.52E+10 & 1.39 &  ... &  ...  & \\
 45 & 159.06448 & 57.65416 & 0.113 & 21.24 & 17.12 & 4.79E+10 & 8.44E+10 & 1.50 &  6.4 & -0.54 & 1\\
 46 & 159.02698 & 57.78402 & 0.044 & 19.06 & 17.05 & 1.17E+10 & 7.09E+09 & 1.20 & 14.7 & -0.61 & \\
 47 & 159.22287 & 57.91185 & 0.102 & 19.89 & 17.42 & 5.33E+10 & 1.88E+10 & 1.17 &  7.9 & -0.50 & 1\\
 48 & 159.98817 & 58.65948 & 0.200 & 20.07 & 17.69 & 1.95E+11 & 6.50E+10 & 1.21 &  9.0 & -0.26 & 1\\
 49 & 159.51942 & 58.04882 & 0.091 & 19.10 & 16.82 & 3.47E+10 & 2.77E+10 & 1.18 & 12.7 & -0.72 & \\
 50 & 159.78503 & 58.23109 & 0.073 & 19.45 & 16.02 & 1.65E+10 & 7.10E+10 & 1.76 &  ... &  ...  & \\
 51 & 159.85828 & 57.96801 & 0.138 & 20.21 & 17.13 & 9.76E+10 & 1.14E+11 & 1.38 & 10.0 & -0.46 & \\
 52 & 160.54201 & 58.66098 & 0.031 & 19.33 & 16.26 & 3.82E+09 & 7.10E+09 & 1.43 &  5.4 & -0.80 & 2\\
 53 & 160.57785 & 58.68312 & 0.120 & 19.99 & 16.54 & 4.48E+10 & 1.22E+11 & 1.66 &  5.2 & -0.95 & 2\\
 54 & 160.41264 & 58.58743 & 0.115 & 20.73 & 16.98 & 1.78E+11 & 1.59E+11 & 1.41 & 19.2 & -0.45 & 1\\
 55 & 160.29353 & 58.25641 & 0.121 & 20.38 & 17.41 & 3.86E+10 & 4.01E+10 & 1.38 &  6.6 & -0.60 & \\
 56 & 160.41617 & 58.31722 & 0.072 & 21.58 & 15.80 & 1.11E+10 & 1.36E+11 & 1.90 &  8.2 & -0.56 & \\
 57 & 160.12233 & 58.16783 & 0.073 & 20.88 & 17.02 & 9.20E+09 & 2.62E+10 & 1.46 &  ... &  ...  & \\
 58 & 160.10104 & 58.15516 & 0.072 &  ...  & 16.74 & 3.49E+10 & 6.44E+10 & 1.65 &  ... &  ...  & \\
 59 & 159.89861 & 57.98557 & 0.075 & 20.10 & 16.55 & 2.55E+10 & 5.13E+10 & 1.50 & 10.8 & -0.32 & \\
 60 & 160.51027 & 57.89706 & 0.116 & 20.24 & 17.13 & 3.30E+10 & 6.96E+10 & 1.47 &  8.6 & -0.53 & \\
 61 & 160.95200 & 58.19655 & 0.073 & 20.48 & 16.48 & 2.19E+10 & 5.19E+10 & 1.42 & 10.1 & -0.34 & 1\\
 62 & 160.91280 & 58.04736 & 0.133 & 19.47 & 17.14 & 1.34E+11 & 6.03E+10 & 1.23 & 16.8 & -0.51 & 1\\
 63 & 160.88481 & 57.67214 & 0.046 & 20.10 & 14.50 & 3.71E+09 & 1.00E+00 & 2.03 &  1.3 & -0.68 & 2\\
 64 & 161.00317 & 58.76030 & 0.073 & 20.42 & 16.59 & 8.43E+10 & 9.93E+10 & 1.49 & 20.2 & -0.58 & 1\\
 65 & 161.37666 & 58.20886 & 0.118 & 19.99 & 16.48 & 1.67E+11 & 1.99E+11 & 1.38 & 21.8 & -0.39 & 1\\
 66 & 161.25533 & 57.77575 & 0.113 & 19.93 & 17.42 & 7.82E+10 & 5.31E+10 & 1.30 & 13.1 & -0.36 & \\
 67 & 161.18829 & 58.45495 & 0.031 &  0.00 & 14.46 & 1.34E+10 & 1.00E+00 & 1.50 & 25.6 & -1.39 & \\
 68 & 163.63458 & 57.15902 & 0.068 & 19.95 & 15.94 & 3.79E+10 & 6.40E+10 & 1.46 & 18.0 & -0.28 & \\
 69 & 162.27214 & 57.62591 & 0.076 & 20.23 & 17.22 & 3.28E+10 & 2.91E+10 & 1.41 &  8.9 & -0.26 & 1\\
 70 & 163.17673 & 57.32074 & 0.090 & 20.54 & 17.73 & 3.34E+10 & 1.82E+10 & 1.26 &  5.0 & -0.15 & 1\\
 71 & 163.21991 & 57.13160 & 0.163 & 21.46 & 17.59 & 1.03E+11 & 1.08E+11 & 1.46 &  7.5 & -0.38 & \\
 72 & 163.25565 & 57.09528 & 0.080 & 19.97 & 17.06 & 6.88E+10 & 3.01E+10 & 1.26 & 19.1 & -0.35 & \\
 73 & 163.35274 & 57.20859 & 0.080 & 20.35 & 17.38 & 1.29E+10 & 1.77E+10 & 1.38 &  4.0 & -0.35 & 1\\
 74 & 161.95050 & 57.57723 & 0.118 & 19.88 & 17.01 & 6.98E+10 & 5.49E+10 & 1.24 & 15.2 & -0.63 & \\
 75 & 162.14607 & 58.02331 & 0.176 & 20.28 & 17.28 & 9.75E+10 & 2.14E+11 & 1.48 &  ... & -0.71 & \\
 76 & 162.02142 & 57.81512 & 0.074 & 20.35 & 17.60 & 4.05E+10 & 2.49E+10 & 1.23 &  6.9 & -0.51 & \\
 77 & 162.10524 & 57.66665 & 0.044 & 19.20 & 17.37 & 5.35E+09 & 1.98E+09 & 1.13 &  4.8 & -0.27 & 1\\
 78 & 162.12204 & 57.89890 & 0.074 & 18.71 & 16.21 & 4.10E+10 & 3.08E+10 & 1.28 & 13.7 & -0.87 & 2\\
 79 & 161.25693 & 57.66116 & 0.045 & 19.00 & 16.48 & 7.28E+09 & 8.57E+09 & 1.28 &  4.2 & -0.97 & \\
 80 & 162.07401 & 57.40280 & 0.075 & 18.67 & 16.81 & 2.50E+10 & 2.97E+10 & 1.29 &  8.2 & -0.82 & \\
 81 & 162.04674 & 57.40856 & 0.075 & 19.63 & 17.12 & 1.91E+10 & 1.67E+10 & 1.38 &  6.1 & -0.60 & \\
 82 & 161.03609 & 57.86136 & 0.121 & 20.11 & 17.14 & 6.62E+10 & 5.52E+10 & 1.22 &  6.7 & -0.53 & \\
 83 & 160.77402 & 58.69774 & 0.119 & 19.73 & 17.13 & 9.17E+10 & 4.23E+10 & 1.29 & 16.0 & -0.69 & 1\\
 84 & 162.48495 & 59.50835 & 0.116 & 22.19 & 17.29 & 4.78E+10 & 1.32E+11 & 1.74 &  6.2 & -0.74 & 2\\
 85 & 162.13954 & 59.01889 & 0.063 & 21.31 & 17.36 & 6.96E+09 & 2.07E+10 & 1.59 &  5.1 & -0.42 & \\
 86 & 162.01674 & 58.86004 & 0.085 & 19.25 & 16.24 & 2.86E+10 & 7.62E+10 & 1.55 &  4.8 & -1.01 & \\
 87 & 161.40736 & 58.19447 & 0.116 & 20.37 & 16.79 & 2.55E+10 & 8.54E+10 & 1.49 &  3.8 & -0.71 & 2\\
 88 & 161.38522 & 58.50156 & 0.116 & 20.89 & 17.41 & 3.59E+10 & 4.11E+10 & 1.40 &  5.1 & -0.34 & \\
 89 & 161.55003 & 58.44921 & 0.050 & 19.21 & 15.61 & 7.08E+09 & 4.94E+10 & 1.68 &  6.2 & -1.84 & 2\\
 90 & 162.64168 & 59.37266 & 0.153 & 20.19 & 17.54 & 1.16E+11 & 5.32E+10 & 1.26 & 17.4 & -0.26 & \\
 91 & 162.53705 & 58.92866 & 0.117 & 19.07 & 16.75 & 6.80E+10 & 5.44E+10 & 1.21 &  8.2 & -0.85 & \\
 92 & 162.65512 & 59.09582 & 0.032 & 18.20 & 15.51 & 1.64E+10 & 1.08E+10 & 1.22 & 34.9 & -0.59 & \\
 93 & 162.79474 & 59.07684 & 0.117 & 22.42 & 17.36 & 4.42E+10 & 8.82E+10 & 1.63 &  9.2 & -0.29 & \\
 94 & 161.80573 & 58.17759 & 0.061 & 21.47 & 17.70 & 1.52E+10 & 1.30E+10 & 1.42 &  3.8 & -0.39 & 1\\
 95 & 163.71245 & 58.39082 & 0.115 & 20.79 & 17.39 & 7.16E+10 & 6.38E+10 & 1.33 & 14.5 & -0.25 & 1\\
 96 & 164.74986 & 58.25686 & 0.132 & 20.42 & 17.39 & 6.45E+10 & 6.72E+10 & 1.40 &  9.0 & -0.29 & 2\\
 97 & 164.77032 & 58.32224 & 0.131 & 19.93 & 17.15 & 7.41E+10 & 8.82E+10 & 1.32 &  7.4 & -0.71 & 2\\
 98 & 164.14571 & 58.79676 & 0.050 & 18.18 & 16.31 & 4.28E+10 & 8.34E+09 & 1.13 & 24.8 & -0.56 & 1\\
 99 & 164.33247 & 57.95170 & 0.077 & 18.56 & 16.42 & 7.35E+10 & 1.88E+10 & 1.19 & 20.2 & -0.67 & 1\\
100 & 164.74155 & 58.13369 & 0.032 & 20.26 & 15.39 & 5.45E+09 & 3.32E+10 & 1.66 &  9.2 & -0.95 & 2\\
101 & 164.46204 & 58.23759 & 0.077 & 21.69 & 16.93 & 1.21E+10 & 1.95E+10 & 1.54 & 11.0 & -0.20 & 2\\
\hline
\multicolumn{12}{l}{Notes}\\
\multicolumn{12}{l}{1. Bright sample target; observed with SH module}\\
\multicolumn{12}{l}{2. Problematic calibration of SL2 module}\\
\multicolumn{12}{l}{3. Affected by solar flare}\\
\multicolumn{12}{l}{a. Stellar mass from \citet{Kauffmann03}}\\
\multicolumn{12}{l}{b. $D_n(4000)$ from \citet{Kauffmann03}}\\
\multicolumn{12}{l}{c. Signal-to-noise ratio of SSGSS low resolution
  spectra at 22~\micron\ (see Sect.~\ref{sn})}\\
\multicolumn{12}{l}{d. IRAC 7.8~\micron\ mag minus synthetic
  7.8~\micron\ mag for aperture correction of SL module (see Sect.~\ref{apcor})}\\
 \enddata
\end{deluxetable}

 \clearpage

\section{New {\it Spitzer} Spectroscopy}
\label{observations}

\subsection{Instrument, Observing Mode, and AORs}
\label{mode}

The full SSGSS sample was observed with the IRS on board the {\it
  Spitzer Space Telescope} as a Cycle-3 Legacy Survey.
All spectroscopic observations were performed in spectral mapping mode.  
Observations were divided between 29 Astronomical Observation Requests
(AORs), each consisting of 4 to 8 galaxies observed in cluster
mode. All galaxies within an AOR were observed consecutively, using
guided telescope shifts to move between targets. 

\subsection{Low-Resolution IRS Spectroscopy}

The entire SSGSS sample was observed with the IRS `low-res'
Short-Low (SL) and Long-Low (LL) modules. The SL module spans 5.2 to 
14.5~\micron\  with resolving power
$R=60$--125 and has slit width 3\farcs6--3\farcs7.
The LL module spans 14 to 38~\micron\  with resolving power
$R=57$--126 and has slit width 10\farcs5--10\farcs7. 
Both the SL and LL modules are divided into two sub-slits---the first
and second orders---abbreviated SL1 (7.4-15.4 \micron), SL1 (5.2-8.7 \micron), LL1 (19.5-38 \micron), and LL2 (14.0-21.3 \micron). Telescope shifts position the
source over each order, and so the first and second orders appears on
opposite halves of the detector. A third, `bonus' order from an
additional sub-slit is obtained with the second order.

Four 
offsets in the direction of the slit were used to
produce a 1-D dither pattern, with two exposures at each position.  
This strategy has been shown to increase the quality of output
spectra, in particular for the handling of rogue pixels,
undersampling, and flat-fielding as described in the IRS instrument
support team document 
{\it Obtaining Spectra of Very Faint Sources with the IRS}, and in \citet{Teplitz07}.

For low-res spectroscopy our sensitivity target was 5$\sigma$ per
resolution element for the continuum and PAH band features, after
moderate smoothing for the faintest sources. To achieve this sensitivity, 
our bright-sample exposure times are 8 minutes in both the SL and
LL modules and our faint-sample exposure times are 8 minutes in
the SL and 16 minutes in the LL module.

\subsection{High-Resolution IRS Spectroscopy}
\label{hiresobs}

The 33 galaxies in the SSGSS bright sample were also observed with the
IRS `hi-res' Short-High (SH) module. This module spans 
9.9 to 19.6~\micron\  with $R \approx 600$, and has slit width 4\farcs7.

For the hi-res spectroscopy our
target line flux was $1\times10^{-18}$~W~m$^{-2}$ at 5$\sigma$,
based on the line fluxes for [Ne~II], [Ne~III], [Ar II] and [S III] for 
star-forming galaxies with infrared luminosities comparable to those
in the bright sample. Our hi-res exposure times are 16 minutes, split
over separate exposures.

For each hi-res AOR, eight separate sky spectra with a total of 16 minute exposure times were also taken.

\subsection{Slit Positions}

A crucial aspect of this program is the comparison of MIR
spectral properties to SDSS optical properties and so it was essential to
ensure overlap between the slit and the 3\arcsec\ SDSS fiber. All
slits are wider than the SDSS aperture, and so the SDSS aperture was
centered within each slit with at least 0\farcs5 accuracy. Acquisition
(``peak-up'') imaging at each AOR enabled us to achieve this accuracy,
as descrived in Section~\ref{peakup}.

Slit orientations were not specified and so are random. As the SL and LL
observations are taken consecutively, the relative orientation of
these slits on each galaxy is the same as their relative orientation on
the spacecraft ($85.9^{\circ}$). SH slit positions are randomly oriented with
respect to the low-res positions.

\begin{figure*}[ht!]
\centering
\vspace{-4cm}
\includegraphics*[width=12cm]{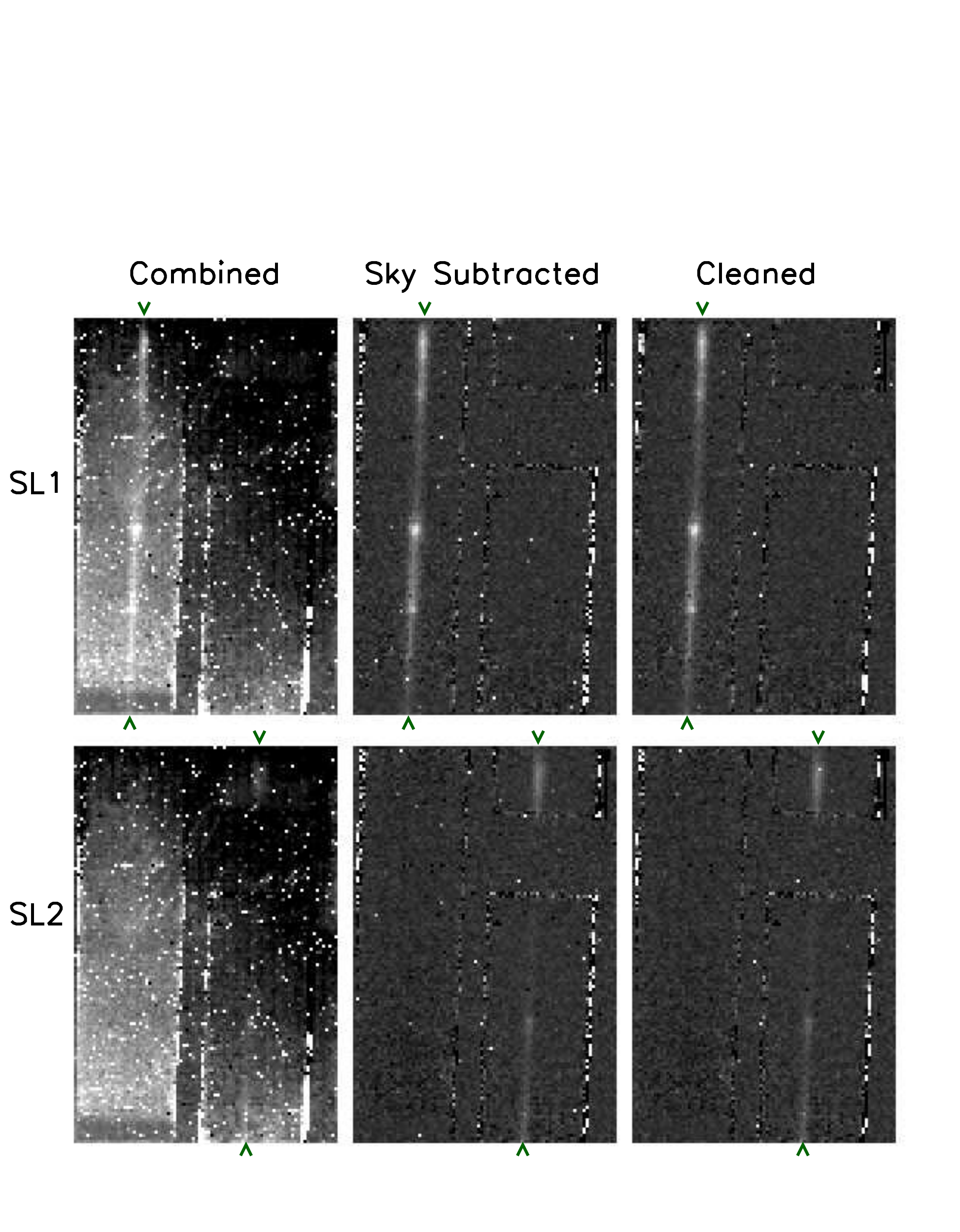}
\caption{Example SL1 ({\it upper}) and SL2 ({\it lower}) frames through the reduction
  process. {\it Left} is the post-pipeline combined frame, {\it middle} is
  the sky-subtracted frame, and {\it right} is the cleaned sky-subtracted
  frame, from which 1-D spectra are extracted. Pointers mark the
  ends of the spectra---to the left of the frames for SL1, and to
  the lower right for SL2. The segments at the upper right
  are the bonus order spectra. 
\label{slreduction}}
\end{figure*}

\subsection{Peak-up Imaging}
\label{peakup}

To achieve the accuracy needed for our slit positioning, we preceeded
the observations in each AOR with a pointing-mode peak-up image. 
The IRS Peak-Up facility uses centroid positioning from an image of
the field in the SL detector array to position the science target in
the slit. In our case, we used bright sources with known offsets to the targets.
Despite the high galactic latitude of our sources, there were
sufficient bright, low-proper-motion sources in each field to 
ensure quality peak-ups. 

In addition, each galaxy was observed with dedicated 4$\times$30 second
peak-up imaging. These peak-ups utilized the blue filter, with spectral coverage of
13.3--18.7~\micron. This $\sim16$~\micron\ coverage gives us photometric
data intermediate to the IRAC 8~\micron\  and MIPS 24~\micron\  imaging.

Figure~\ref{mosaic} (Appendix A) shows the dedicated peak-up images for all galaxies in
the sample, including slit positioning.

\subsection{Solar Flare}

A solar flare during one AOR resulted in very high cosmic ray counts
for five targets. The affected galaxies are SSGSS 20, 22, 23, 24, and 25.

\section{Data Reduction}
\label{reduction}

\subsection{Low-Resolution Spectroscopy}
\label{lowresreduction}

\begin{figure*}[ht!]
\centering
\includegraphics*[width=12cm]{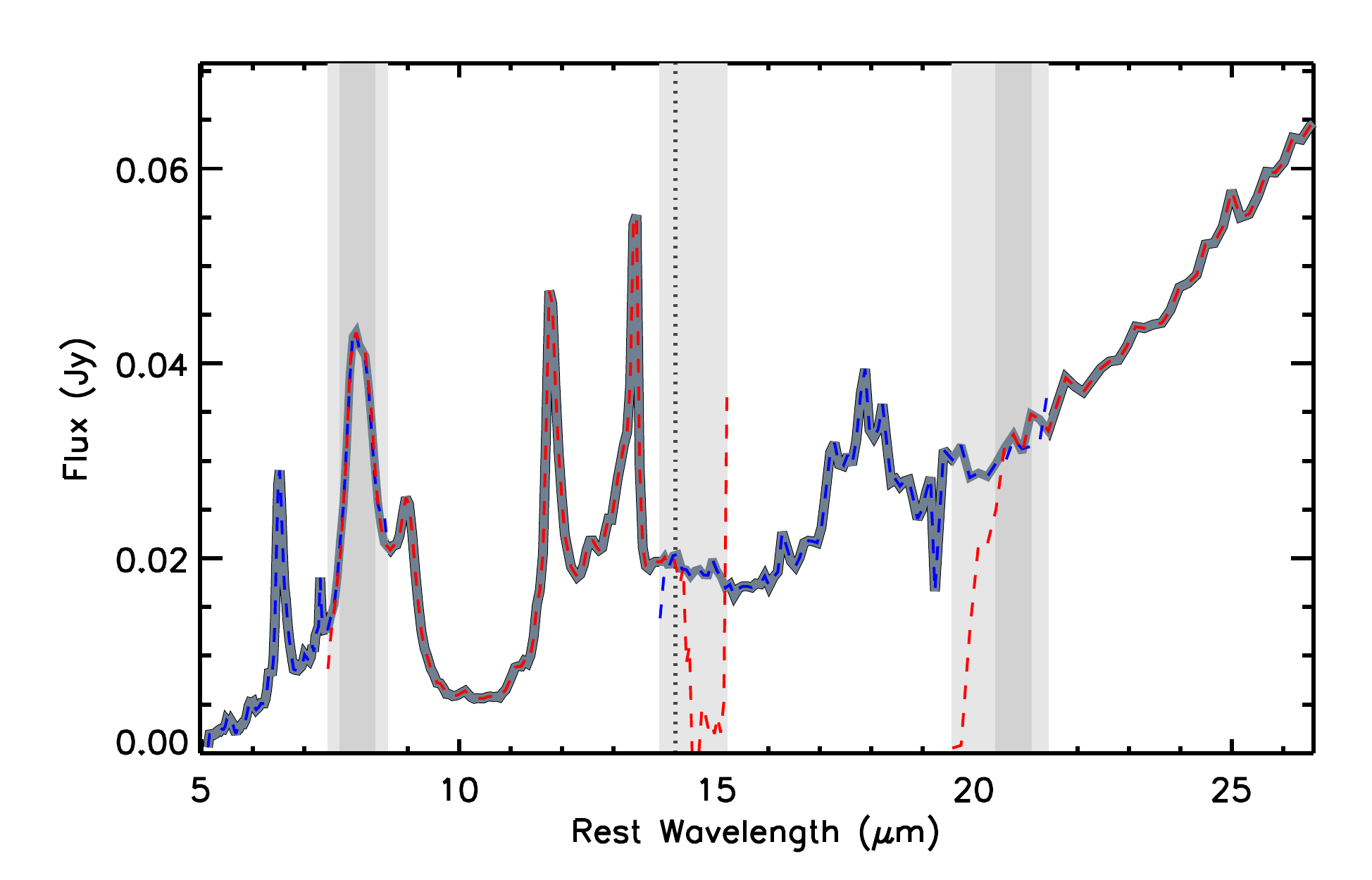}
\caption{Example of method used to stitch the low-res spectra. The
  colored dashed spectra are from individual channels, from
  left to right: SL2 (blue), SL1 (red), LL2 (blue), LL1 (red). The
  dark gray spectrum is stitched from these channels. Dark gray
  shaded areas represent the channel-overlap regions where two
  channels are medium-combined. The light gray shaded areas are the
  channel-overlap regions where the stitched spectrum is set to the
  values of the nearest channel. The default medium-combined region is
  the central 50\% of the overlap region, however this is adjusted
  manually to avoid end-of-channel transmission drops. The dotted line
  is the stitch point for the SL1--LL2 junction, which defaults to the
  middle of the overlap region but in many cases is set manually due to
  transmission drops.
\label{stitch}}
\end{figure*}


Standard IRS processing was performed by the Spitzer Pipeline
version S15.3.0. Pipeline steps included 
ramp fitting, dark subtraction, droop, linearity
correction and distortion correction, flat fielding, masking and
interpolation, and wavelength calibration.


No separate sky images were taken with the low-res modules and so sky 
frames were constructed from combined science frames. Between the
first- and second-order exposures, the source is shifted to
opposite sides of the detector. This results in a sky exposure for the 
order not currently containing the source.
Within each AOR, all such sky exposures were median
combined to produce a master sky frame for each order.

Bad and rogue pixels masks for each exposure were created with
IRSCLEAN (version 1.9), with AGGRESSIVE keyword set to 2, and with
negative pixels also masked.
We also applied the campaign rogue pixel masks released 
by Spitzer. 

Figure~\ref{slreduction} shows example frames for the SL module through
the reduction process, including the initial post-pipeline frame with 
exposures combined, the sky-subtracted frame, and the cleaned frame.


Extraction of 1-D spectra was performed using SPICE version 2.0.1. 
The default setting was used to create a wavelength collapsed
spatial profile, and the center of the profile was determined manually
due to excess failures of the auto-center option. SPICE then performed
spectral extraction along this peak. The point source tune module
applied flux corrections for changes in the wavelength-dependent Point
Spread Function (PSF) width and aperture size, and assumed emission by
a point source. 

\subsubsection{Low-res Combining and Stitching}

The extracted 1-D spectra were combined by weighted mean. First, the
spectra at the four dither positions along the slit were were combined
for each order. Then
the first, second and bonus orders were stitched to produce separate SL and LL
spectra using the following semi-manual process: 
by default, the central 50\% of each overlap region was combined by
weighted mean and the outer quartiles of the overlap region were set to the
nearest order. In select cases the combined region was adjusted to
avoid obvious drops in transmission at the ends of either order. The SL and LL
modules were combined by a clean cut between the spectra, determined
manually to avoid end-of-channel transmission drops. 
Figure~\ref{stitch} illustrates the method used for stitching spectra.


The overlap region of the SL and LL modules is
$\sim14$--$14.5$~\micron, and in this region the slit widths are 
3$\farcs$7 and 10$\farcs$5 for SL and LL respectively. The approximate
LL-to-SL ratio of admitted light from centrally-aligned point source
in the overlap region is $\sim$1.3, based on a simple model of the
{\it Spitzer} PSF \citep{Smith04}, and the overall LL sensitivity 
is a factor of $\sim$5 higher than the SL module in this region  ({\it
  Spitzer Observer's Manual, p.160}). 
While these differences are accounted for by the pipeline reduction
assuming a centrally-aligned point source,  for extended sources we
may expect an offset at their intersection when the spectra are stitched.

In fact no galaxies exhibit noticeable discontinuities across the modules in
this region, indicating that deviations from the point-source
approximation in the overlap region are small, even for the SL module
at $\sim$14--15\micron.  No re-scaling was necessary or performed.


\subsection{High-Resolution Spectroscopy}
\label{hiresreduction}

\begin{figure*}[ht!]
\centering
\includegraphics*[width=12cm]{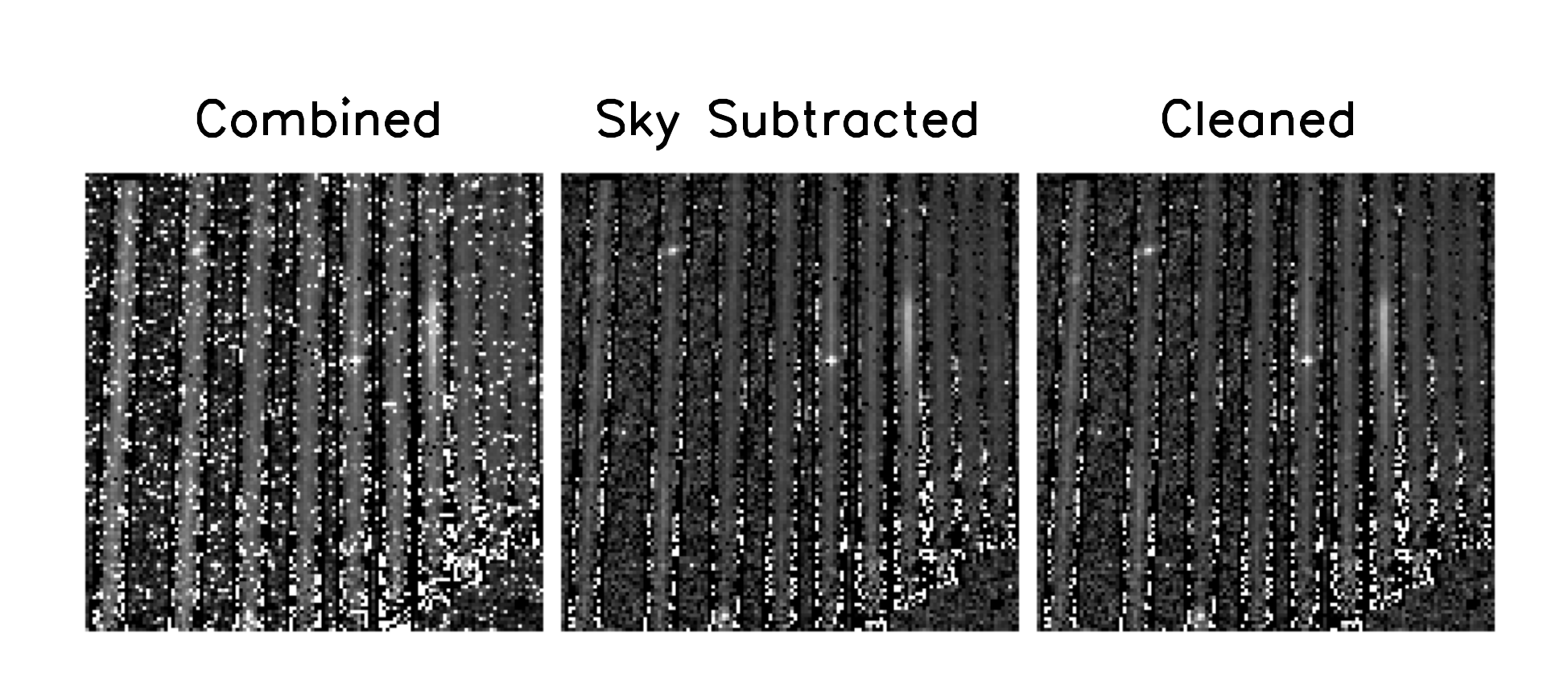}
\caption{Example SH frames through the reduction
  process. {\it Left} is the post-pipeline combined frame, {\it middle} is
  the sky-subtracted frame, and {\it right} is the cleaned sky-subtracted
  frame, from which 1-D spectra are extracted.
\label{shreduction}}
\end{figure*}


As with the low-resolution spectroscopy (Sect. \ref{lowresreduction}),
the Spitzer Pipeline version S15.3.0 was used to perform standard IRS
processing. Unlike the low-res data, for the hi-res observations we
combined the eight separate exposures of each galaxy before
extraction. This improved the signal to allow higher quality extractions.


As noted in Section~\ref{hiresobs}, separate hi-res sky frames were
taken for each AOR, median combined from eight 16~minute exposures.
Masking followed the same method used for the low-res
spectroscopy, utilizing IRSCLEAN (version 1.9) and the Spitzer
campaign rogue pixel mask. 
Likewise, extraction was performed as for the low-res data, using SPICE v.2.0.1.
The hi-res orders were stitched into a single spectrum using a similar
method to the low-res spectra: order-overlap regions were combined by
taking the weighted mean within the central 50\% or the overlap
regions, while for the outer quartiles the nearest order was used.

Figure~\ref{shreduction} shows example frames for the SH module through
the reduction process, including the initial post-pipeline frame with 
exposures combined, the sky-subtracted frame, and the cleaned frame.

\subsection{Achieved Sensitivity}
\label{sn}

\begin{figure}[ht!]
\centering
\includegraphics*[width=7cm]{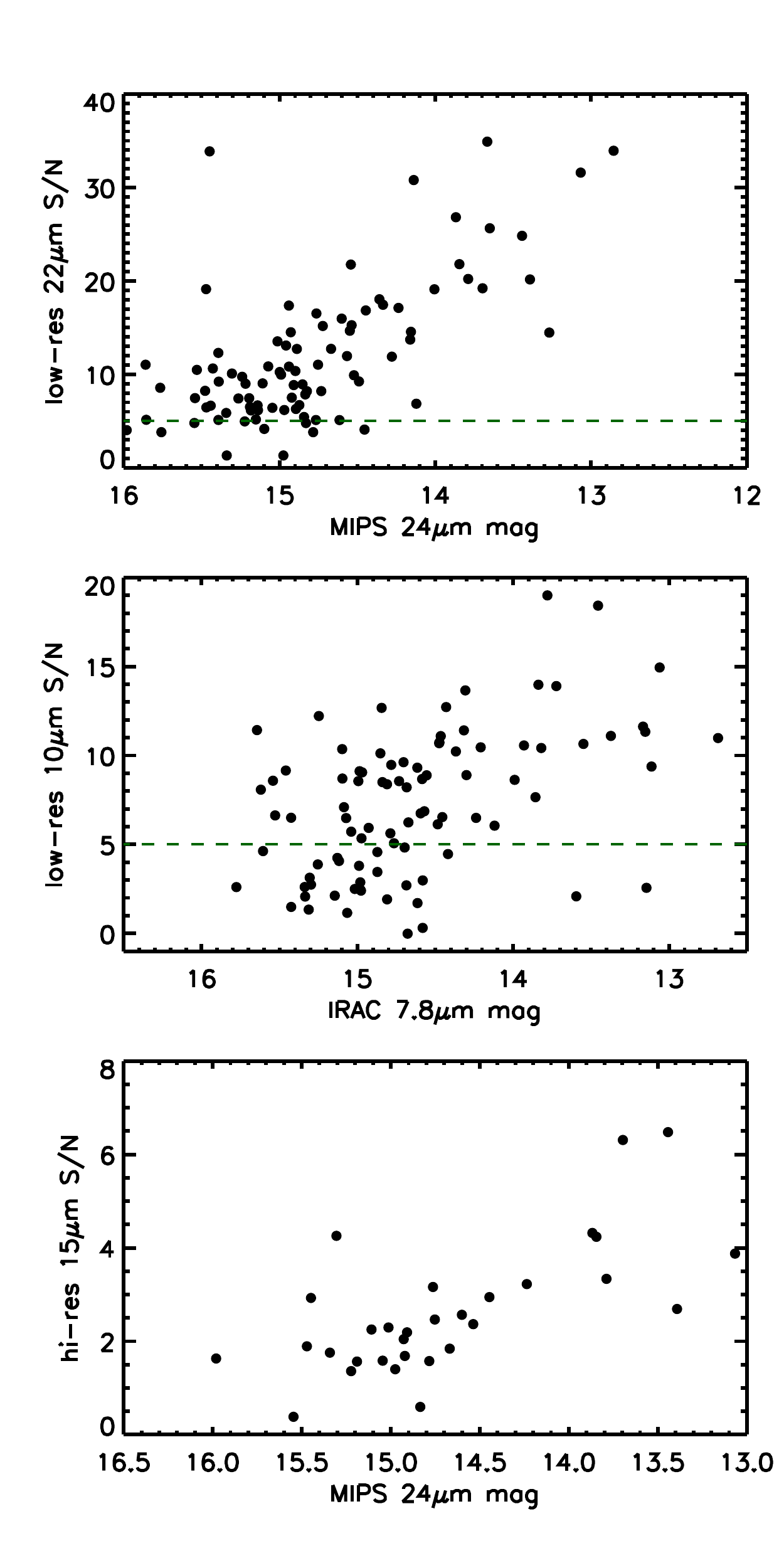}
\caption{Signal-to-noise ratios of low-res spectra at 10~\micron\
  (upper) and 22~\micron\ (middle) versus IRAC 7.8~\micron\ and
  MIPS 24~\micron\ mag respectively, and of the hi-res spectra at
  15~\micron\ versus MIPS 24~\micron\ mag. Dashed lines indicate
  target S/N. 
\label{contsn}}
\end{figure}

\begin{figure}[ht!]
\centering
\includegraphics*[width=7cm]{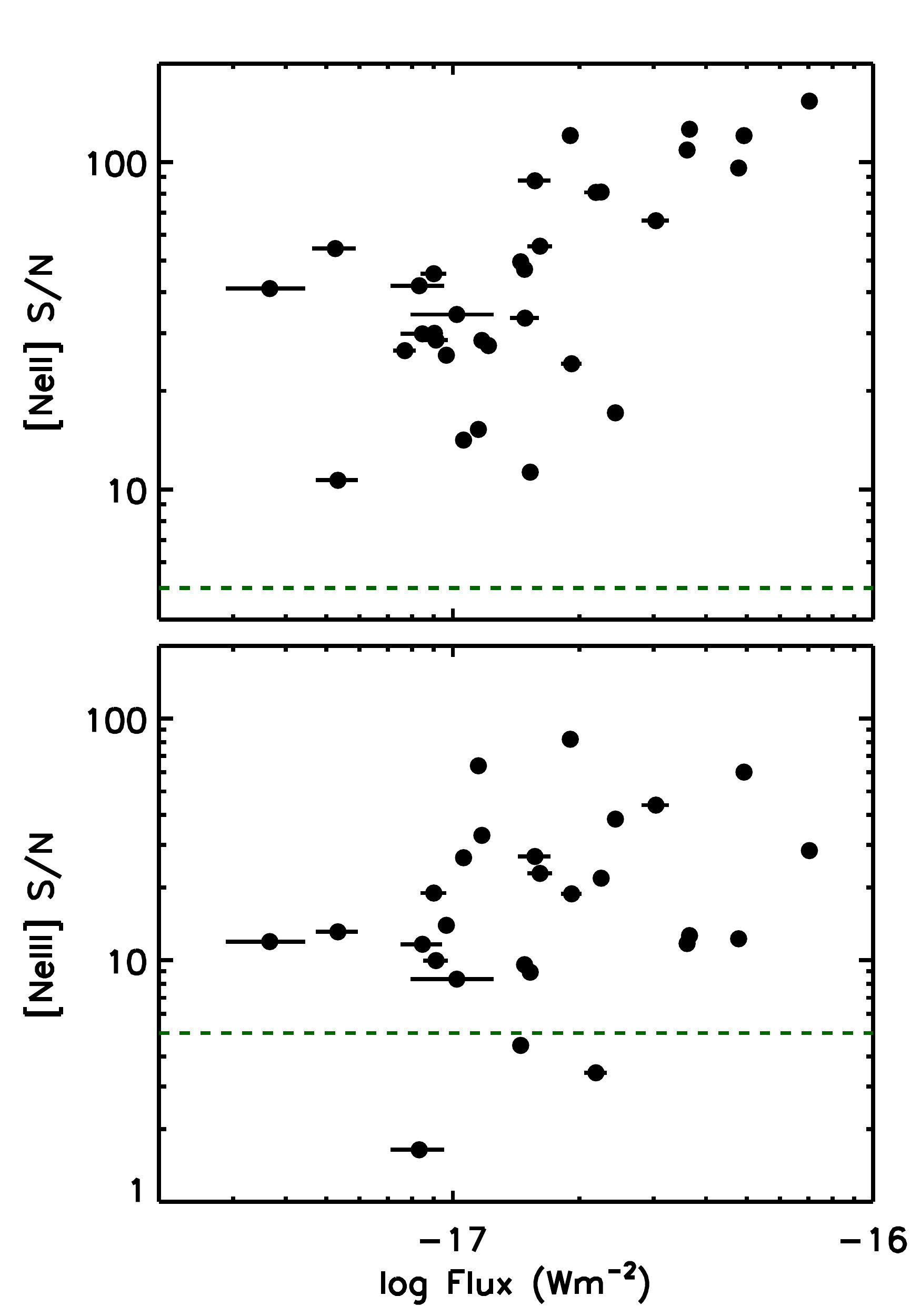}
\caption{Signal-to-noise ratios of the [Ne~II] (upper) and [Ne~III]
  (lower) emission lines from the hi-res spectra versus the
  PAHFIT-measured fluxes. Dashed lines indicate
  target S/N. 
\label{linesn}}
\end{figure}

We estimate the continuum sensitivity achieved by determining the
signal-to-noise ratio (S/N) in the processed spectra at relatively
featureless regions of the spectra.
We compare the mean flux to the corresponding noise in the regions
9--10~\micron\ and 20--24~\micron\ for the low-res spectra and
14.7--15.3~\micron\ for the hi-res spectra. The noise is taken to be the standard 
deviation in these regions after subraction of a power law fit. 
Figure~\ref{contsn} shows the continuum S/N for low-res and
hi-res spectra versus IRAC 7.8~\micron\ and MIPS 24~\micron\ mag.
At $\sim22$~\micron\ nearly all sources exceed the target
S/N $=$ 5, with a minimum S/N of 4, and have the expected tight relationship with 24~\micron\ mag. 
At $\sim10$~\micron\ the variable silicate absorption results in a number of
sources with S/N $<$ 5. In this case there is greater variability of S/N with photometric
magnitude due to this absorption and the overlap of more PAH bands
with the IRAC 7.8~\micron\ band. 22~\micron\ S/N is provided in Table~\ref{sampletable}.

We estimate the S/N for the [Ne~II]$_{12.8}$ and [Ne~III]$_{15.6}$ emission lines in the
hi-res spectra by comparing the integrated flux of the lines to the
standard deviation in the relatively smooth, blueward regions at
12.25--12.75~\micron\ and 15--15.5~\micron. Figure~\ref{linesn} shows
line S/N versus the line flux (see Sect.~\ref{linestrengths}). 
[Ne~II] and [Ne~III] are detected in all galaxies with S/N $>$ 5. In most cases, both [Ne~II] and [Ne~III] 
are detected with S/N $>$ 10. There is strong scatter in the trend between line
flux and S/N due to the weakness of the trend between continuum and
line luminosity, and this is especially pronounced in [Ne~III], where the
line emission may be linked more to an AGN components than
to star formation.

The flux uncertainties reported by SPICE, which are included in the
released data, have been checked for the low-res modules by comparing
the multiple dithered observations for each galaxy. These uncertainties
were found to be accurate to within a factor of 1.5.

\subsection{Aperture Effects}
\label{apcor}

\begin{figure}[ht!]
\includegraphics*[width=8cm]{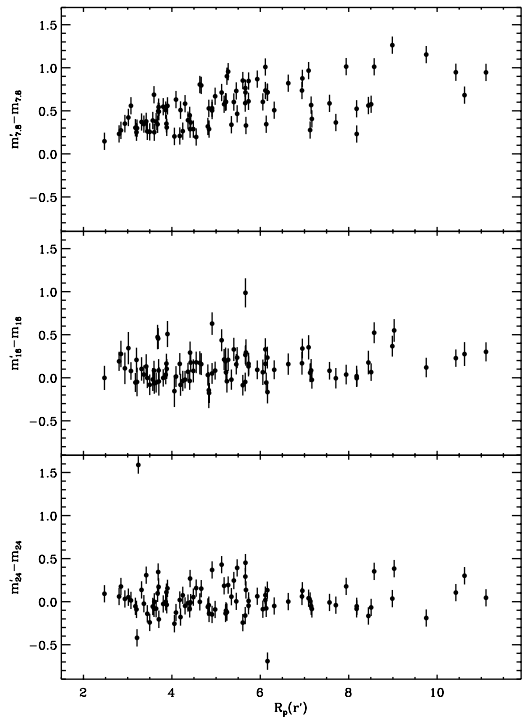}
\caption{Synthetic minus SWIRE photometry as a function of r-band petrosian radius
  8~\micron, 16~\micron, 24~\micron. The trend between this magnitude
  difference and radius observed in the 8~\micron\ band suggests slit
  losses in the SL module, but not at longer wavelengths.
\label{fig:c1}}
\end{figure}

\begin{figure}[ht!]
\vspace{1cm}
\includegraphics*[width=8cm]{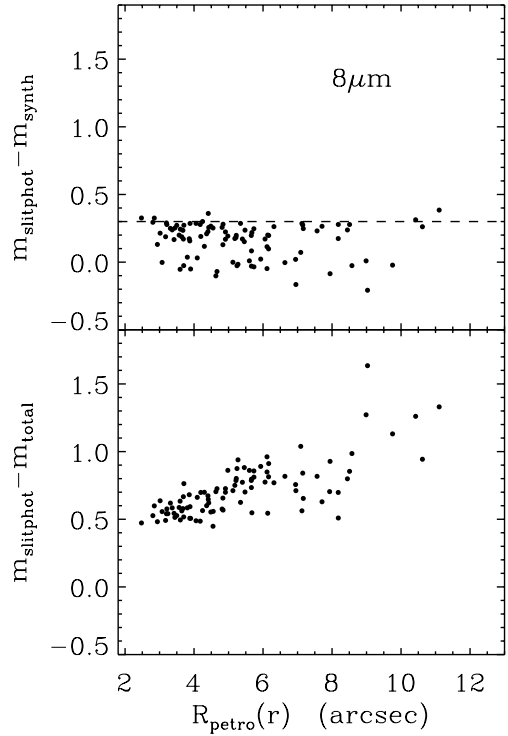}
\caption{SL slit aperture magnitude minus synthesized and circular
  aperture magnitudes as a function of r band petrosian radius.
\label{fig:c2}}
\end{figure}

An increasing fraction of the wavelength-dependent \emph{Spitzer} PSF 
falls outside a given IRS slit with increasing wavelength. The
standard spectral extraction with SPICE corrects for this slit loss
assuming emission by a point source. At $>$~16~\micron\ the
combination of the larger Spitzer PSF and the wider IRS 
slits means that all galaxies are well-approximated as point
sources (Fig.~\ref{mosaic}, Appendix A).
However, a number of the SSGSS galaxies have sufficient
spatial extent that they are not well-approximated by point sources in  
the narrower SL and SH modules. 

We investigate slit loss in the low resolution
modules by comparing synthetic photometry from the reduced, calibrated
low-res spectra to large-aperture photometry obtained from SWIRE imaging
(IRAC channel 4 7.8~\micron\ and MIPS 24~\micron), and from the SSGSS
16~\micron\ peak-up imaging.
Synthetic photometry is obtained by integrating the spectra using
k\_correct (v4.14; Blanton03b).
Figure \ref{fig:c1} shows the difference in
these magnitudes as a function of the $r'$ band Petrosian radius ($R_P$),
with aperture photometry performed in 7\arcsec\ and 12\arcsec\ radius
apertures for 8~\micron\ and 24~\micron\ respectively, as
described in \citet{Johnson07}. 
Differences may be attributed to a number of sources (e.g., source
confusion in the imaging aperture photometry, relative calibration
errors) but any trend with the angular size of the galaxy is
suggestive of slit losses. No such trend is observed at 16~\micron\ and
24~\micron. However, at 8~\micron\ a significant trend is apparent,
indicating slit losses in the SL module.


To isolate the effect of slit loss from any calibration uncertainties, 
we perform photometry of the 8~\micron\ IRAC images within the region
corresponding to the SL slit aperture (a 3\farcs65$\times$9\farcs6 rectangle),
and compare this to the synthetic and circular aperture magnitudes.
Figure \ref{fig:c2} (upper) shows the difference between slit aperture
and synthetic magnitudes. A 0.3 mag aperture correction is applied by 
SPICE at 8~\micron\ and so the synthetic magnitudes are higher than the 
derived slit aperture magnitudes. Taking this offset into account, the
slit aperture samples more flux than the extracted spectra, with an
average offset of $\sim 0.2$~mag and similar scatter. The magnitude
difference is independent of galaxy size, indicating that the trend
observed in Figure \ref{fig:c1} does not result from calibration
effects. 
Figure \ref{fig:c2} (lower) shows slit aperture versus 
circular aperture magnitudes, revealing a similar trend to that
observed with the 8~\micron\ synthetic magnitudes. The maximum
difference is 1-1.5 mag for $R_p(r') \gtrsim 9\arcsec$, 

The apparent slit losses in the SL module mean that care is necessary
when considering the shape of the 
($\lesssim 16~\micron$) Spectra Energy Distribution (SED) of large galaxies. The difference between the
synthetic and photometric magnitudes at 8~\micron\ should provide a
reasonable estimate of the aperture correction. These corrections are
provided Table~\ref{sampletable}.

\section{SSGSS Data}
\label{data}

\begin{figure*}
\centering
\vspace{-2cm}
\includegraphics*[width=15cm]{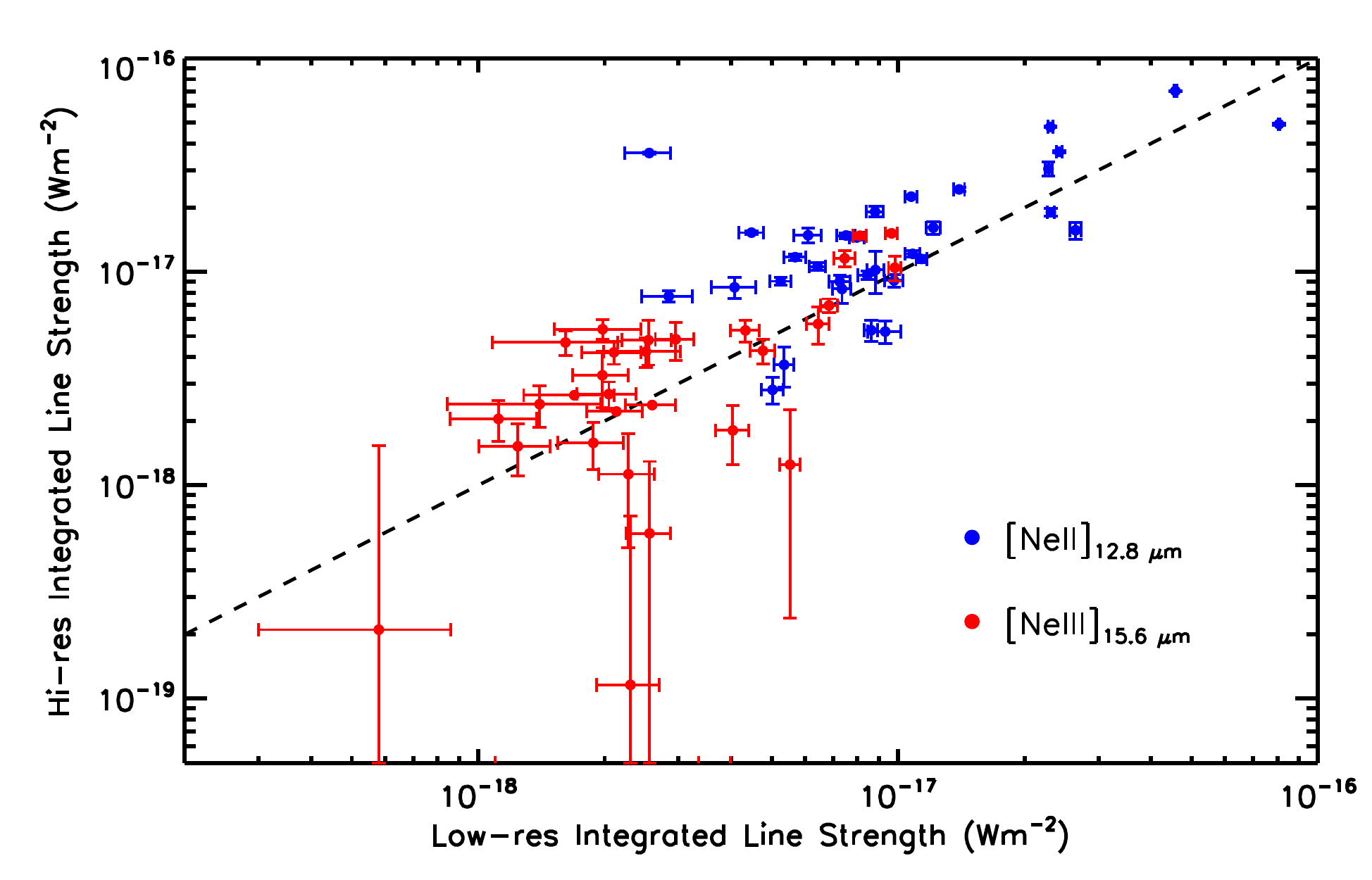}
\caption{Comparison of the [Ne~II]$_{12.8}$~\micron\ (blue) and
  [Ne~III]$_{15.6}$~\micron\ (red) emission line integrated strengths as
  measured with PAHFIT using the low-res and hi-res spectra. Error
  bars represent the 1-$\sigma$ uncertainties in the fits.
\label{lohiline}}
\end{figure*}

Figure~\ref{mosaic} shows multi-wavelength imaging for all 101
galaxies in the SSGSS sample, including GALEX FUV and NUV, SDSS
multi-colour, r, H, IRAC 3.6~\micron\ and 8~\micron, 16~\micron\ from SSGSS
peakup images, and MIPS 25~\micron\ and 70~\micron. Overlaid are the
slit positions for the SL, LL, and SH modules.

\begin{figure*}[!ht]
\centering
\includegraphics*[width=12cm]{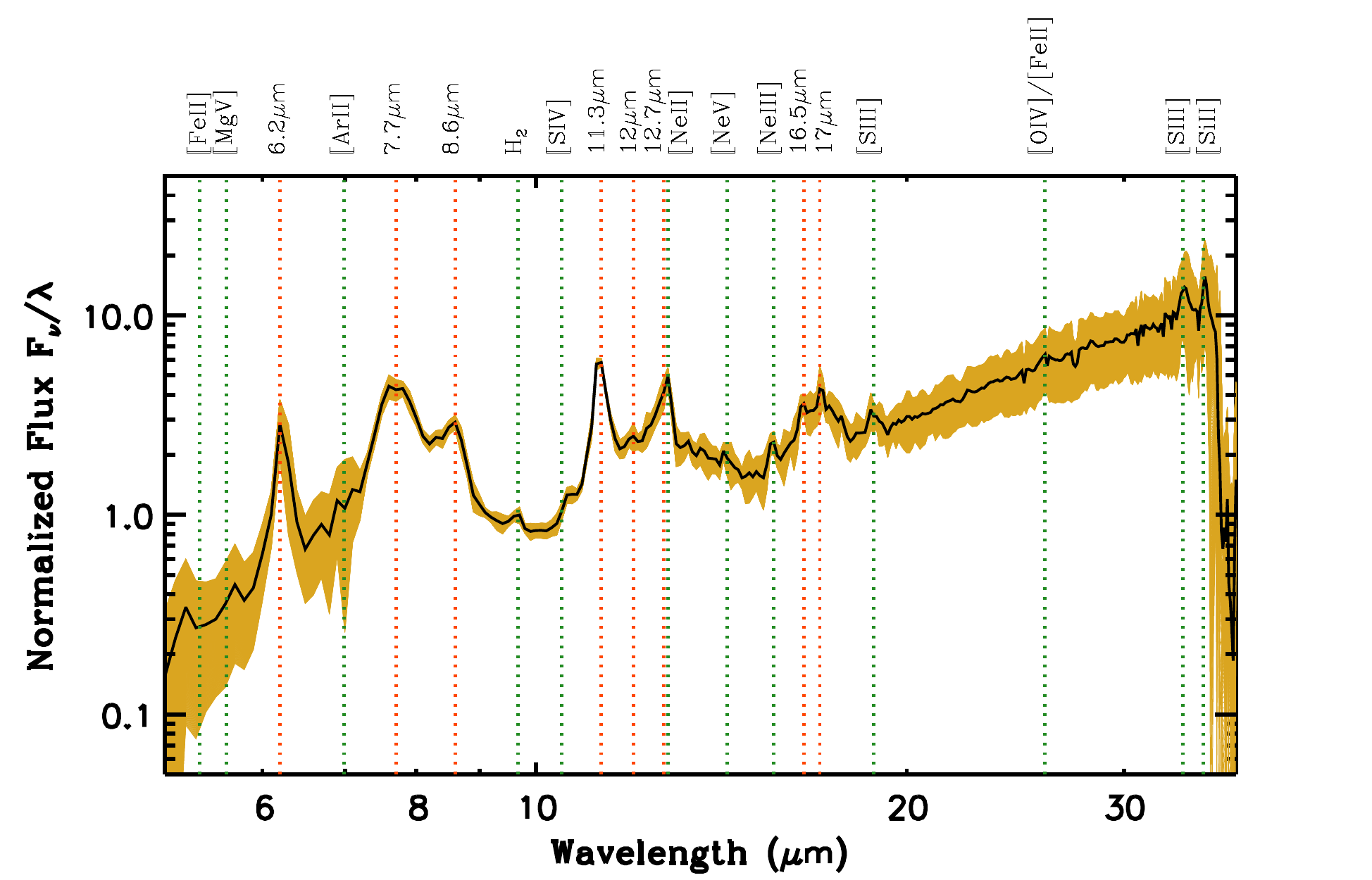}
\includegraphics*[width=12cm]{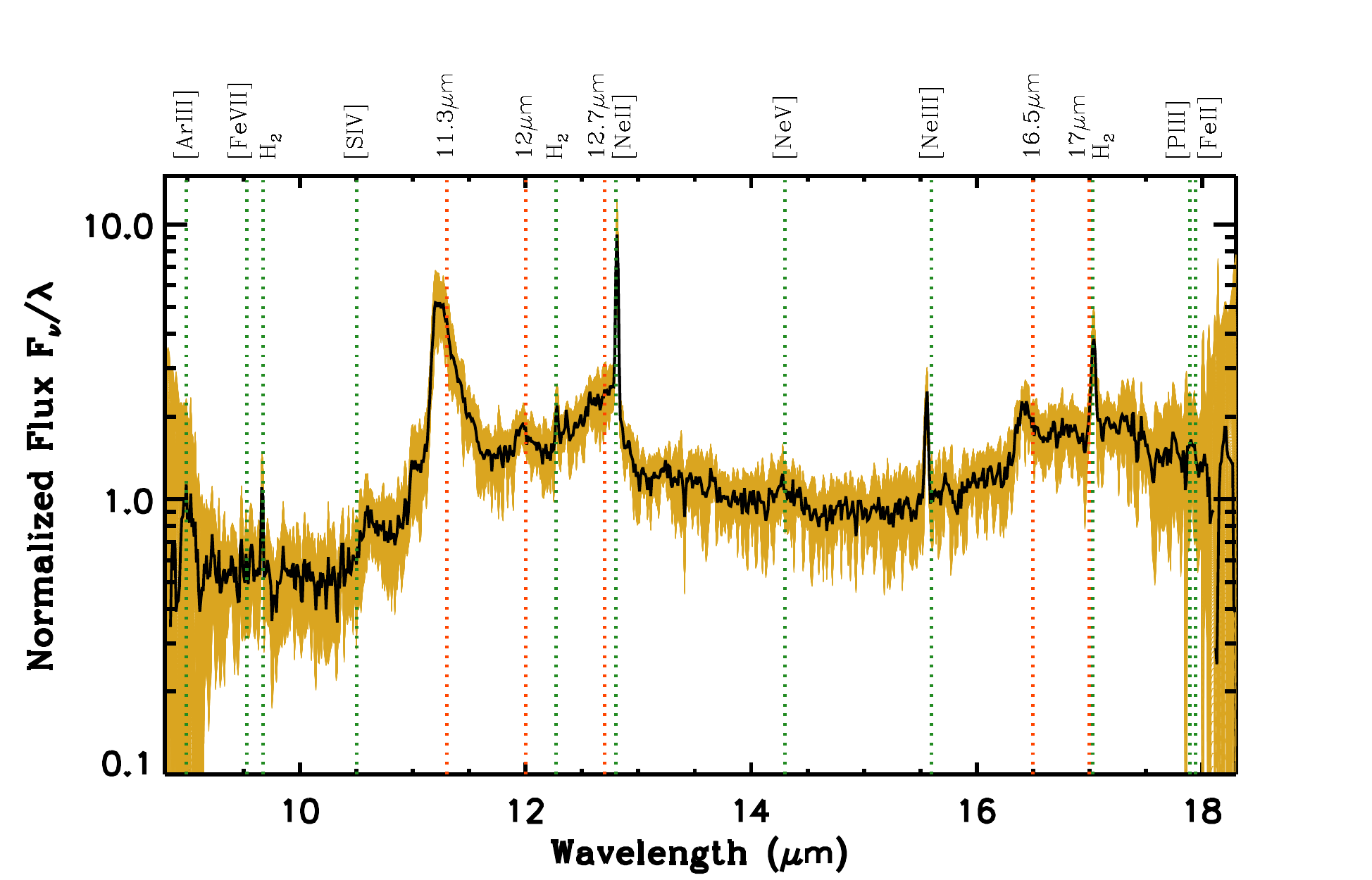}
\caption{{\it Upper panel:} composite SL and LL low-res spectrum of the 100 galaxies of
  the current SSGSS data release (solid black line). Low-res spectra are
  normalized between 
  20--24~\micron. The shaded region represents 1-sigma scatter of the
  normalized spectra about the mean spectrum.
  {\it Lower panel:} composite SH hi-res spectrum of the 33 galaxies of
  the SSGSS bright sample. Hi-res spectra are normalized between 14--15~\micron. 
\label{comp}}
\end{figure*}

\subsection{Low-Resolution Spectra}

\begin{figure*}[!ht]
\centering
\vspace{-2cm}
\includegraphics*[width=12cm]{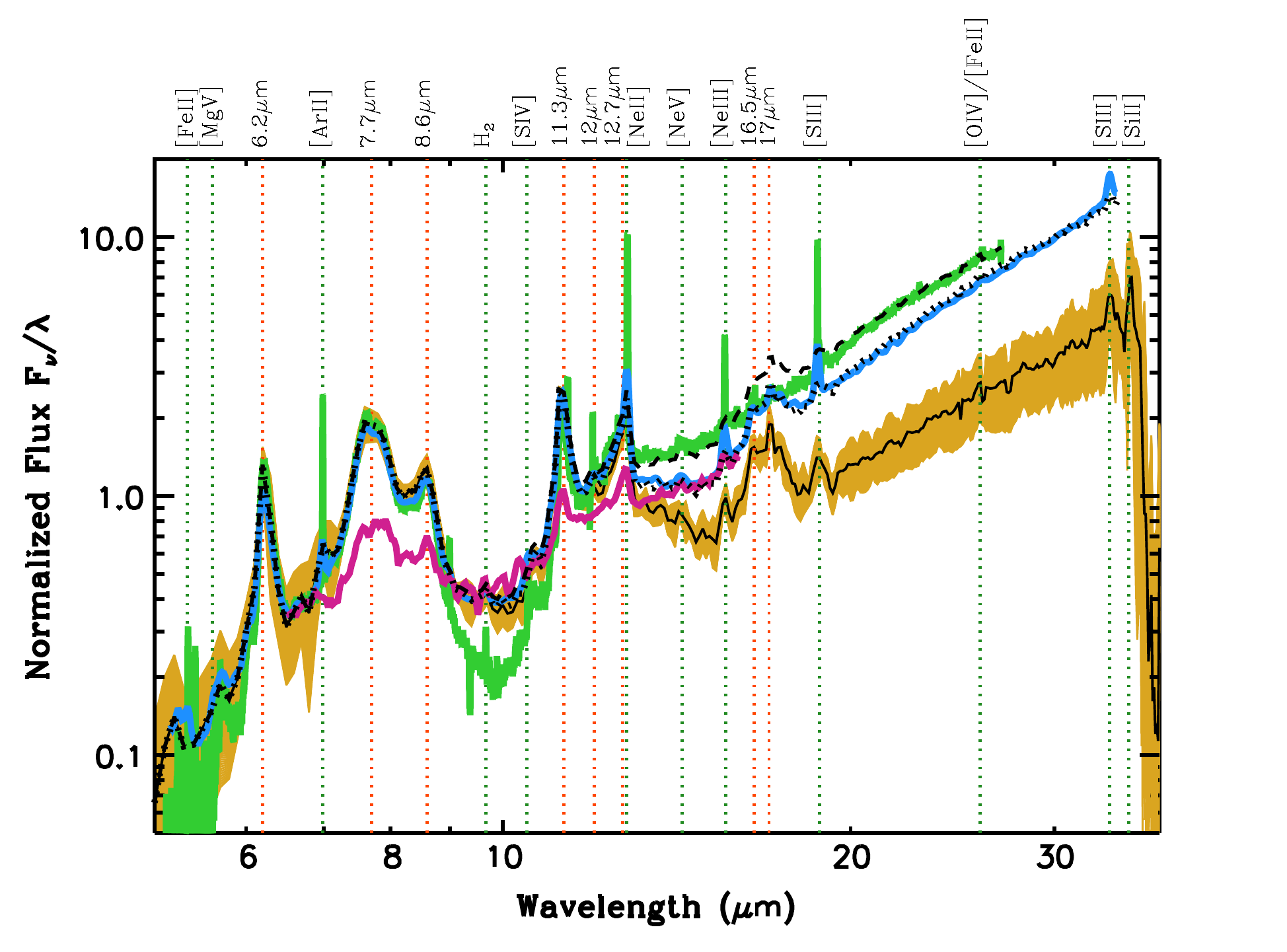}
\includegraphics*[width=12cm]{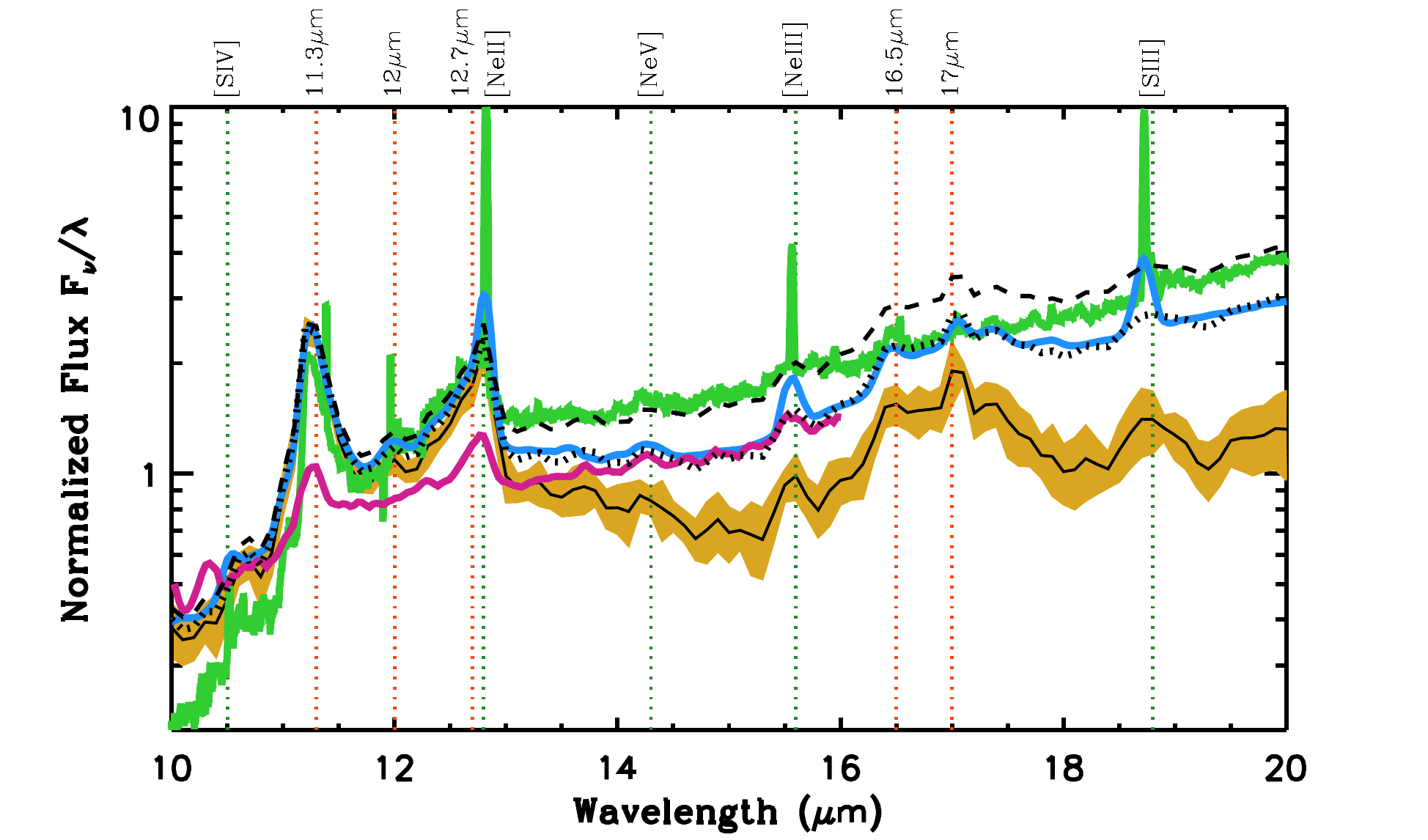}
\caption{{\it Top:} Composite low-res spectrum of the SSGSS sample
  (solid black line), with the spectrum of M82 (green; \citealt{Sturm}),
  the composite spectrum of local starbursts (blue;
  \citealt{Brandl}), and the
  composite of $z\sim1$ ULIRGS (magenta; \citealt{Dasyra}). 
  The shaded region represents the 1-sigma scatter of the
  normalized spectra about the mean spectrum.
  The dotted black lines show best fits
  to M82 and starburst composites of a model consisting of the SSGSS composite plus
  a dual-temperature blackbody. The fits indicate that starburst spectra
  are well represented by that of a normal, star-forming
  galaxy plus an enhanced thermal dust component at $T\lesssim120K$.
{\it Bottom:} 10--20\micron\ region of top panel.
\label{ssgss_starburst}}
\end{figure*}

Figure~\ref{lowresspectra} (Appendix B) shows the reduced, combined, and stitched
SL and LL spectra for the 100 galaxies in the current data release. For each
galaxy the SSGSS catalog number, BPT type \citep{BPT,Kewley,KauffmannAGN03}, redshift, stellar mass,
NUV-R color, and $D_n(4000)$ are given.

The relatively low sensitivity of the 5.25--7.58~\micron\ SL2 module,
coupled with the relative faintness of many galaxies in this range, results in 
very low SL2 signal in 10 sources. This spectral region should be used
with caution in these sources, which are noted in Table~\ref{sampletable}.

For one galaxy, SSGSS 56, the SL--LL overlap region falls directly on the
redshifted 12\micron\ PAH complex and the [Ne~II]$_{12.8}$ line. 
End-of-order transmission drops in both the SL and LL modules affect
these features, and so both are lost in this spectrum.

\subsection{High-Resolution Spectra}

Figure~\ref{hiresspectra} (Appendix B) shows the reduced, combined, and stitched
SH spectra for the 33 galaxies in the SSGSS bright sample. For each
galaxy the SSGSS catalog number, BPT type, redshift, stellar mass,
NUV-R color, and $D_n(4000)$ are given.

\subsection{Emission Line Strengths}
\label{linestrengths}

We measure emission line strengths by integrating line regions after
subtraction of the continuum. Here, the continuum refers to all
non-emission line features, including PAH features and the thermal
continuum. 
The continuum is determined using the PAHFIT code of
\citet{Smith07}, which fits a physically-motivated model to each
spectrum using $\chi^2$ minimization. We describe our use of PAHFIT in
detail in \citet{odowd09} (OD09). When applied to the hi-res spectra, we
remove emission line fitting from the process and mask the emission
line regions. We are not concerned with accurate measurement of
continuum features---only that we find a reasonable extrapolation of
the continuum in the region of the emission lines.

The emission line strengths measured by
PAHFIT from the hi-res spectroscopy are highly reliable, as the
continuum near most lines is reasonably smooth, allowing robust
subtraction. These line strengths compare closely
to those measured using a simple linear fit to the continuum
surrounding each line. In the case of the [Ne~II]$_{12.8}$ line, PAHFIT
is significantly more reliable than cruder continuum-fitting methods
because of the close proximity of this line to the peak of the
12.7~\micron\ PAH feature.

Figure~\ref{lohiline} compares [Ne~II]$_{12.8}$ and [Ne~III]$_{15.6}$
lines strengths between the low- and hi-res spectra. Although there is
general agreement, on average the low-res spectroscopy shows weaker
line strengths by $\sim$20--50\%, with similar scatter compared to the hi-res spectra. In four
galaxies, PAHFIT claims detection of [Ne~III] lines in the low-res
spectra where they are undetected at high resolution. The
uncertainties reported by PAHFIT for the low-res [Ne~II] lines are 
significantly underestimated, suggesting that there is more degeneracy 
between this line and the 12.6~\micron PAH complex than estimated by 
in the fit.

Table~\ref{hilinetable} gives emission line strengths measured from the
hi-res spectra. Line strengths from the low-res spectra can be found
on the SSGSS website (see Sect.~\ref{cproducts}).

\subsection{Composite Spectra}
\label{composite}

\begin{deluxetable}{cccccc}
\tabletypesize{\small}
\tablewidth{0pt}
\tablecaption{
\label{hilinetable}}
\tablehead{
\colhead{SSGSS} & \colhead{[S IV]$_{10.5 \micron}$} &
\colhead{[Ne~II]$_{12.8 \micron}$} & \colhead{[Ne~III]$_{15.6 \micron}$}
& \colhead{H$_2$ (9.7~\micron)} & \colhead{H$_2$ (12.2~\micron)}
}
\startdata
1  &  1.71 $\pm$  0.59 & 11.50 $\pm$  0.41 &  1.25 $\pm$  1.01 &        ...        &  1.64 $\pm$  0.55 \\
 2  &  1.89 $\pm$  1.28 &  3.67 $\pm$  0.79 &  4.79 $\pm$  1.12 &  1.54 $\pm$  0.62 &  0.97 $\pm$  0.39 \\
 11 &  0.84 $\pm$  0.91 &  2.80 $\pm$  0.40 &  0.21 $\pm$  1.32 &  2.82 $\pm$  0.62 &  1.44 $\pm$  0.49 \\
 12 &  3.58 $\pm$  0.57 & 36.55 $\pm$  0.42 &  4.20 $\pm$  0.51 &  3.52 $\pm$  0.89 &  4.90 $\pm$  0.77 \\
 14 &  0.77 $\pm$  0.48 & 12.15 $\pm$  0.43 &  2.22 $\pm$  0.00 &  3.88 $\pm$  0.82 &  4.28 $\pm$  1.21 \\
 15 &        ...        &  9.65 $\pm$  0.43 &        ...        &  3.30 $\pm$  0.78 &  1.40 $\pm$  0.98 \\
 16 &  3.63 $\pm$  0.47 & 14.50 $\pm$  0.38 &  1.52 $\pm$  0.41 &  1.13 $\pm$  0.78 &  4.70 $\pm$  0.77 \\
 17 &  4.27 $\pm$  1.25 & 49.26 $\pm$  0.43 & 11.58 $\pm$  1.04 & 10.85 $\pm$  0.62 &  9.81 $\pm$  0.49 \\
 27 &  6.60 $\pm$  0.51 & 70.46 $\pm$  0.39 & 14.78 $\pm$  0.47 &  3.99 $\pm$  0.91 & 13.77 $\pm$  0.82 \\
 28 &        ...        &  8.47 $\pm$  0.96 &  4.67 $\pm$  0.63 &  6.14 $\pm$  0.78 &  2.45 $\pm$  0.33 \\
 30 &  2.57 $\pm$  0.84 &  5.33 $\pm$  0.61 &  4.84 $\pm$  0.98 &  4.45 $\pm$  0.71 &  1.24 $\pm$  0.47 \\
 32 &  2.34 $\pm$  0.65 & 10.22 $\pm$  2.29 &  5.32 $\pm$  0.63 &  2.67 $\pm$  0.49 &  2.74 $\pm$  0.52 \\
 33 &  5.15 $\pm$  1.13 &  8.32 $\pm$  1.21 &  2.64 $\pm$  0.00 &  4.22 $\pm$  0.53 &        ...        \\
 34 &  3.28 $\pm$  0.80 & 10.60 $\pm$  0.42 &  3.28 $\pm$  0.97 &  1.16 $\pm$  0.73 &  2.55 $\pm$  0.44 \\
 35 &        ...        & 9.66 $\pm$ 0.43 &           ...               & 3.35 $\pm$ 0.78 & 1.39 $\pm$ 0.98 \\
 39 &        ...        &  9.04 $\pm$  0.38 &  1.58 $\pm$  0.39 &        ...        &        ...        \\
 45 &        ...        & 19.16 $\pm$  1.09 &  1.13 $\pm$  0.62 &  1.29 $\pm$  0.54 &  1.26 $\pm$  0.37 \\
 47 &        ...        & 16.12 $\pm$  1.09 &  2.40 $\pm$  0.53 &  1.71 $\pm$  0.52 &  2.49 $\pm$  0.33 \\
 48 &        ...        &  9.01 $\pm$  0.64 &  2.38 $\pm$  0.00 &  1.03 $\pm$  1.16 &  1.91 $\pm$  0.48 \\
 54 &  4.60 $\pm$  0.96 & 15.67 $\pm$  1.42 &  1.81 $\pm$  0.56 &  4.46 $\pm$  0.52 &  4.61 $\pm$  0.50 \\
 61 &  3.51 $\pm$  0.58 & 36.07 $\pm$  0.45 &  4.27 $\pm$  0.55 &  3.62 $\pm$  0.87 &  4.64 $\pm$  0.76 \\
 62 &  3.93 $\pm$  0.71 & 24.36 $\pm$  0.53 &        ...        &  4.31 $\pm$  0.97 &  5.32 $\pm$  0.81 \\
 64 &  3.84 $\pm$  0.57 & 47.83 $\pm$  0.46 &  6.96 $\pm$  0.51 &  3.98 $\pm$  0.73 &  7.26 $\pm$  0.95 \\
 65 &  2.54 $\pm$  0.70 & 30.40 $\pm$  2.29 &  5.39 $\pm$  0.57 &  4.46 $\pm$  0.58 &  4.16 $\pm$  0.42 \\
 69 &        ...        & 14.81 $\pm$  0.45 &  2.05 $\pm$  0.44 &  1.03 $\pm$  1.17 &  2.54 $\pm$  0.82 \\
 70 &  1.08 $\pm$  0.53 &  9.12 $\pm$  0.63 &  2.68 $\pm$  0.37 &        ...        &  1.45 $\pm$  0.43 \\
 73 &  1.33 $\pm$  0.58 &  7.69 $\pm$  0.48 &        ...        &  4.64 $\pm$  0.95 &  1.54 $\pm$  0.53 \\
 77 &  4.99 $\pm$  1.44 & 11.72 $\pm$  0.41 &  5.70 $\pm$  1.14 &        ...        &  4.74 $\pm$  0.38 \\
 83 &        ...        &  5.25 $\pm$  0.63 &  4.23 $\pm$  0.68 &  5.20 $\pm$  0.57 &  2.59 $\pm$  0.50 \\
 94 &  1.85 $\pm$  0.65 & 15.28 $\pm$  0.33 &  0.60 $\pm$  0.70 &  2.10 $\pm$  0.64 &  2.59 $\pm$  0.50 \\
 95 &        ...        & 14.85 $\pm$  1.16 &  0.12 $\pm$  0.60 &  2.65 $\pm$  0.66 &  2.17 $\pm$  0.43 \\
 98 &  4.34 $\pm$  1.33 & 19.02 $\pm$  0.80 & 10.46 $\pm$  1.35 &  5.25 $\pm$  0.72 &  7.06 $\pm$  0.64 \\
 99 &  4.71 $\pm$  0.54 & 22.53 $\pm$  0.41 & 15.16 $\pm$  0.39 &  2.39 $\pm$  1.26 &        ...        \\
\hline
\multicolumn{6}{c}{Line fluxes in $10^{-18}$~Wm$^{-2}$}\\
\enddata
\end{deluxetable}

Figure~\ref{comp} (upper) shows the composite of the low-res SL and LL
spectra for the entire SSGSS sample, normalized between 9--11~\micron.
Primary PAH features are prominent, including a pronounced 17~\micron\
complex. Several emission lines are also distinct.

Figure~\ref{comp} (lower) shows the composite of the SH spectra for
the SSGSS bright sample, normalized between 14--15~\micron.
PAH features have significantly more detailed line profiles than in
the low-res spectra, including
clear resolution of the 16.5~\micron\ feature from the 17~\micron\ complex
and interesting structure in both the 11.3~\micron\ and 12.7~\micron\
features. In addition, the [Ne~III]$_{15.6\micron}$ line is well-resolved from the 
12.7~\micron\ feature.

\section{Results and Discussion}

\subsection{Comparison to Starburst Spectra}

The SSGSS sample spans the range of physical properties of
`normal' star-forming galaxies, as described in Section~\ref{sample}. We investigate the
difference between the composite SSGSS spectra and galaxy spectra of
previous IR surveys which have typically been dominated by more
extreme starburst galaxies. 

Figure~\ref{ssgss_starburst} shows the low-res SSGSS composite spectrum
(normalized between 9--11~\micron\ for composite construction) overplotted with an ISO spectrum of M82 \citep{Sturm} and 
the {\it Spitzer} local starburst composite of \citet{Brandl}. These spectra are
normalized according to the results of the fits described below. 
Also plotted is the $0.8<z<1.2$ LIRG composite of \citet{Dasyra}, 
normalized at the blue end of this spectrum, 6.5-6.8~\micron. 

M82 and the local starburst composite have similar EWs
for PAH features at 11.3~\micron\ and blue-ward. However the EW
of the 12.7~\micron\ feature is lower in M82, and the EW of the 17~\micron\
feature is lower in both. These starburst galaxies appear to have
stronger long wavelength continuum emission than the galaxies of the SSGSS sample.
Note that the MIR spectral shape of M82 is within the range seen for the \citet{Brandl} starbursts.
The composite for $z\sim2$ ULIRGs shows significantly reduced EWs for
all PAH bands, and a thermal dust component slope
consistent with that of the local starburst template, however the
short wavelength range of the ULIRG composite makes this difficult to establish from these
data. 

To determine whether the difference in spectral shape between local
starbursts and SSGSS galaxies is consistent
with a difference in small-grain dust temperature distribution, 
we fit the M82 and local starburst composite spectra with a model consisting
of the SSGSS low-res composite spectrum plus a dual-temperature
blackbody dust component, in the case of M82 masking a 10~\micron\
window around the 9.7~\micron\ silicate absorption feature. 
Both are well fit with the addition of a 120K component plus a single
cooler component at around 80K,
and are poorly fit with components warmer than 130K.
The lower EWs of the 12.7~\micron\ and 17~\micron\ features are 
reproduced in the fit to the starburst composite, while only the
12.7~\micron\ feature is accurately reproduced in M82; the 17~\micron\
feature has lower EW than can be explained by a heightened
continuum. In OD09 it was seen that the 17~\micron\ feature
is relatively weak compared to the 7.7~\micron\ feature in galaxies
undergoing more intense star formation. This is consistent with our
observation of the weaker feature in M82. 
The results of these fits are also shown in Figure
\ref{ssgss_starburst}. The relative normalization of all spectra in
this figure are determined by the best fits.

If we assume that the emission between 20~\micron\ and
30~\micron\ in the SSGSS composite arises solely from a 120K thermal
dust continuum component, then we estimate the additional flux from this
component as a factor of two higher in the local starburst composite
and a factor of eight higher in M82 compared to the SSGSS average.
Although the additional continuum light apparent at the longer
wavelengths of the starburst spectra is likely to be due to dust at a
range of temperatures,
this simple model shows that the difference in spectra is
consistent with starburst galaxies having a higher proportion of
emission from $\lesssim$~120K dust than more quiescent galaxies.

\subsection{Comparison of Absorption and Continuum Properties}
\label{absorb}

\begin{figure*}[!ht]
\centering
\includegraphics*[width=15cm]{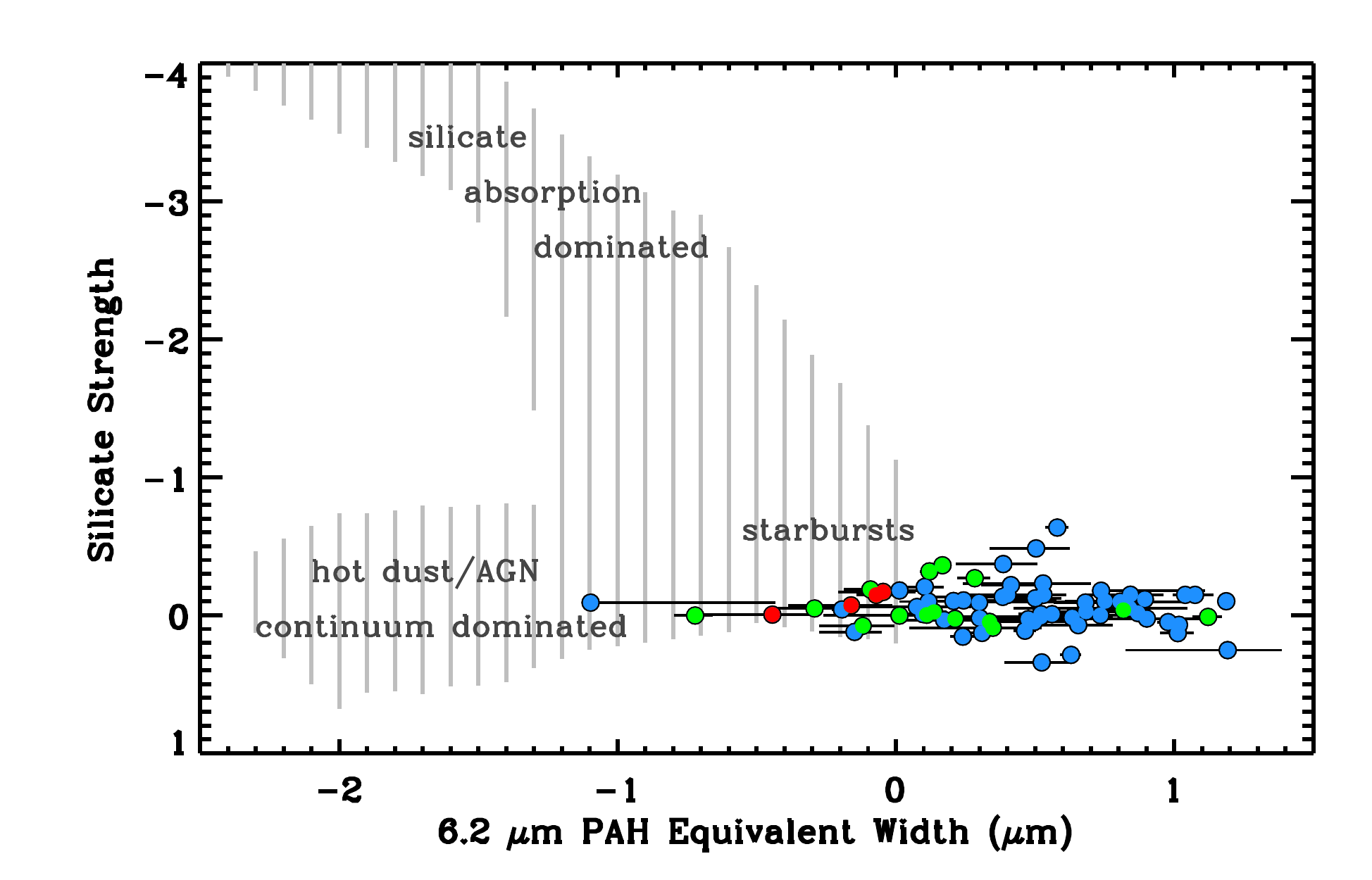}
\caption{Equivalent width of the 6.2~\micron\ PAH feature versus a
  silicate absorption strength, defined as the log of the ratio of the
  minimum in the 9.7~\micron\ absorption trough to the interpolated
  value at this point from a linear fit to the 4.5--5~\micron\ and
14.5--15~\micron\ regions. 
SSGSS galaxies are colour-coded according to BPT designation
\citep{Kewley,KauffmannAGN03}; blue points are star-forming galaxies, green points are
composite sources, and red points are AGN.
 The shaded region shows the positions on
this plot of LIRGs, ULIRGS, starbursts and AGN from
\citet{Spoon07}. These more extreme galaxies tend to have stronger
silicate absorption and brighter continua than the more quiescent
star-forming galaxies of the SSGSS sample.
\label{spoonplot}}
\end{figure*}

The shape of the MIR spectra of the most extreme IR emitters 
tends to be dominated by either a bright continuum --- from either AGN
or hot dust --- or by heavy
silicate absorption centered at $\sim9.7$~\micron. \citet{Spoon07} (S07)
showed that this gives rise to a bimodality in the space of  6.2~\micron\ EW
versus silicate absorption strength for ULIRGs and other IR-luminous
galaxies and for AGN, from analysis of sources observed with {\it Spitzer} and {\it ISO}.

Figure~\ref{spoonplot} shows the location of the SSGSS galaxies in
this space, with the regions occupied by the S07 sources shaded.
We calculate our indicator of absorption strength similarly to the method used by S07
for starburst galaxies. We take the ratio of the 9.5--10.5~\micron\
minimum to the ``unabsorbed'' flux at this point, the latter determined by
interpolation using a linear fit between 4.5--5~\micron\ and
14.5--15~\micron\ regions, which are relatively unaffected by PAH
emission or silicate absorption. 6.2~\micron\ PAH EW is calculated using PAHFIT
to measure the feature strength, as described in \citet{Treyer10}
(hereafter T10). This
method differs from that used by S07, who determine the continuum, and
hence the EW, by fitting a spline between the edges of the
6.2~\micron\ feature. 

Methods for EW determination based on estimation of the
continuum from the boundaries of the 6.2~\micron\ feature result in saturation at widths greater than
$\sim$0.5--1~\micron. This is due to the fact that, at higher line
strengths, the continuum estimation includes increasing amounts of
light from the 6.2~\micron\ feature and the neighboring
7.7~\micron\ PAH complex in the continuum estimation. At EW
$\gtrsim 0.5$--1~\micron\ the flux at the points typically used to estimate
the continuum is in fact dominated by PAH emission.
\citet{Sargsyan09} report 6.2~\micron\ EWs for SSGSS galaxies,
finding $EW < 1$~\micron\ for the entire sample. They
calculate EW using a continuum fit anchored at 5.5~\micron\ and
6.9~\micron. This approach, like the spline method, is vulnerable to
underestimates and saturation at high EWs due to the presence of
significant PAH flux in the regions used for continuum estimates.
T10 presents an analysis of this effect, and a comparison with the 
\citet{Sargsyan09} results.
As many SSGSS galaxies have strong 6.2~\micron\ features, it is
important that their EWs be measured using full continuum+PAH profile fitting.

In Figure~\ref{spoonplot} we can see that SSGSS galaxies have silicate
absorption indicators consistent with the low absorption locus. A
number of our sources are consistent with the starburst galaxies in
S07's analysis, but much of our sample extends to significantly higher
EWs, in part because of the EW saturation issue. Sources with AGN
components by BPT designation \citep{BPT,Kewley,KauffmannAGN03} have brighter continua,
and hence lower EWs, than purely Star-Forming (SF) galaxies. These EWs may
also have been reduce by destruction of PAH by the AGN.

\subsection{PAH Contribution to the IR Luminosity}
\label{SF}

\begin{figure*}[!ht]
\centering
\includegraphics*[width=17cm]{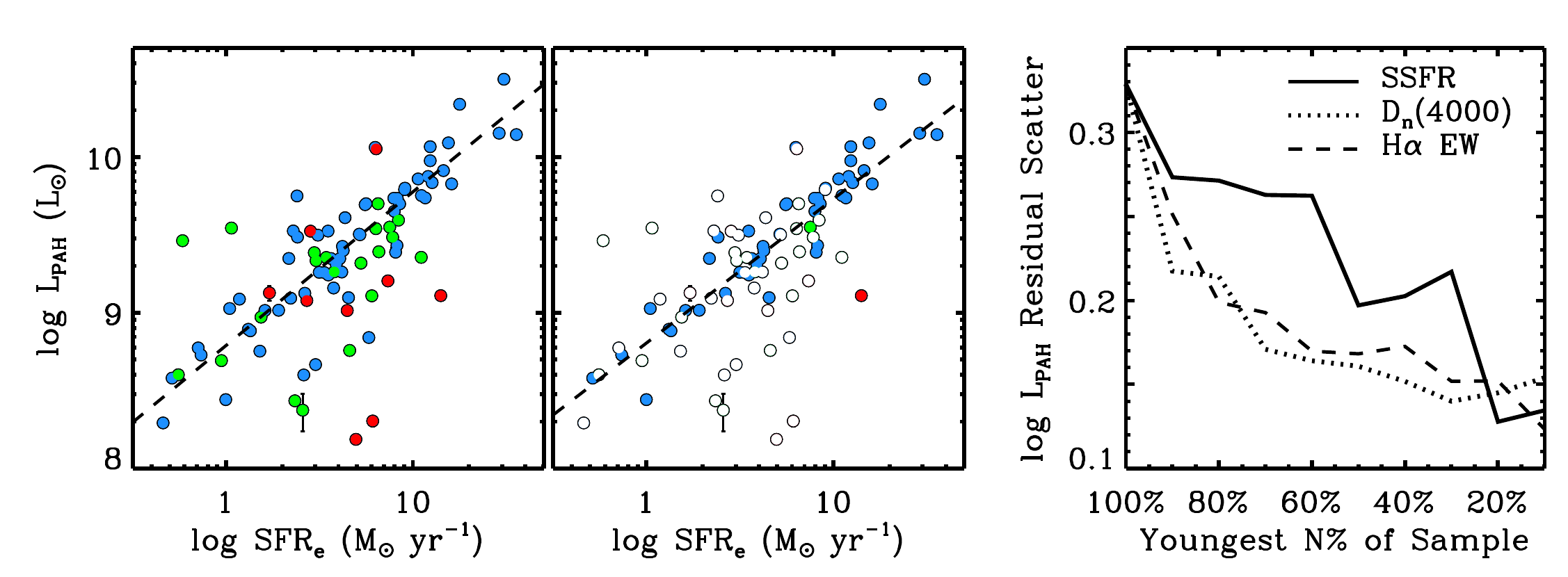}
\caption{{\it Left:} total PAH luminosity versus SFR$_e$. {\it Middle:}
 as with the right panel, but, only galaxies with 
 SSFR $< 1\times10^{-10}$~yr$^{-1}$ are filled, revealing an 
 especially tight trend for galaxies dominated by young
 populations. Dashed lines show the best regression fits to
 star-forming galaxies (left) and the filled circles (middle).
{\it Right:} standard deviation in the fractional residuals in
infrared luminosity after subtraction of the best linear fit to the plot of 
$L_{PAH}$ $SFR_e$. For each point in this plot, the fit was made to the
subset of the sample with specific star formation indicator (SSFR,
$D_n(4000)$, or H$\alpha$~EW) below the given value. The
trend indicates that, while young populations exhibit a very
tight relationship between PAH or total IR luminosity and SFR, this relationship
becomes increasingly scattered as populations age.
\label{LPAHvsSFR}}
\end{figure*}

\begin{figure*}[!ht]
\centering
\includegraphics*[width=15cm]{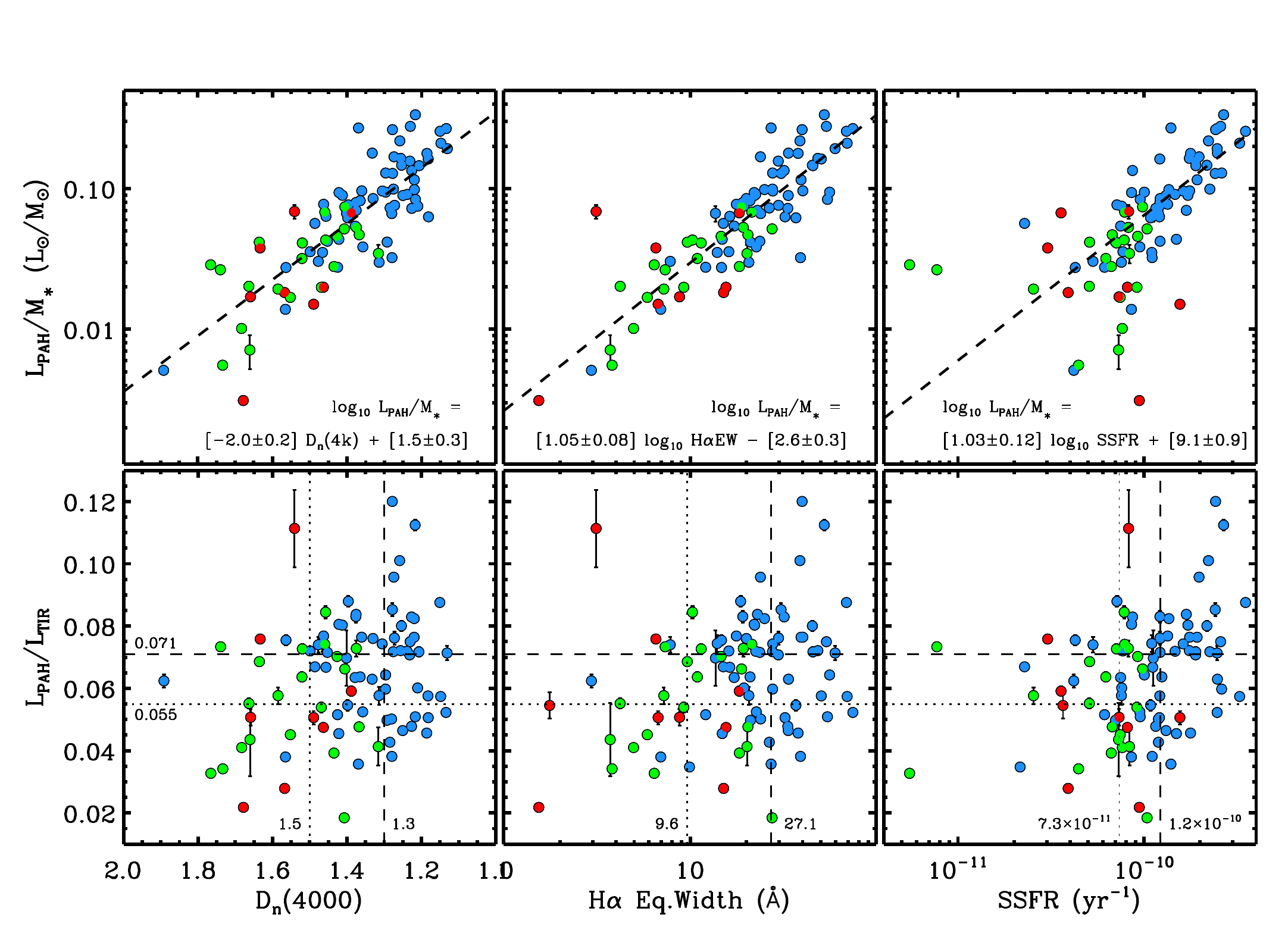}
\caption{{\it Upper:} $L_{PAH}/M_*$ versus $D_n(4000)$ ({\it left}), H$\alpha$~EW
  ({\it middle}), and SSFR ({\it right}).
Galaxies are colour-coded according to BPT type as in
Figure~\ref{spoonplot}. AGNs and composite sources sit on the same loci as SF galaxies, and
have systematically older stellar populations. Dashed lines show
regression fits for the SF population (blue points), with AGNs/composites excluded. 
{\it Lower:} As above, but for $L_{PAH}/L_{TIR}$. Dashed and dotted
lines show the median values for SF galaxies and AGNs/composites
respectively. 
\label{LPAHvsSFR_specific}}
\end{figure*}

For star-forming galaxies, total infrared luminosity ($L_{TIR}$) 
is an good proxy for the rate of current star formation, with much of the starlight in dusty starburst 
environments being reprocessed into the IR. SSGSS
has already shown that PAH emission is also strongly tied to a
galaxy's star formation history; the relative intensities of PAH bands trace specific
star formation 
indicators (OD09), while the luminosities of individual PAH
bands correlate strongly with total SFR (T10). 

Understanding this latter relation, and in general the relative strength of the PAH
contribution to galaxies' MIR emission, is important for modeling the
IR SEDs of higher redshift galaxies where the PAH bands cannot be
so clearly resolved, and for SED modeling in large photometric surveys such as
SWIRE \citep{Lonsdale03}. T10 demonstrated the near-linearity of the
relation between the intensity of individual PAH features and both $L_{TIR}$
and SFR. For the purpose of model fitting to MIR data
in which individual PAH features are not well resolved, it is also valuable
to determine the link between total PAH luminosity ($L_{PAH}$) and
SFR and $L_{TIR}$. 

We measure $L_{PAH}$ using PAHFIT
(\citealt{Smith07}; see Sect.~\ref{linestrengths}), summing the 
integrated strengths of all PAH bands between, and including, the
6.2~\micron\ to 17~\micron\ features. The relatively weak 3.6~\micron\
feature falls outside our spectral range and is not included in our
determination of $L_{PAH}$. We correct $L_{PAH}$ for aperture effects, 
comparing SWIRE and SSGSS peak-up photometry to synthetic photometry
(see Sect.~\ref{apcor}),
and interpolating corrections to PAH peak wavelengths. 
$L_{TIR}$ is determined by fitting \citet{DraineLi} model SEDs to the
{\it Spitzer} (IRAC, IRS blue peak up, and MIPS) photometry, and then
integrating between 3 and 1100~\micron.~
SFR is from \citet{Brinchmann04}, who fit the models of
\citet{CharlotLonghetti01} to strong emission 
lines of galaxies in the SDSS spectroscopic sample, using a Kroupa
IMF \citep{Kroupa2001} and the \citet{CharlotFall2000} dust
model. We use the aperture-corrected median of the derived
SFR likelihood distributions, designated $SFR_e$.

Figure~\ref{LPAHvsSFR} (left) shows $L_{PAH}$ versus
$SFR_e$. These quantities are strongly correlated for the full SSGSS
sample, however we are interested in calibrating $L_{PAH}$ to star
formation, and so we perform regression analysis for SF galaxies
only:

\begin{eqnarray}
log_{10}~L_{PAH}(L_{\odot})=\nonumber\\
{ }[0.99\pm0.05]~log_{10}~SFR_e(M_{\odot}yr^{-1}) + [8.8 \pm 0.1]
\end{eqnarray}

The slope consistent with that presented in T10 for the 7.7~\micron\ complex.
As also found by T10, galaxies with AGN components generally fall on the same locus as
purely star-forming galaxies, although have significantly greater
scatter. This is primarily due to the fact that SFR is determined with
much lower accuracy for the more evolved populations in which
BPT-designated AGN are exclusively found.

This is further explored in Figure~\ref{LPAHvsSFR} (middle), which
highlights the youngest 50\% of the SSGSS sample, as defined by 
specific $SFR_e$ (SSFR; SSFR $> 10^{-10}$~yr$^{-1}$), chosen simply to
divide the sample in two. The correlation is significantly tighter when
only these more actively star-forming galaxies  are included. 
Fig.~\ref{LPAHvsSFR} (right) shows
the standard deviation in the log residuals for $L_{PAH}$ after
subtraction of the best linear fit for a range of subsample
sizes. The subsamples are sorted to contain the youngest 
stellar populations, selected by three
different measures: SSFR, $D_n(4000)$, and H$\alpha$~EW. 
In all cases, scatter decreases as the dominance of the young population
increases; almost monotonically for $D_n(4000)$
and H$\alpha$~EW, and for these measures the scatter flattens
significantly for the subsample containing the youngest $\sim$60\% of
sources, and for smaller ``youngest'' subsamples. The
diagnostics corresponding to the $\sim50^{th}$--$60^{th}$ percentile transition are
$D_n(4000) = 1.4$, H$\alpha$~EW~$= 18$\AA, and
SSFR~$=10^{-10}$~yr$^{-1}$. 
For more evolved sources, the exceptionally tight trend between
log~$L_{PAH}$ and log~$SFR_e$ ($\sigma < 0.2$) gains 
significant scatter.
Again, this is primarily due to the inaccuracy with which SFR is measured for evolved populations.

Figure~\ref{LPAHvsSFR_specific} (upper panels) show the 
mass-weighted relationships. The ratio of $L_{PAH}$ to stellar mass ($M_*$ is
plotted versus optical star formation diagnostics from SDSS: the narrow band 4000\AA\
break measure $D_n(4000)$ \citep{Kauffmann03} and H$\alpha$~EW
(left and middle respectively), and also SSFR (right), which is
described above.
There are tight linear correlations between all quantities, although
again we see the increased scatter in SFR for evolved galaxies. The
best-fit regression lines for the SF galaxies are also shown. The
mean 1-$\sigma$ residual scatter in $L_{PAH}/M_*$ for SF galaxies
after subtraction of the best-fit lines is $\sim$20\% for all three
cases. For $D_n(4000)$  and H$\alpha$~EW, the residual is $\sim$20\%
even including galaxies with AGN components, indicating that these age
diagnostics are good predictors of mass-weighted PAH intensity for
most galaxies. This emphasizes that the increased scatter seen in
$L_{PAH}$ and $L_{TIR}$ versus SFR in evolved populations is due to
inaccuracy in measuring SFR rather than, say, a blurring due to
evolved stars contributing more significantly to dust heating. 
As noted by \citet{Calzetti2010}, the correlation between PAH
emission and metallicity leads to another source of increased
uncertainty in the link between star formation and IR dust emission in
evolved sources.

Total IR luminosity is also closely related to these SF diagnostics,
and has a very similar slope and scatter as
$L_{PAH}$, for all three parameters and for both the absolute and
mass-weighted relations.  For the purpose of modeling high-z and
photometric SEDS, it is useful to understand how the PAH-to-total IR
ratio changes with these properties. Figure~\ref{LPAHvsSFR_specific}
(lower panels) show $L_{PAH}/L_{TIR}$ versus these same SF
diagnostics, with dashed and dotted lines showing the medians for the SF and
AGN/composite subsamples respectively.
In all SF diagnostics there is a significant but highly scattered trend, with
more evolved galaxies and AGN showing weaker PAH features for a given
$L_{TIR}$. It is not possible to extricate the effects of age and AGN
status on $L_{PAH}/L_{TIR}$ from these data, although as shown in OD09
both seem to play a part. More specifically, short wavelength PAH
features (6.2~\micron\ and 7.7~\micron) are diminished in more evolved
galaxies and AGN, while longer wavelength features are not. This is
seen in the PAH ratios (OD09) and with respect to $L_{TIR}$ (T10).
The trend is between $L_{PAH}/L_{TIR}$ and star formation is still
weakly significant for SF galaxies alone, with AGN 
excluded, although this is entirely driven by a small number of
heavily star-forming sources with very high $L_{PAH}/L_{TIR}$.

\subsection{Dust Mass}
\label{dustmasssect}

\begin{figure*}[!ht]
\centering
\includegraphics*[width=16cm]{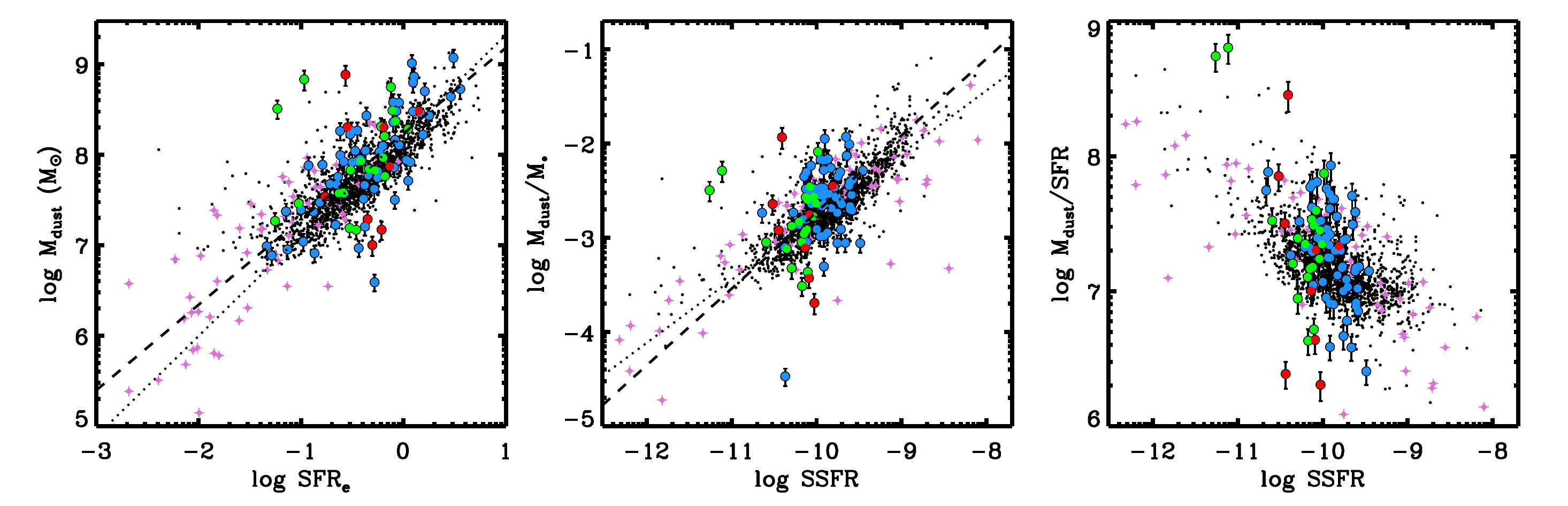}
\caption{{\it Left:} dust masses for SSGSS galaxies from models of
  \citet{DraineLi2} versus SFR$_e$ \citep{Brinchmann04} (coloured
  circles, coded by BPT designation; see Fig.~\ref{spoonplot}). Also
  plotted are the results of the dust model of \citet{daCunha10} for
  SDSS galaxies (black points) and SINGS galaxies (pink crosses).  The
  best log-log regression lines are shown for SSGSS (dashed) and SDSS
  (dotted). {\it Middle:} dust-to-stellar mass fraction versus
  SSFR, and {\it right:} dust mass-to-SFR ratio.
\label{dustmassSFR}}
\end{figure*}

\begin{figure*}[!ht]
\centering
\includegraphics*[width=16cm]{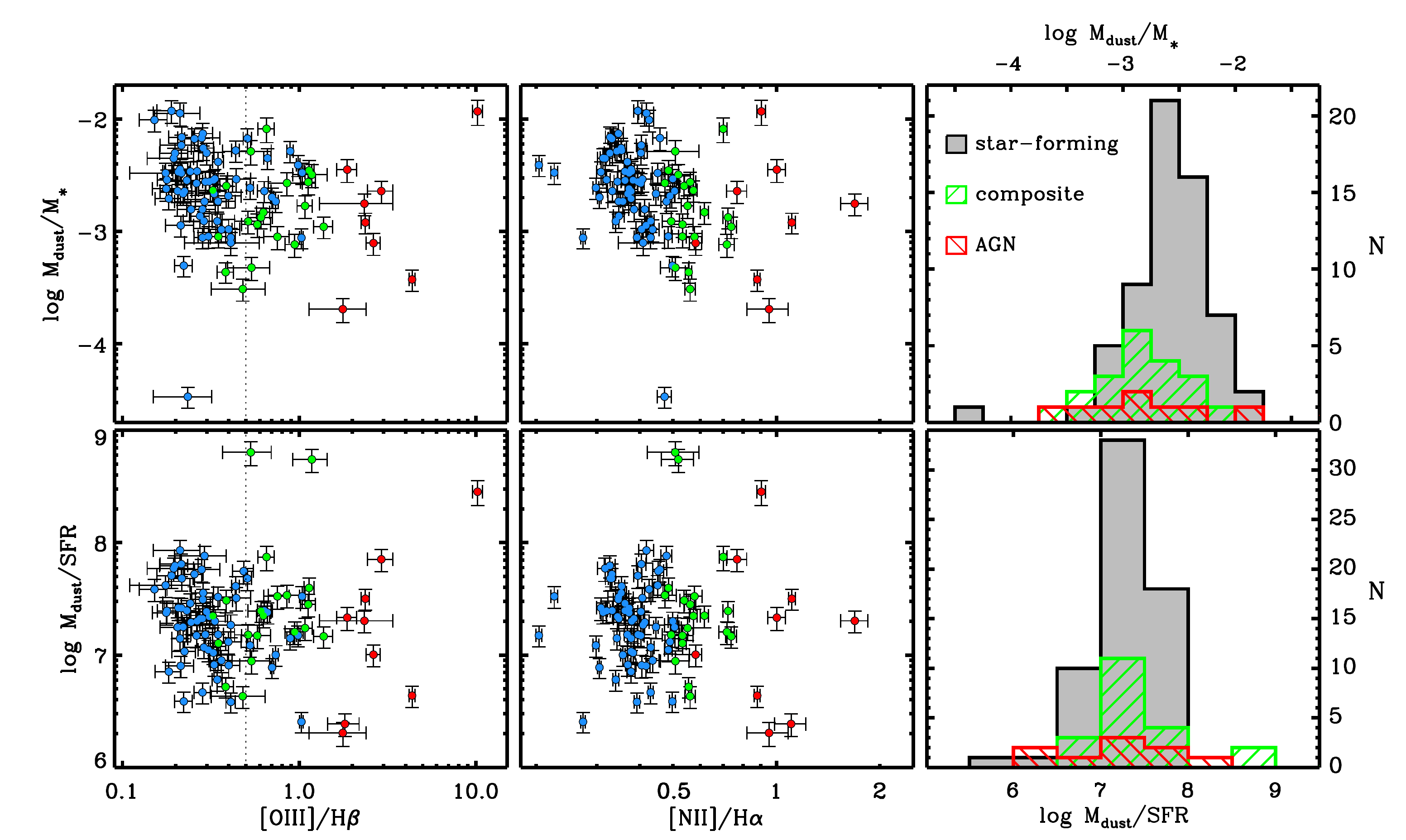}
\caption{{\it Upper Panels:} dust-to-stellar mass fraction versus the optical
  emission line ratios [OIII]/H$\beta$ ({\it left}) and [NII]/H$\alpha$
  ({\it middle}). These line ratios define BPT AGN type, which is designated
  by the same colour coding as Fig.~\ref{spoonplot}. Also shown is the
  distribution of dust fractions by AGN type ({\it right}). AGN have
 significantly lower dust fractions than SF galaxies. SF galaxies with
  [OIII]/H$\beta > 0.5$ (dotted line) are young galaxies with
  starburst-source hard radiation fields; these galaxies have
  significantly higher dust fractions than AGN with similar  [OIII]/H$\beta$.
  {\it Lower Panels: } same as above, except with the ratio of dust mass to
  SFR$_e$.
\label{dustmassBPT}}
\end{figure*}

The SSGSS IR spectra, with its coverage of all major PAH features,
combined with  MIPS 70~\micron\ and 160~\micron\ photometry, enables us
to fit detailed dust models to determine dust mass. We use the models
of \citet{DraineLi2}, which calculate IR emission spectra for
starlight-heated dust consisting of amorphous silicate and graphite grains,
and of PAH grains. SSGSS spectra and photometry are fit to template libraries derived
from these models (Draine \& Li, private communication) to
estimate the distribution of illuminating starlight intensity and
the relative abundance of PAH. These are combined with MIPS fluxes and the
relations detailed in \citet{DraineLi2} to determine
the total dust mass.

Based on these models, the median dust mass for the sample is
$M_{dust} = 8.5\times10^7$~M$_{\odot}$, with 90\% of the sample in the range
$10^7$---$10^{9}$~M$_{\odot}$. 

Figure~\ref{dustmassSFR} (left) shows the calculated dust masses for the
SSGSS sample versus SFR$_e$, with colour coding to indicate AGN
type by BPT designation. Also plotted are the results of 
\citet{daCunha10} (D10), who fit the GALEX-SDSS-2MASS-{\it IRAS}
photometry  of SDSS and SINGS galaxies with SEDs derived from population
synthesis and dust attenuation models \citep{daCunha08} to calculate
both dust mass and SFR. SSGSS dust masses are very similar to the D10
SDSS sample. The SINGS dust masses are roughly consistent with
the range of dust masses calculated using the \citet{DraineLi2} models
for a sub-sample of SINGS galaxies (\citet{Draineetal07}), which
reinforces the comparibility of the two models. 

M$_{dust}$ is highly correlated with SFR$_e$, with an almost linear
relationship that is similar to that found by D10. The best-fit regression line is:

\begin{eqnarray}
M_{dust}(M_{\odot}) = \nonumber\\
{ }(2.0 \pm 0.3) \times 10^7 ~ SFR_e(M_{\odot}yr^{-1})^{(0.95 \pm 0.11)}
\end{eqnarray}


We note that D10 find a tighter relation between these two parameters
for SDSS galaxies. This may indicate a more
accurate SFR estimation by D10 over the optical
emission line estimate, due to their use of the full FUV-to-FIR SED. 

Figure~\ref{dustmassSFR} (middle) shows dust-to-stellar mass fraction
versus SSFR for SSGSS, again including the results of D10. There is a
significant positive correlation between dust fraction and SSFR
($\tau=0.24$; correlation significance: $p=0.001$),
although here there is significantly more scatter than observed by D10
in their sample. The best-fit regression line is:

\begin{equation}
M_{dust}/M_{\odot} = (2.5 \pm 1.0) \times 10^5 ~ SSFR(yr^{-1})^{(0.8 \pm 0.2)}
\end{equation}

Figure~\ref{dustmassSFR} (right) shows the ratio of dust mass to SFR,
which, given the tight correlation between SFR and gas mass, may be
considered a proxy for the dust-to-gas mass ratio, and so traces
the enrichment of the interstellar medium. The SSGSS results follow
the general trend of the D10 sample, 
($\tau=-0.175$; correlation significance: $p=0.015$),
however the careful sample
selection of SSGSS means that it contains fewer of the rare, extreme
SF galaxies, and the UV selection eliminates the oldest red, dead
galaxies; the result is a more restricted range of SSFR that 
makes it difficult to verify the slope of this relation.

These strong correlations between galaxys' dust content and star
formation rate --- and especially between their absolute values --- supports
the idea that SFR provides a useful proxy for dust mass, as found by
D10. The careful sample selection of SSGSS adds weight to this result.

As noted by D10, these relations can be interpreted in evolutionary
terms. The ISM is enriched by dust with ongoing star formation, and at
the same time gas mass decreases, resulting in a drop in SFR. With
this drop, the production of dust is quickly superseded by destruction
of grains in the ISM, resulting in the positive correlation between
dust fraction and SSFR (Fig.~\ref{dustmassSFR}, middle). At the same
time, this reduction in dust content with galaxy age appears to be 
outpaced by the reduction in gas content and the associated drop in
SSFR, resulting in the anticorrelation between dust-to-gas mass ratio
vs. SSFR (Fig.~\ref{dustmassSFR}, right).

In this scenario, the balance between ISM enrichment and the
destruction of dust by shocks and by the ambient radiation field
defines the direction and slope of these relations. 
The BPT emission line ratios, [OIII]/H$\beta$ and
[NII]/H$\alpha$, are sensitive to the hardness of the ambient radiation field
and the level of ISM enrichment respectively, and so may shed light on the
nature of this balance. Additionally, as seen in Figure~\ref{dustmassSFR}, galaxies with
BPT-designated AGN components have lower dust fractions than SF
galaxies. Although this may be purely due to the increased incidence (and
observability) of AGN in older galaxies, the difference in dust
fraction with SSFR seems to be divided surprisingly cleanly down
AGN lines. It is possible that AGN shocks and hard UV photons are
important in destroying small dust grains, given their potential affect
on PAH grain size distributions (\citealt{Smith07}; OD09; T10)

Figure~\ref{dustmassBPT} (upper panels) shows dust-to-stellar mass
fraction versus [OIII]/H$\beta$ and
[NII]/H$\alpha$, as well as the distribution of dust fractions by
BPT designation. Both ratios show a scattered but significant inverse correlation, and in both
cases there is no significant trend taking either SF galaxies or AGN
separately. 

[OIII]/H$\beta$ provides a good proxy for the hardness of
the UV radiation field, and so it is tempting to link dust grain
destruction with an increase in hard UV photons, although it is
not possible to extricate this effect from the drop in dust
production with diminishing star formation. Notably,
SF galaxies with hard UV fields ([OIII]/H$\beta \gtrsim$ 0.5) all have 
have significantly higher dust fractions than AGN with similar 
[OIII]/H$\beta$. These same galaxies have
among the highest SSFRs, implying that either the radiation field has
little to do with grain destruction, or that dust production due to heavy
star formation in these galaxies more than compensates for such depletion.

[NII]/H$\alpha$ is sensitive to gas-phase metallicity, and so may be
expected to have a positive correlation with dust-to-stellar mass
fraction. The observed negative correlation supports the idea
that the drop in dust content of these galaxies ($M_{dust}/M_*$)
outpaces the increase in the gas-phase metallicity (and corresponding
dust-to-gas mass fraction) as these galaxies evolve.  

Figure~\ref{dustmassBPT} (lower panels) shows the ratio of dust mass
to SFR vs. BPT line ratios. There is no correlation between this and any of the line
ratios, and no trend with AGN status. This is expected;
while dust-to-stellar mass fraction decreases with stellar population
age, gas content and the associated SFR probably decreases more
quickly, eliminating any trend.

Dust masses for the entire SSGSS sample
are available as part of the data release (see Sect.~\ref{cproducts})

\section{Data Products and Value-Added Catalogs}
\label{products}

\subsection{Spitzer Data Archive}

SSGSS is a Spitzer Legacy Program, and all spectroscopy is
available for download from the Spitzer Data Archive. The archive is
accessed via Spitzer's Leopard software. The web interface for the
archive and link to the Leopard download page can be found at:
http://archive.spitzer.caltech.edu/.

The data release includes low-resolution spectra of the 
100 galaxies in the full SSGSS sample and high-resolution spectra of
the 33 galaxies in the bright SSGSS sample. These are 
reduced as described in Section~\ref{reduction}. We include spectra
with slit positions both combined and uncombined, and with orders and
(for low-res) SL and LL modules both stitched and unstitched.

\subsection{Complementary Data Products}
\label{cproducts}

In addition to the spectroscopic data, a range
of complementary data products are available via the SSGSS website:\\

\noindent
http://astro.columbia.edu/ssgss\\

\noindent
{\it UV through IR imaging data}: including FUV and NUV from
GALEX; ugriz from SDSS; IRAC 3.6, 4.5, 5.8, and 7.8~\micron\ bands;
MIPS 24, 70, and 160~\micron\ bands; and
16~\micron\ IRS Peakup from SSGSS, with corresponding uncertainty
and/or background images. \\

\noindent
{\it Matched catalogs of SDSS properties}: these are from the SDSS
studies at MPA/JHU group\footnote{see www.mpa-garching.mpg.de/SDSS/}
\citep{Brinchmann04}, and are drawn from data 
release 4 of their catalogs. These include
measured and derived parameters such as  magnitudes, line strengths,
Petrosian radii, SFRs, and index strengths, as well as
magnitudes K-corrected using the method of \citet{Blanton06}, 
stellar masses, mass-to-light ratios, dust attenuations, etc. from
\citet{Kauffmann03}, and gas-phase metallicities from \citet{Tremonti}.\\

\noindent
{\it PAH strengths}: integrated intensities for the most prominent PAH
features and complexes at 6.2~\micron, 7.7~\micron, 8.6~\micron,
11.3~\micron, and 17~\micron, as well as total PAH intensity, 
measured by the PAHFIT \citep{Smith07} decompositions of SSGSS spectra
as described in OD09.\\ 

\noindent
{\it Emission line strengths}: line strengths from both the
low-res and hi-res spectra from PAHFIT decompositions.\\

\noindent
{\it Derived properties:} dust masses (see Sect.~\ref{dustmasssect}), silicate strength
 (see Sect.~\ref{absorb}), and silicate optical depth from
 PAHFIT decompositions.\\

\section{Summary}

We have presented SSGSS, a new Spitzer spectroscopic survey of
galaxies, colour-selected to be representative of the normal,
star-forming galaxy population. 
SSGSS targets 101 galaxies in the Lockman Hole, all with extensive
ancilliary imaging and spectroscopy from the FUV to the FIR. 
All galaxies were observed with IRS low resolution modules from 5---40~\micron\ and a
subsample of the 33 brightest were observed with at high resolution
from 10---20~\micron. 

For the low resolution spectra, our target S/N of 5 per resolution element was
exceeded except in some SL2 spectra and in regions
of high silicate absorption around 10\micron. At high resolution, S/N $>$ 10 was
achieved for all but one [Ne~II] lines and S/N $>$ 10 for $\sim$90\% of
[Ne~III] lines.
Of the 101 observed galaxies, 100 resulted in
useful spectra, which are available from the Spitzer archive.

Comparison of SSGSS galaxies to starbursts shows that the starburst spectra
differ primarily in the shape of the warm dust continuum, consistent
with having elevated emission at small-grain dust
temperatures of $\lesssim 120$~K. Silicate absorption levels of SSGSS
galaxies are uniformly lower than those often observed in LIRGs and
ULIRGs, and are similar to those observed in starbursts.

Using PAH strengths measured with the PAHFIT code of \citet{Smith07}, we
have investigated the link between total PAH emission and both IR luminosity and star 
formation indicators. PAH luminosity correlates with star formation
indicators $D_n(4000)$ and H$\alpha$~EW very tightly, except for the
oldest $\sim$20\% of the sample. PAH-to-total IR luminosity ratio
changes slightly between SF galaxies and those with AGN components,
dropping from $\sim$0.07 to $\sim$0.05. This approximately reflects
the change between actively star forming galaxies and more evolved galaxies.

Dust masses calculated from the models of \citet{DraineLi2} reveal a
close relationship between dust mass and SFR, and are consistent with
the evolutionary scenario described by D10 in which dust that is produced quickly in a
rapidly star-forming phase is destroyed more quickly than it is
replenished as star formation drops.

SSGSS offer great scope for further study.
In OD09, the sample has been used to conduct a detailed 
analysis of the physical drivers of PAH spectral shapes, and 
\citet{Treyer10} investigates the use
of MIR diagnostics for SF and AGN activity.
Future work will include the development of template PAH spectra
calibrated to a wide range of observables, analysis of the role of
the size, shape, and composition (via metallicity) of small dust
grains in determining the continuum shape.


\appendix

\section*{Appendix A: Multi-Wavelength Thumbnails}

\begin{figure*}[h!]
\centering
\includegraphics*[width=14cm]{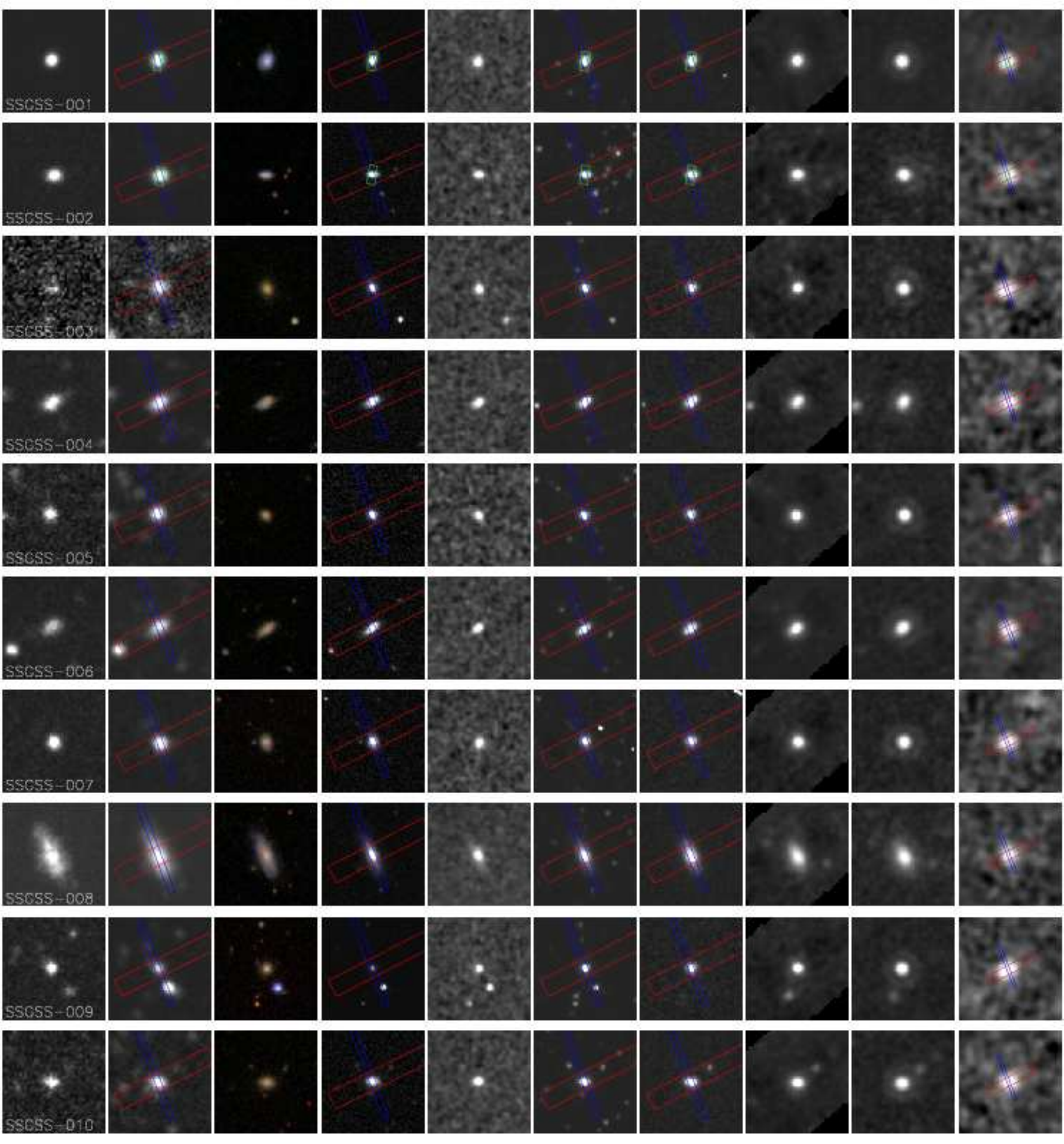}
~~FUV~~~~~NUV~~~~~SDSS~~~~~~~~r~~~~~~~~~~H~~~~~~~3.6~\micron~~~~8~\micron~~~~~16~\micron~~~~25~\micron~~~70~\micron\\
\caption{Multi-wavelength images of the 101 galaxies of the SSGSS
  sample. Each horizontal strip shows a given galaxy in, respectively: GALEX FUV and NUV, SDSS
multi-colour, r, H, IRAC 3.6~\micron\ and 8~\micron, 16~\micron\ from SSGSS
peakup images, and MIPS 25~\micron\ and 70~\micron. All boxes are
60\arcsec\ on a side except for the 70~\micron\ band, which is 120\arcsec.
\label{mosaic}}
\end{figure*}
\begin{figure*}
\centering
\includegraphics*[width=14cm]{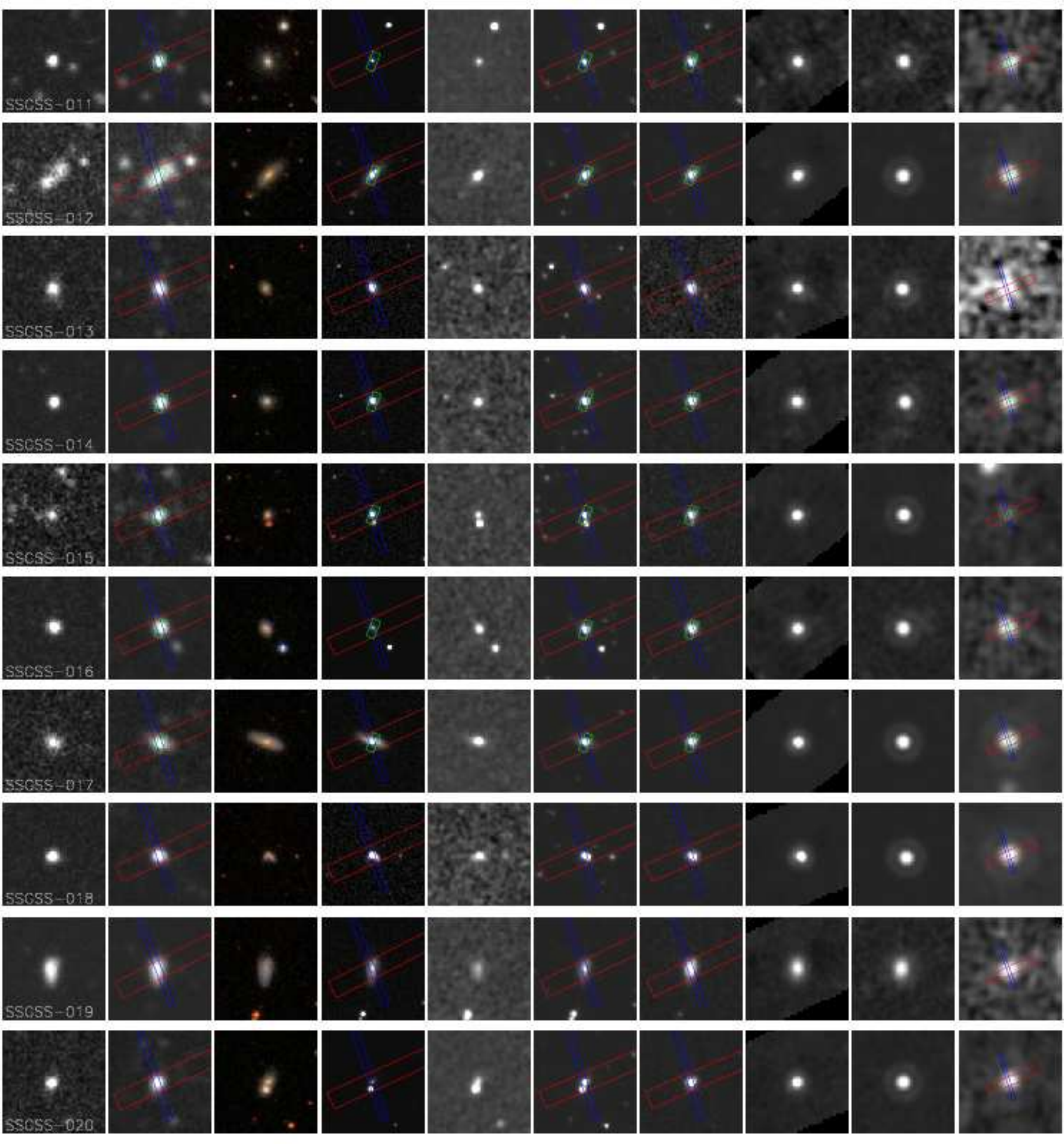}
\\
~~FUV~~~~~NUV~~~~~SDSS~~~~~~~~r~~~~~~~~~~H~~~~~~~3.6~\micron~~~~8~\micron~~~~~16~\micron~~~~25~\micron~~~70~\micron\\
\smallskip
Figure~\ref{mosaic} (continued)
\end{figure*}
\begin{figure*}
\centering
\includegraphics*[width=14cm]{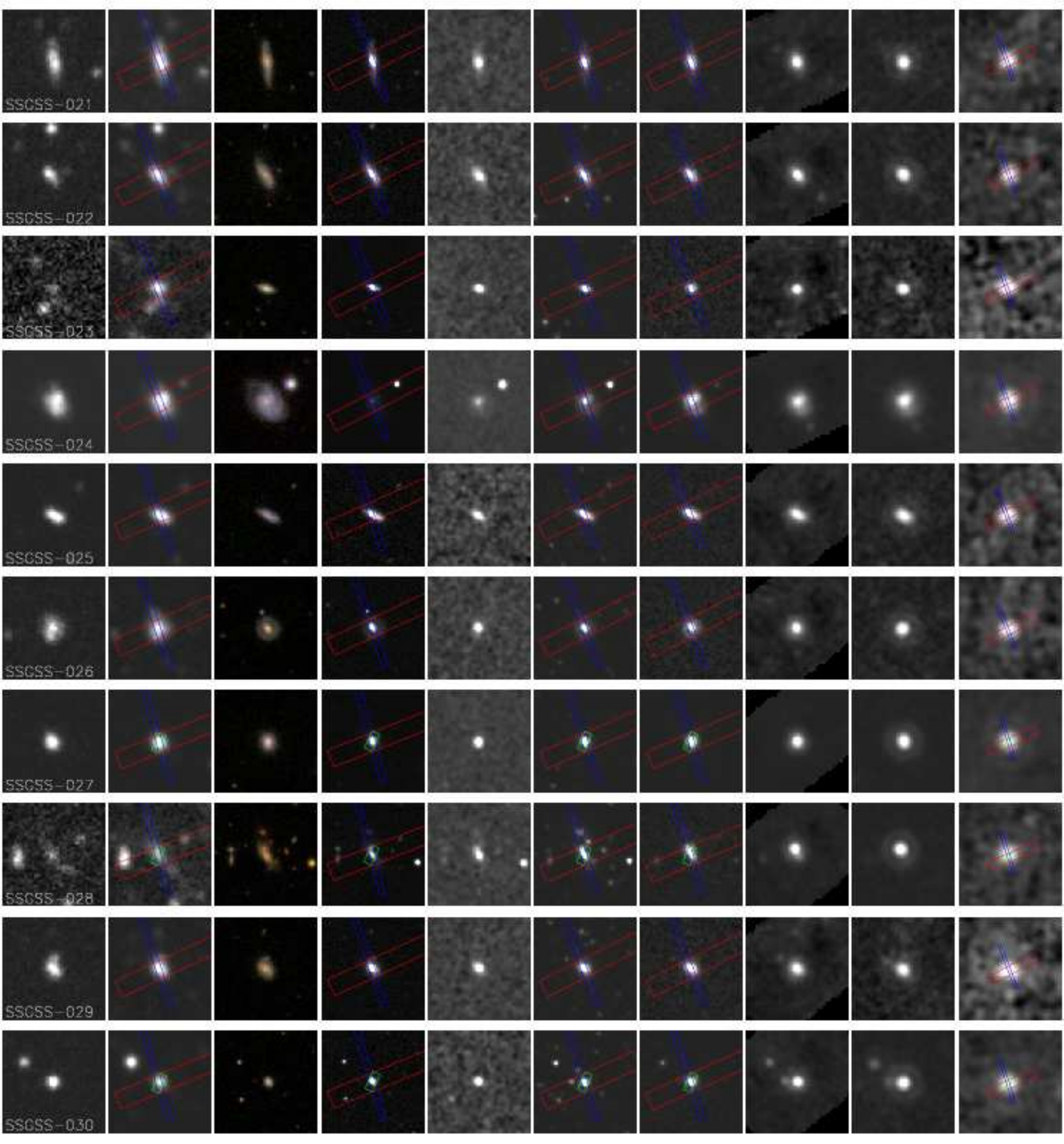}
\\
~~FUV~~~~~NUV~~~~~SDSS~~~~~~~~r~~~~~~~~~~H~~~~~~~3.6~\micron~~~~8~\micron~~~~~16~\micron~~~~25~\micron~~~70~\micron\\
\smallskip
Figure~\ref{mosaic} (continued)
\end{figure*}
\begin{figure*}
\centering
\includegraphics*[width=14cm]{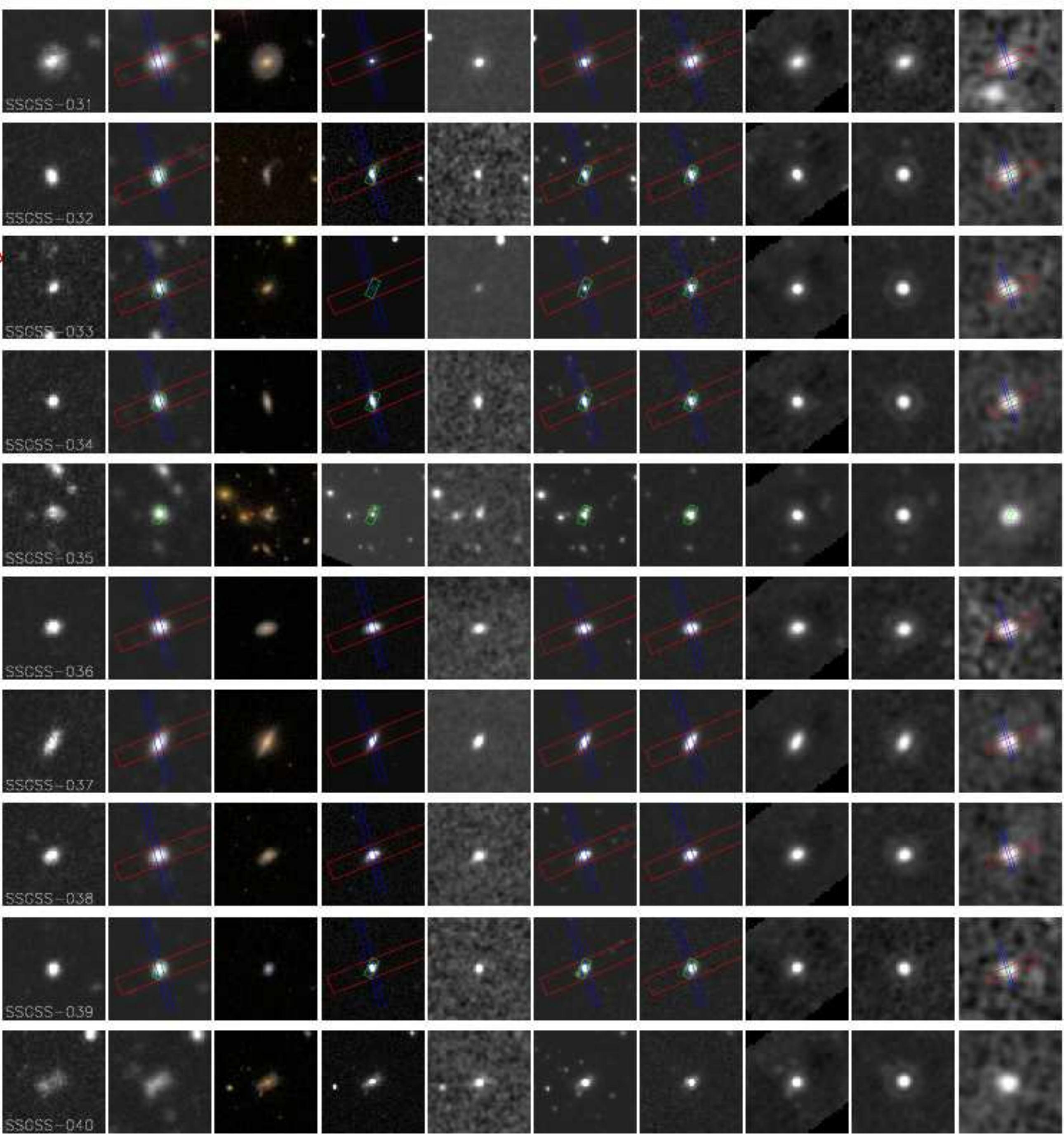}
\\
~~FUV~~~~~NUV~~~~~SDSS~~~~~~~~r~~~~~~~~~~H~~~~~~~3.6~\micron~~~~8~\micron~~~~~16~\micron~~~~25~\micron~~~70~\micron\\
\smallskip
Figure~\ref{mosaic} (continued)
\end{figure*}
\begin{figure*}
\centering
\includegraphics*[width=14cm]{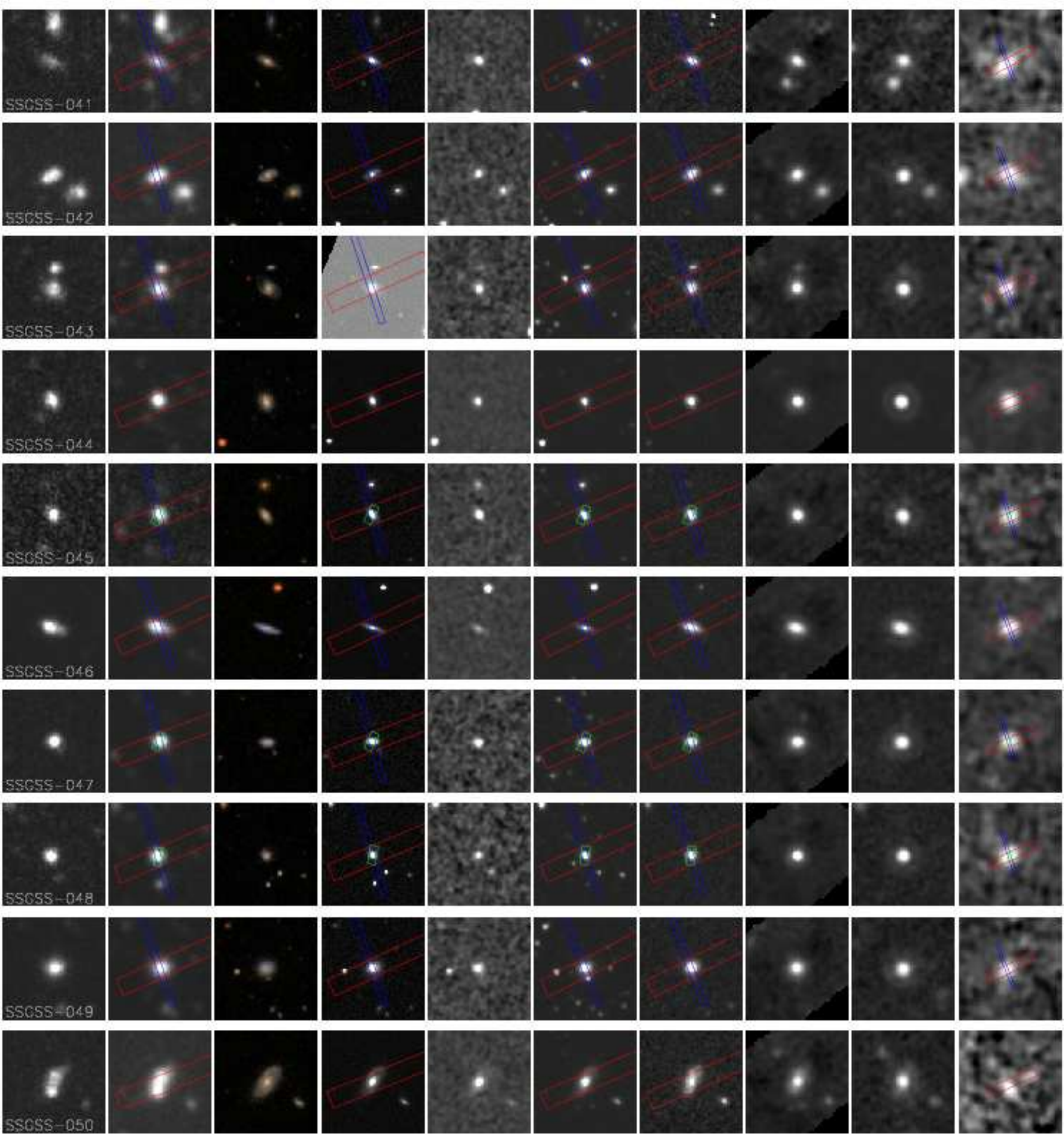}
\\
~~FUV~~~~~NUV~~~~~SDSS~~~~~~~~r~~~~~~~~~~H~~~~~~~3.6~\micron~~~~8~\micron~~~~~16~\micron~~~~25~\micron~~~70~\micron\\
\smallskip
Figure~\ref{mosaic} (continued)
\end{figure*}
\begin{figure*}
\centering
\includegraphics*[width=14cm]{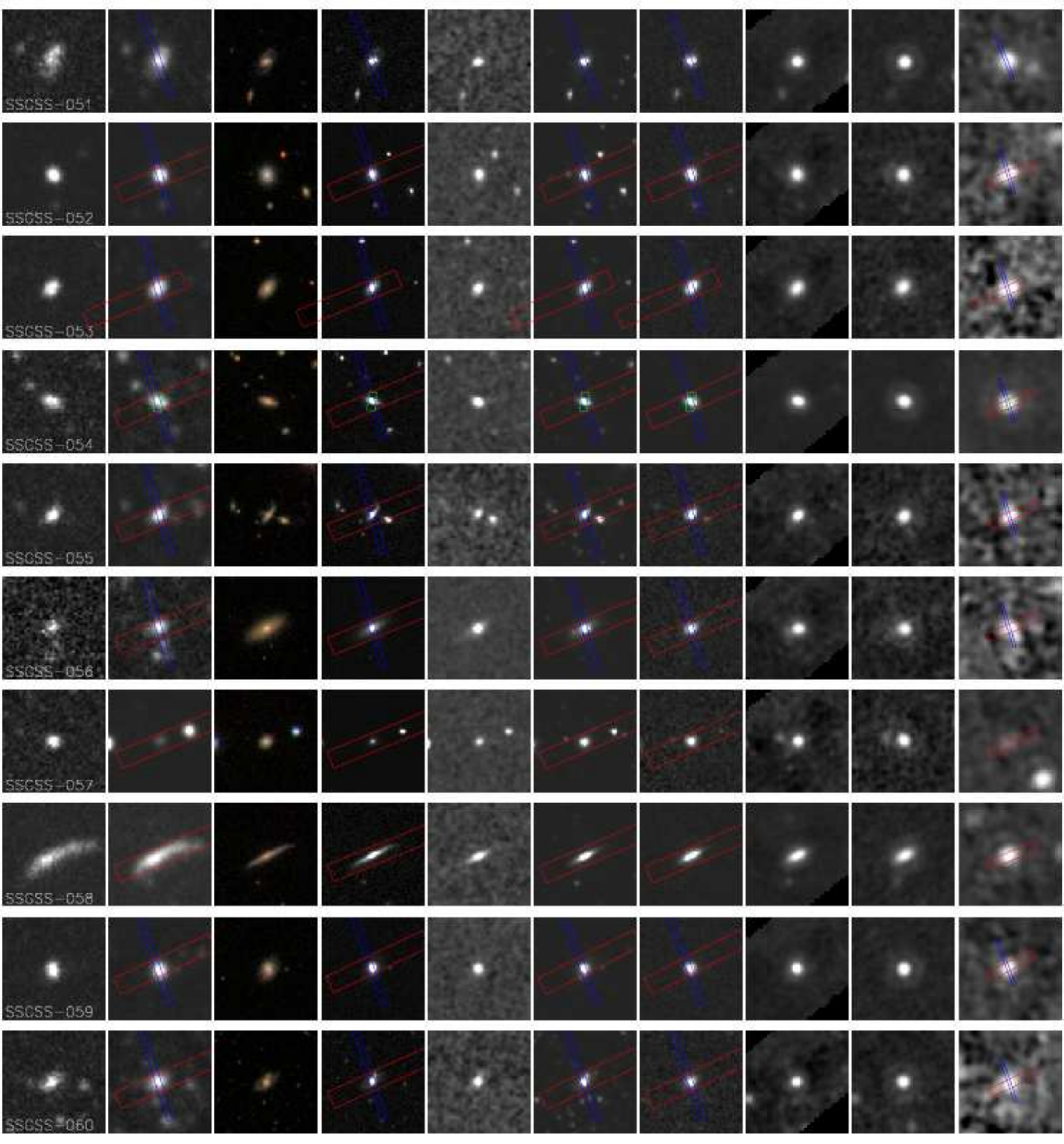}
\\
~~FUV~~~~~NUV~~~~~SDSS~~~~~~~~r~~~~~~~~~~H~~~~~~~3.6~\micron~~~~8~\micron~~~~~16~\micron~~~~25~\micron~~~70~\micron\\
\smallskip
Figure~\ref{mosaic} (continued)
\end{figure*}
\begin{figure*}
\centering
\includegraphics*[width=14cm]{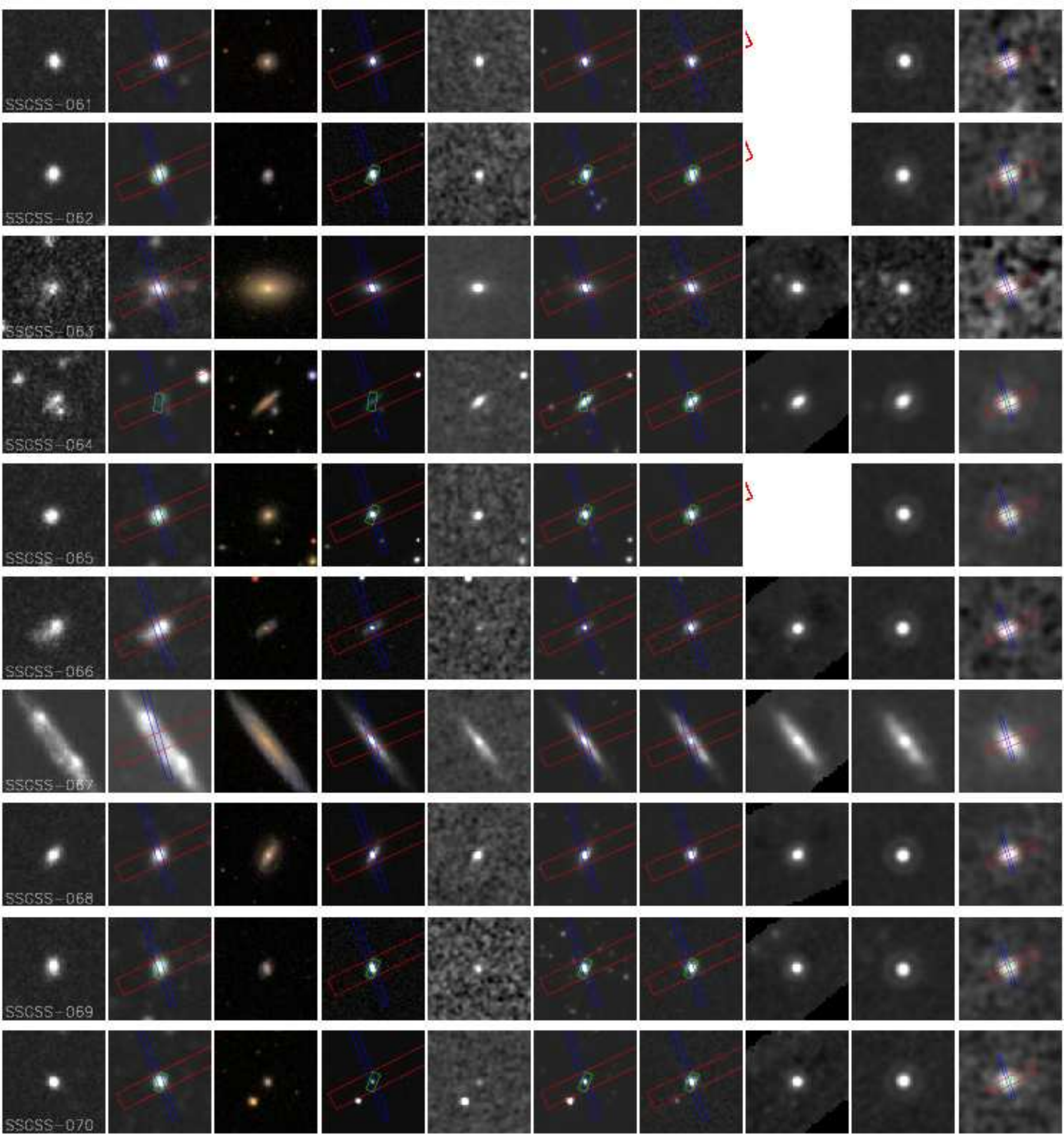}
\\
~~FUV~~~~~NUV~~~~~SDSS~~~~~~~~r~~~~~~~~~~H~~~~~~~3.6~\micron~~~~8~\micron~~~~~16~\micron~~~~25~\micron~~~70~\micron\\
\smallskip
Figure~\ref{mosaic} (continued)
\end{figure*}
\begin{figure*}
\centering
\includegraphics*[width=14cm]{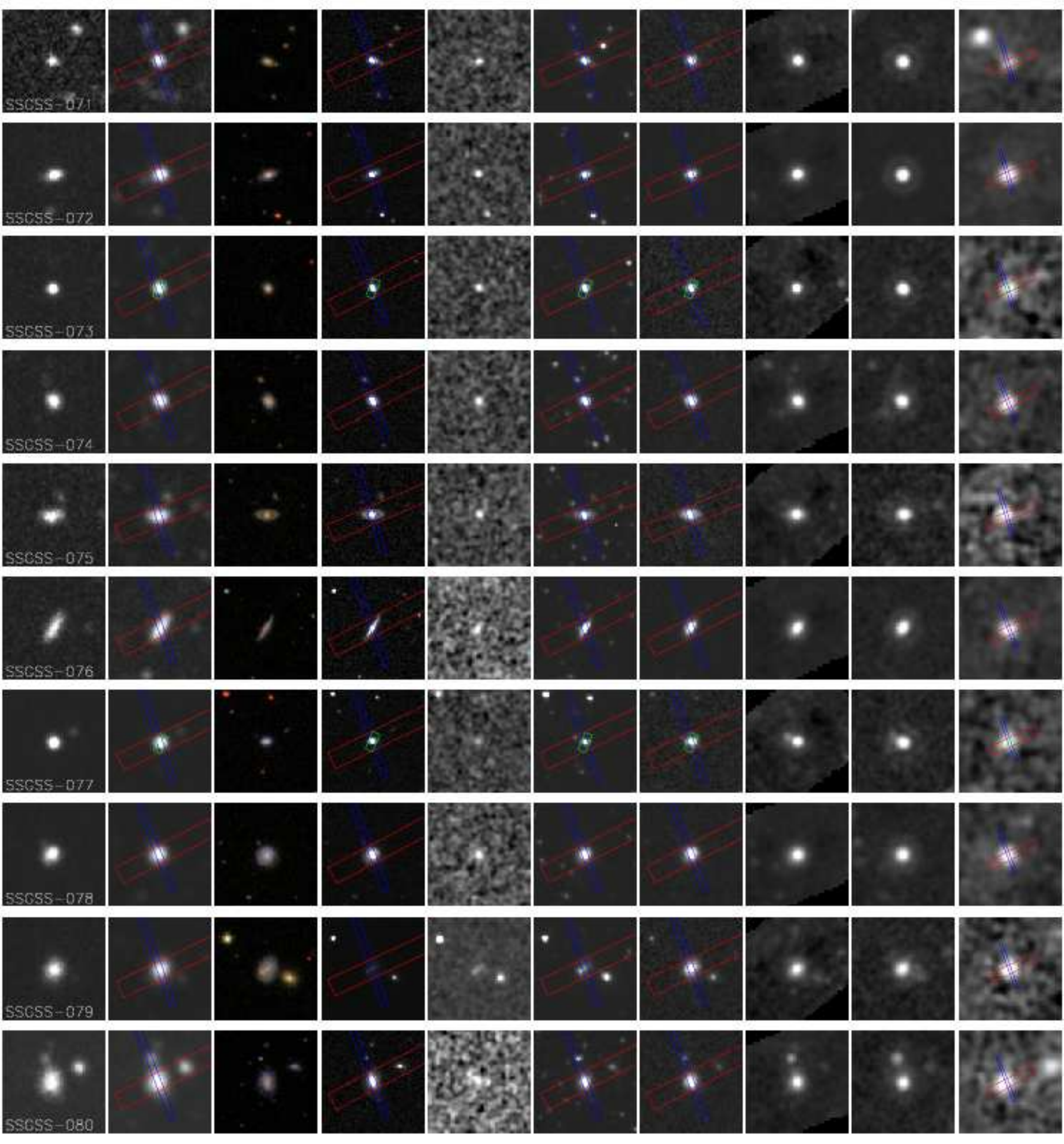}
\\
~~FUV~~~~~NUV~~~~~SDSS~~~~~~~~r~~~~~~~~~~H~~~~~~~3.6~\micron~~~~8~\micron~~~~~16~\micron~~~~25~\micron~~~70~\micron\\
\smallskip
Figure~\ref{mosaic} (continued)
\end{figure*}
\begin{figure*}
\centering
\includegraphics*[width=14cm]{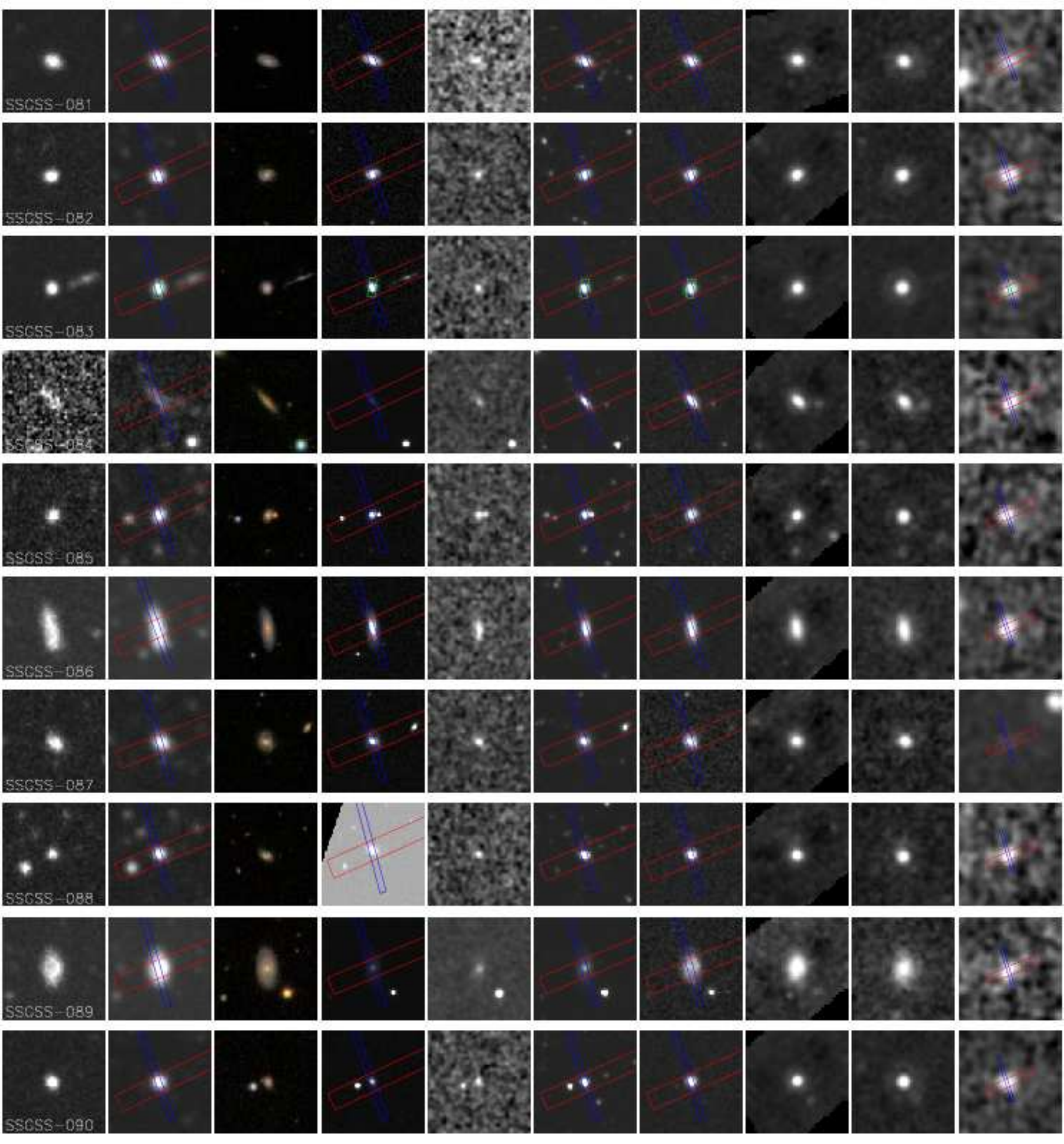}
\\
~~FUV~~~~~NUV~~~~~SDSS~~~~~~~~r~~~~~~~~~~H~~~~~~~3.6~\micron~~~~8~\micron~~~~~16~\micron~~~~25~\micron~~~70~\micron\\
\smallskip
Figure~\ref{mosaic} (continued)
\end{figure*}
\begin{figure*}
\centering
\includegraphics*[width=14cm]{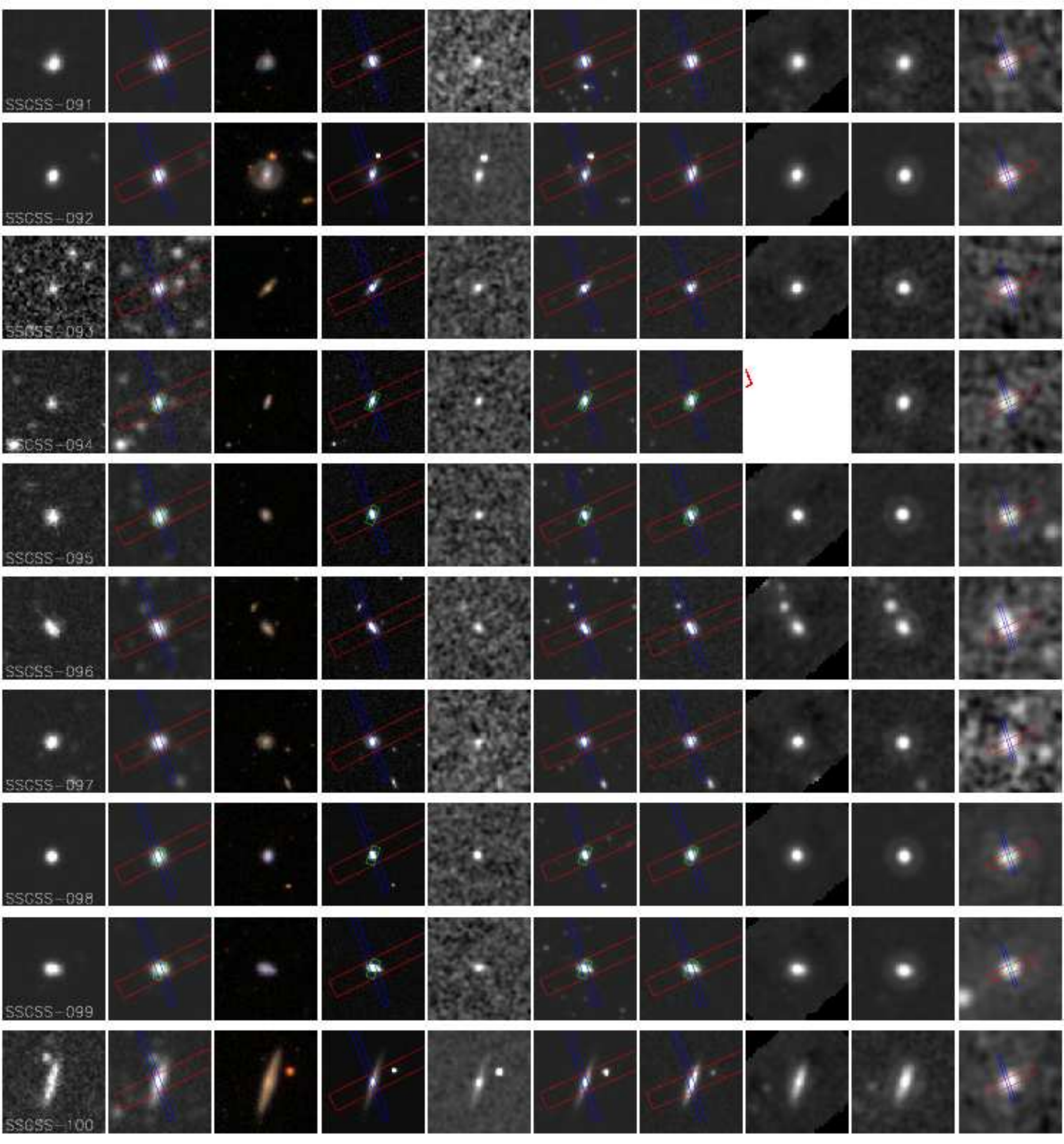}
\\
~~FUV~~~~~NUV~~~~~SDSS~~~~~~~~r~~~~~~~~~~H~~~~~~~3.6~\micron~~~~8~\micron~~~~~16~\micron~~~~25~\micron~~~70~\micron\\
\smallskip
Figure~\ref{mosaic} (continued)
\end{figure*}

\clearpage

\section*{Appendix B: Reduced IRS Spectra}


\begin{figure*}[h!]
\centering
\includegraphics*[width=13cm]{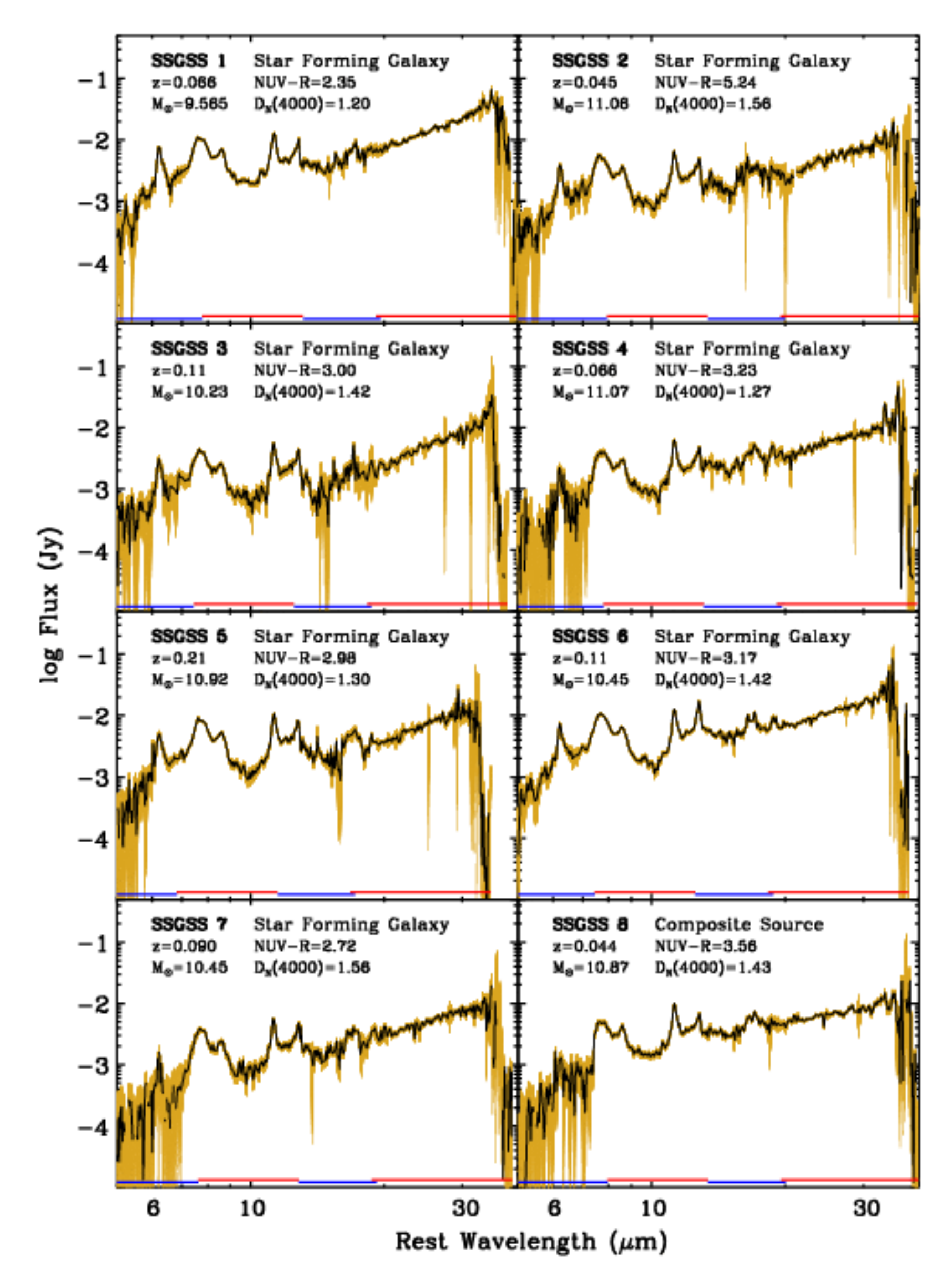}
\caption{Short-low and long-low IRS spectra, reduced and stitched as
  described in Section~\ref{lowresreduction}. The yellow shaded region shows
  the $\pm 1$-sigma errors. The wavelength range of each IRS module is
  shown at the base of each panel, from left to right: SL2 (blue), SL1
  (red), LL2 (blue), LL1 (red). Also given are the SSGSS catalog
  number, BPT type \citep{BPT}, redshift, stellar mass, 
 NUV-R color, and $D_n(4000)$ age diagnostic.
\label{lowresspectra}}
\end{figure*}
\begin{figure*}
\centering
\includegraphics*[width=13cm]{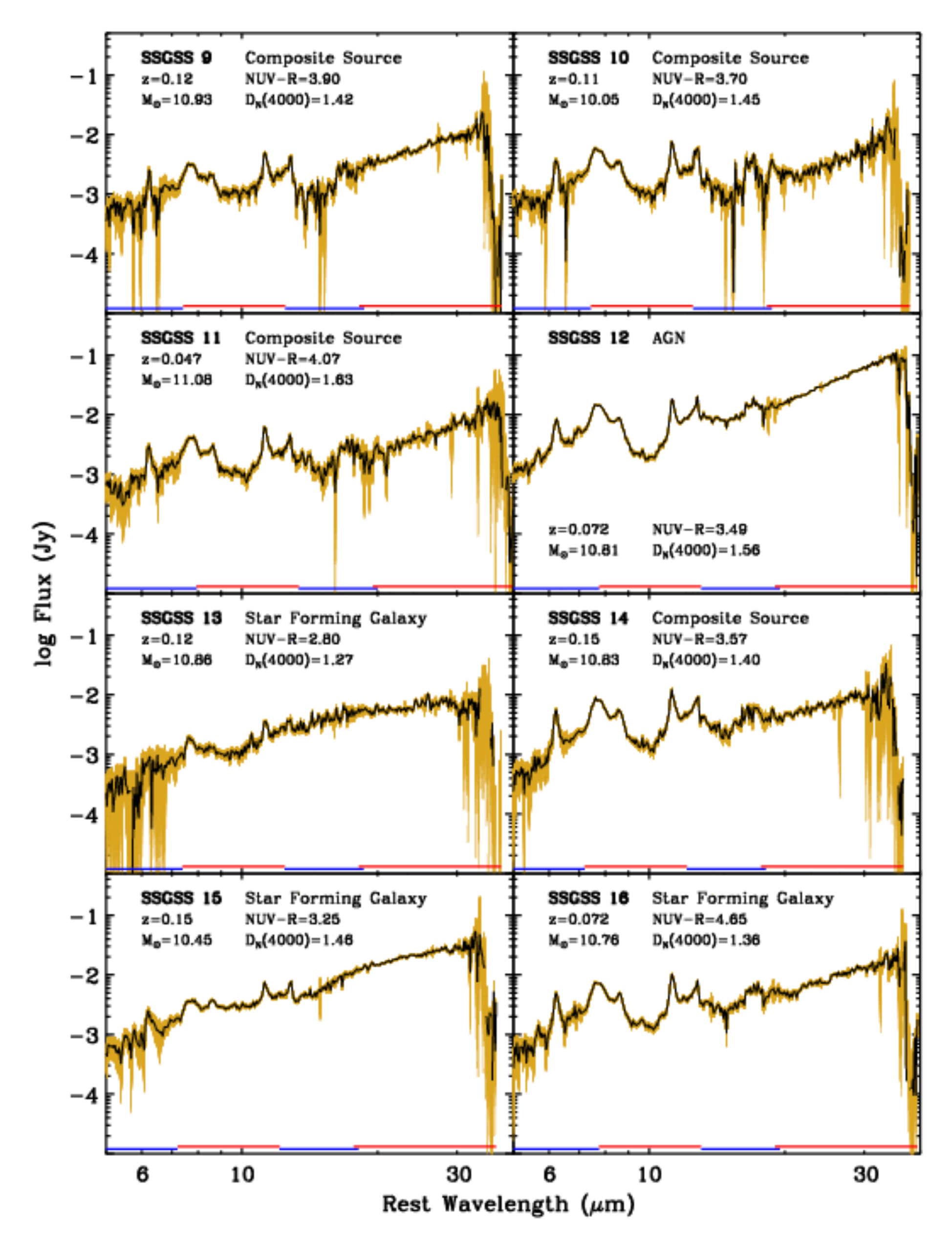}\\
Figure~\ref{lowresspectra} (continued)
\end{figure*}
\begin{figure*}
\centering
\includegraphics*[width=13cm]{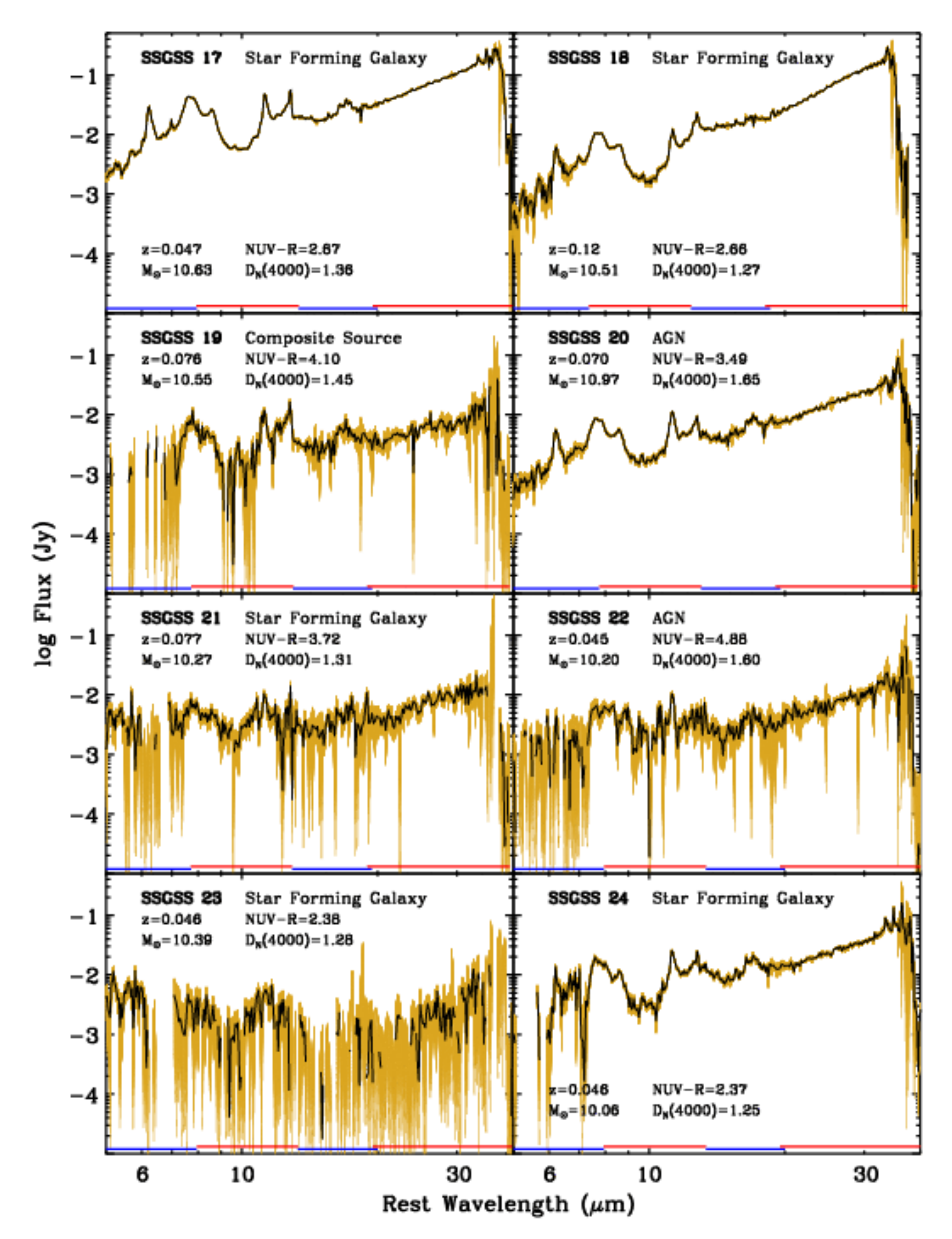}\\
Figure~\ref{lowresspectra} (continued)
\end{figure*}
\begin{figure*}
\centering
\includegraphics*[width=13cm]{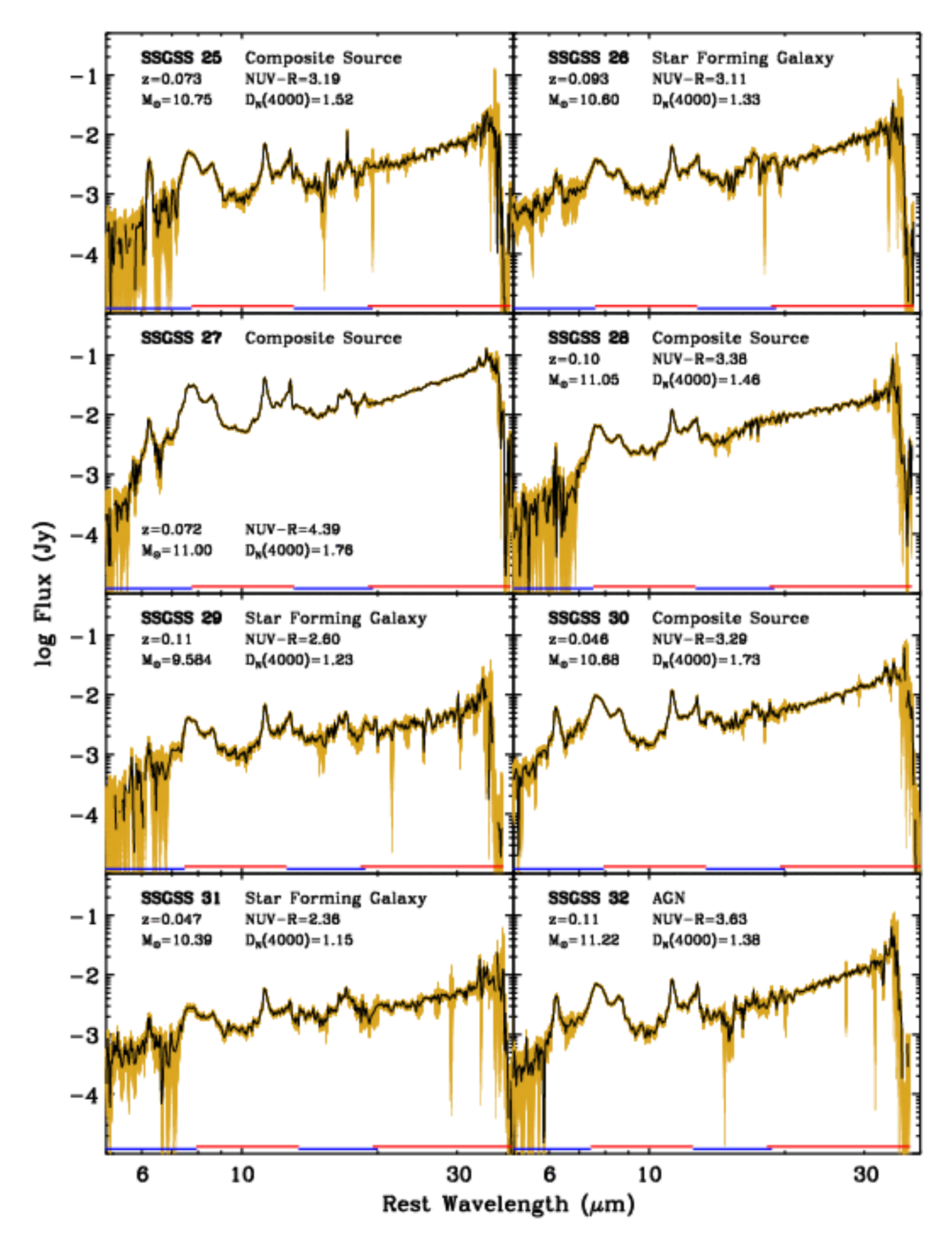}\\
Figure~\ref{lowresspectra} (continued)
\end{figure*}
\begin{figure*}
\centering
\includegraphics*[width=13cm]{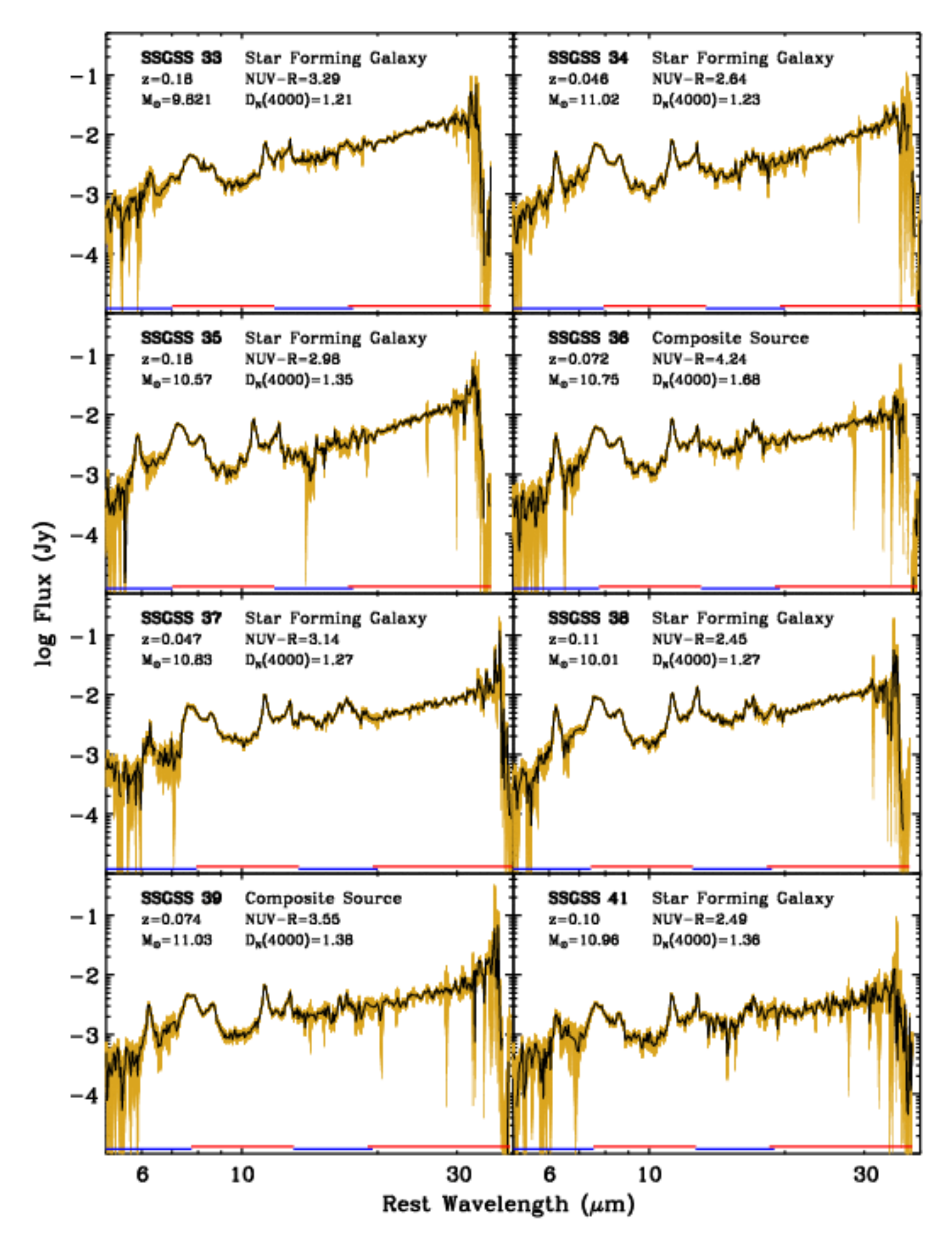}\\
Figure~\ref{lowresspectra} (continued)
\end{figure*}
\begin{figure*}
\centering
\includegraphics*[width=13cm]{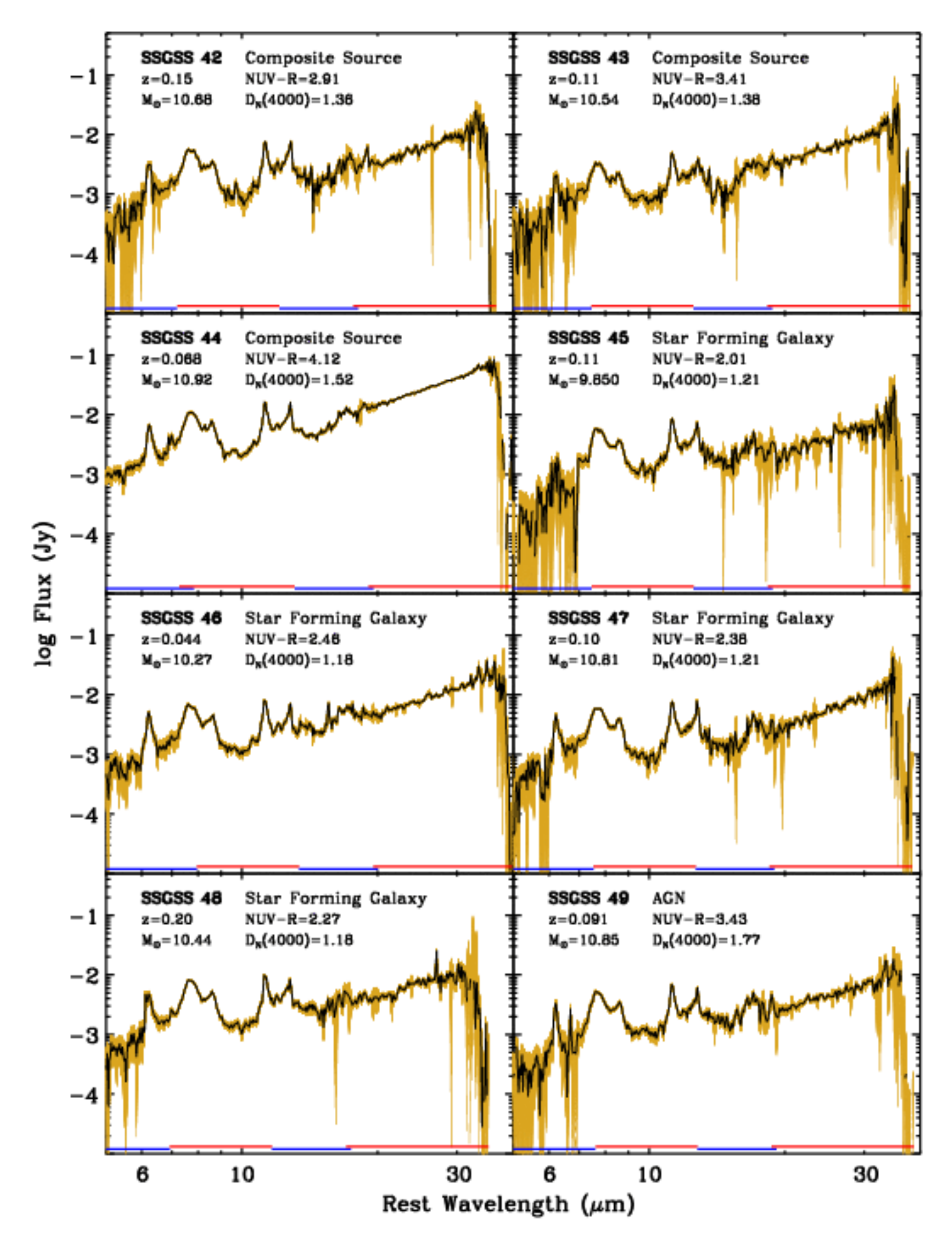}\\
Figure~\ref{lowresspectra} (continued)
\end{figure*}
\begin{figure*}
\centering
\includegraphics*[width=13cm]{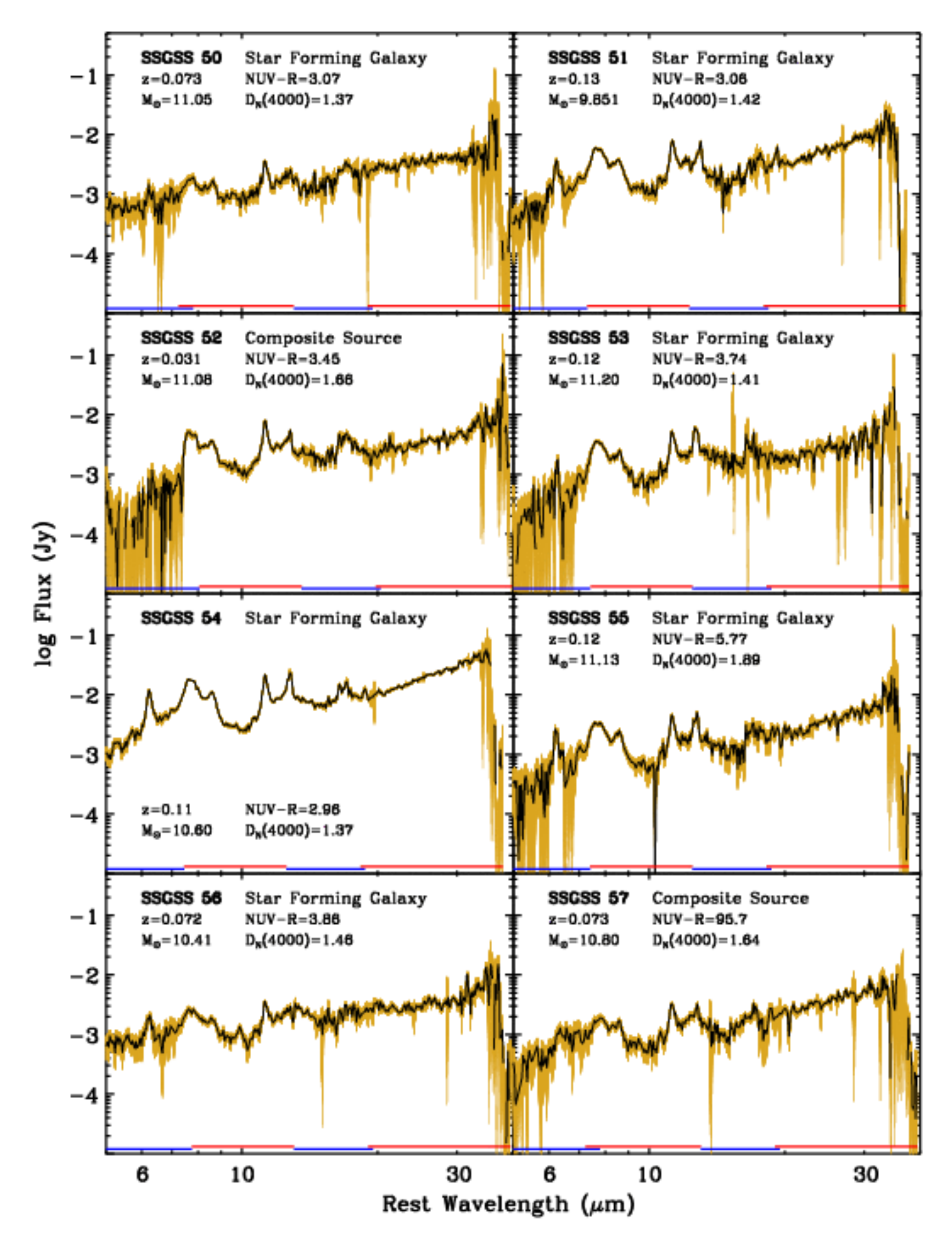}\\
Figure~\ref{lowresspectra} (continued)
\end{figure*}
\begin{figure*}
\centering
\includegraphics*[width=13cm]{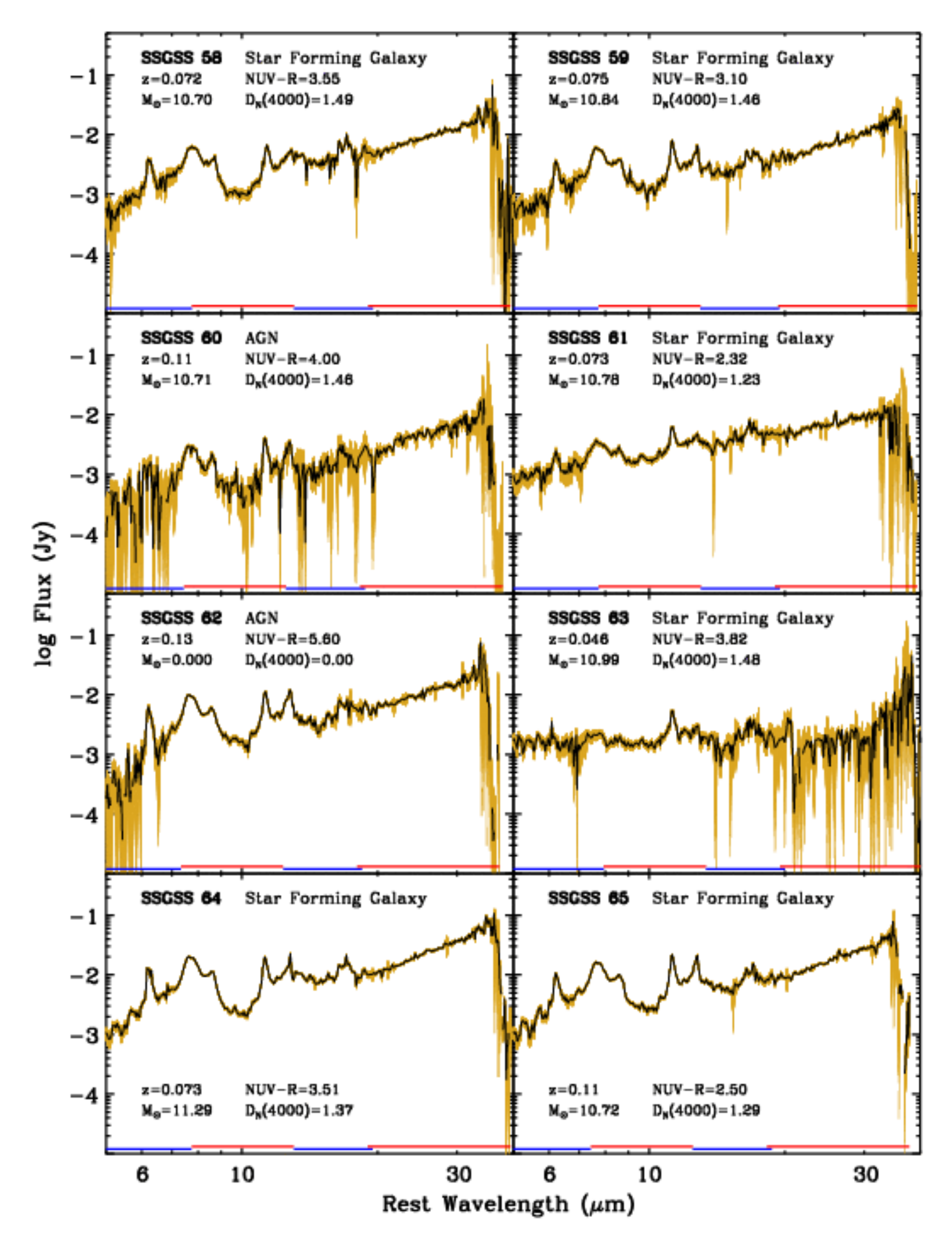}\\
Figure~\ref{lowresspectra} (continued)
\end{figure*}
\begin{figure*}
\centering
\includegraphics*[width=13cm]{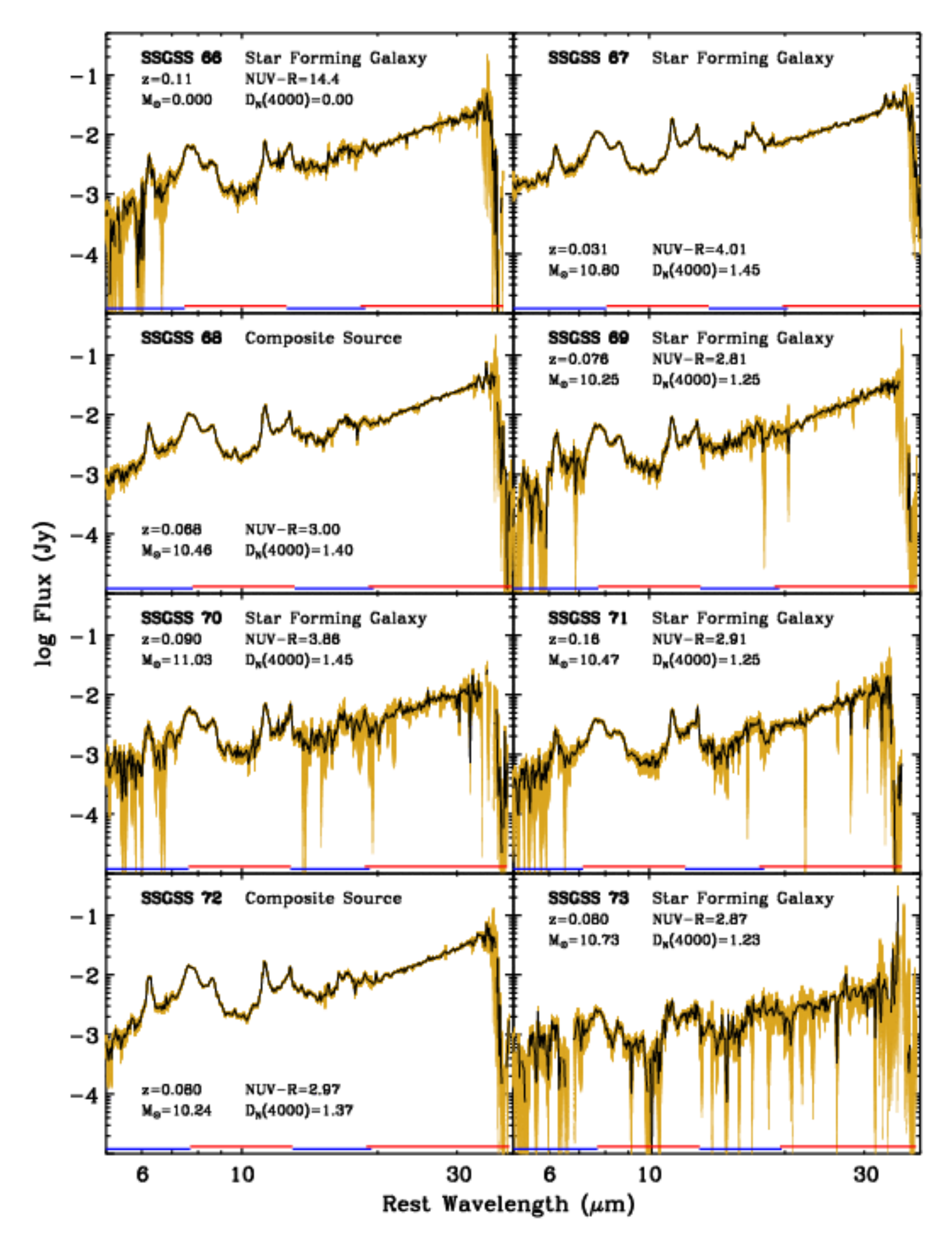}\\
Figure~\ref{lowresspectra} (continued)
\end{figure*}
\begin{figure*}
\centering
\includegraphics*[width=13cm]{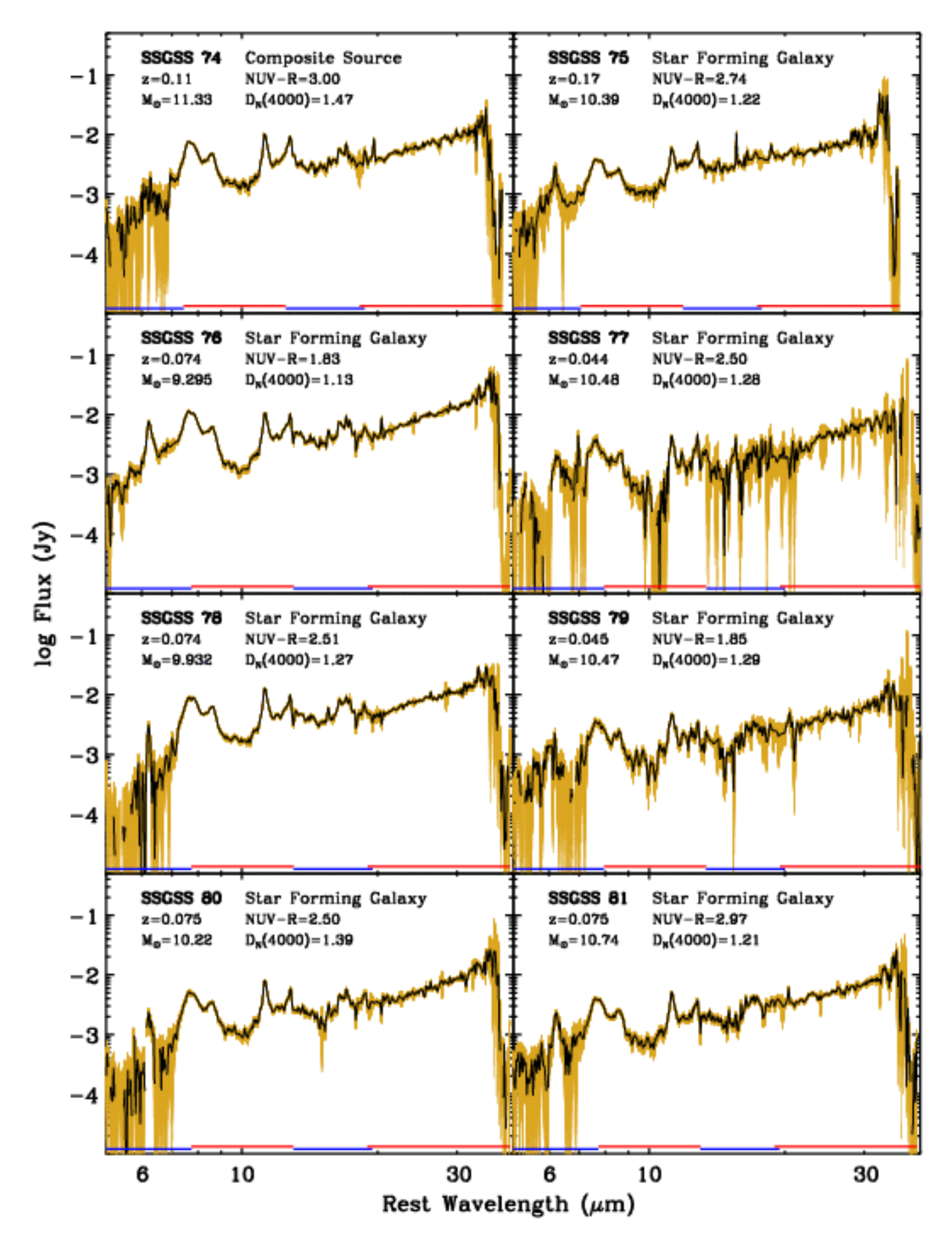}\\
Figure~\ref{lowresspectra} (continued)
\end{figure*}
\begin{figure*}
\centering
\includegraphics*[width=13cm]{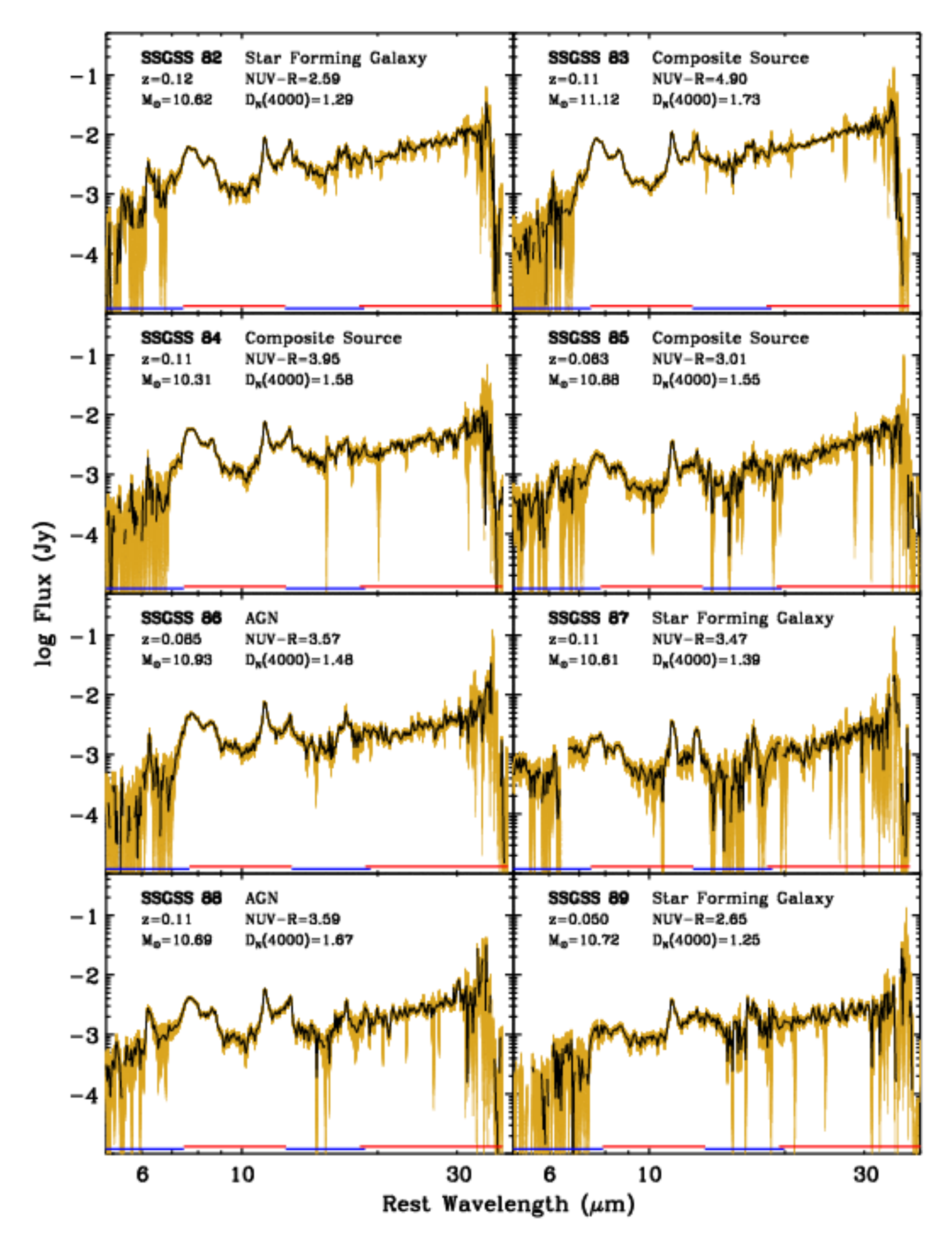}\\
Figure~\ref{lowresspectra} (continued)
\end{figure*}
\begin{figure*}
\centering
\includegraphics*[width=13cm]{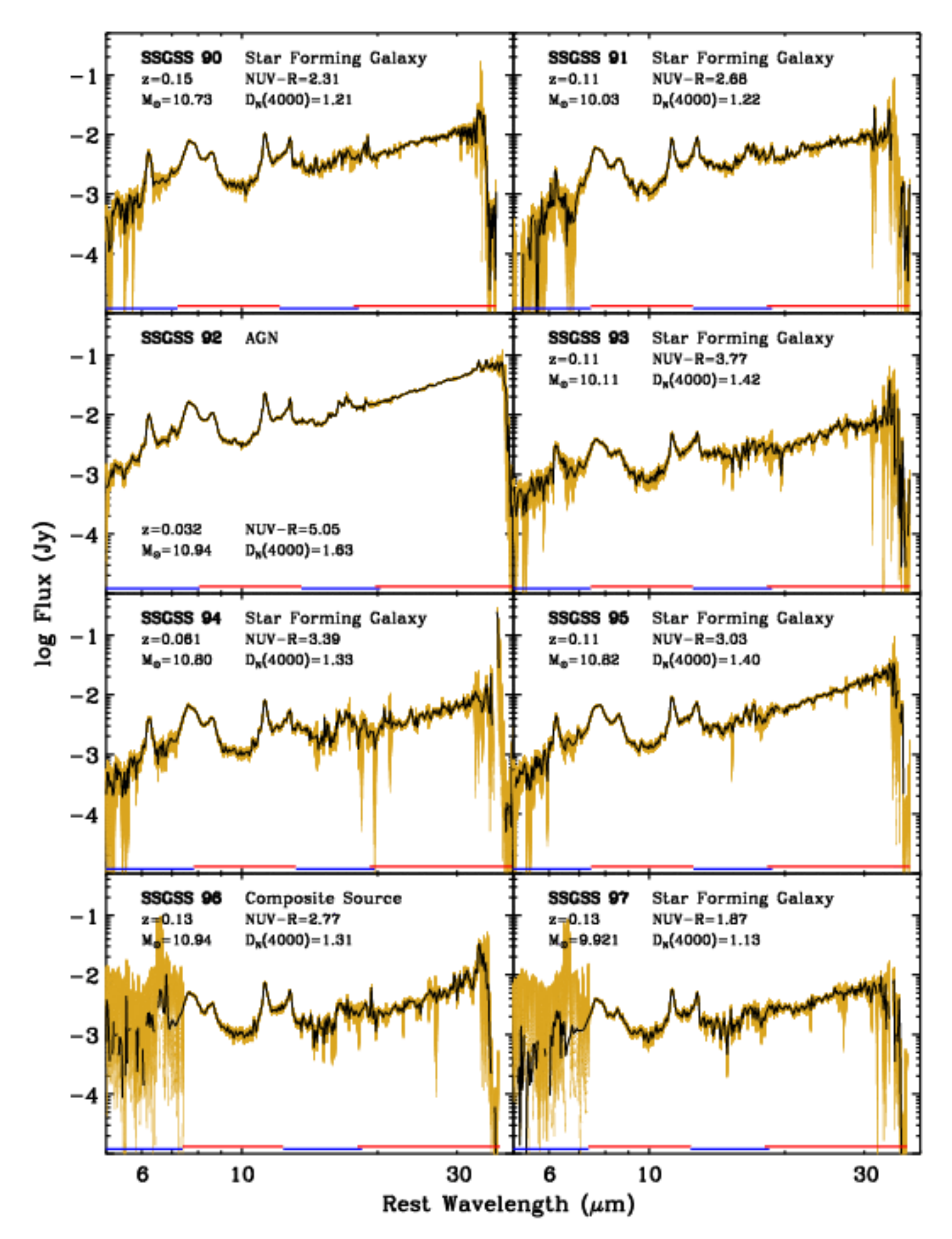}\\
Figure~\ref{lowresspectra} (continued)
\end{figure*}
\begin{figure*}
\centering
\includegraphics*[width=13cm]{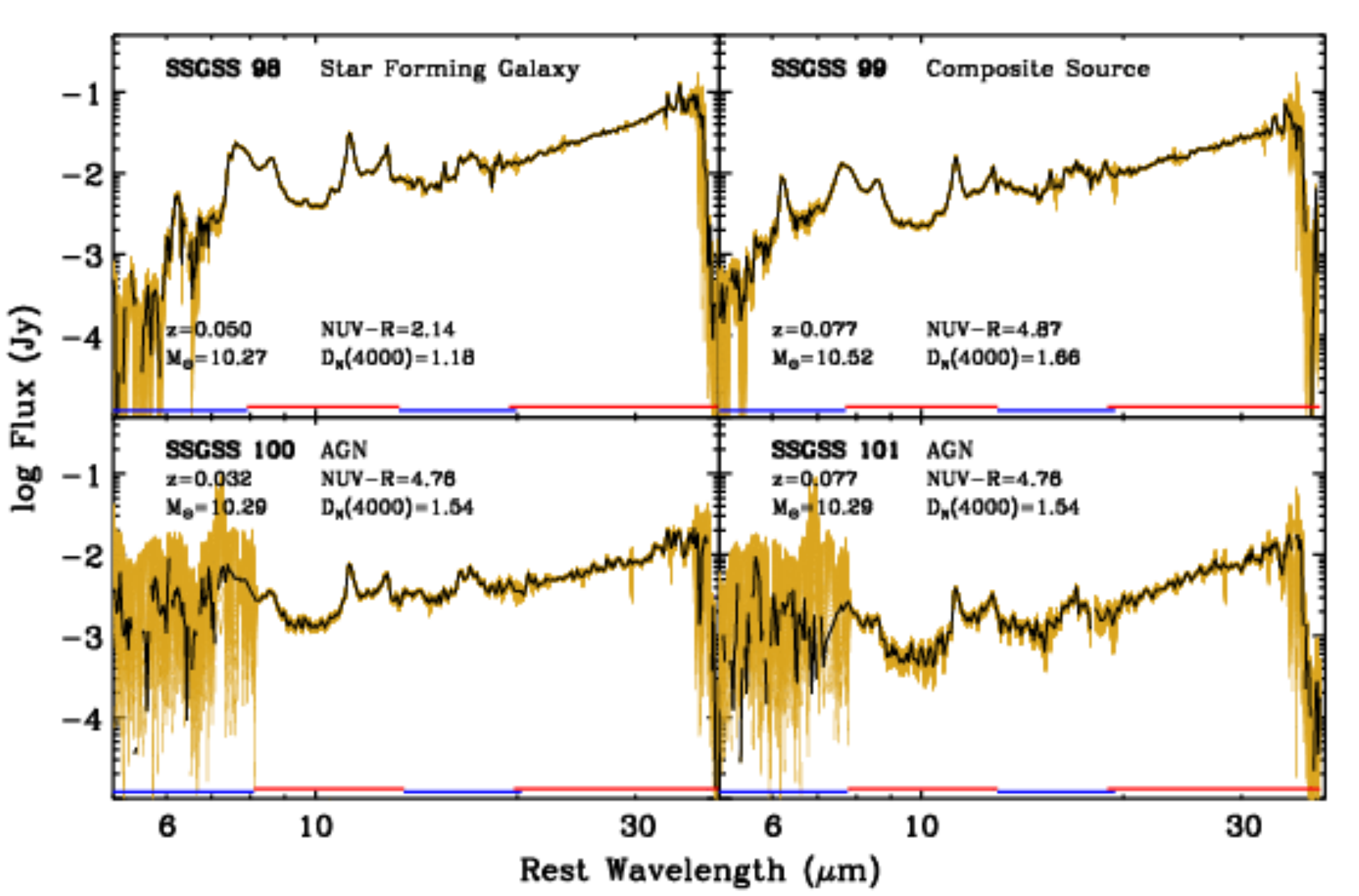}\\
Figure~\ref{lowresspectra} (continued)
\end{figure*}

\clearpage


\begin{figure*}
\centering
\includegraphics*[width=14cm]{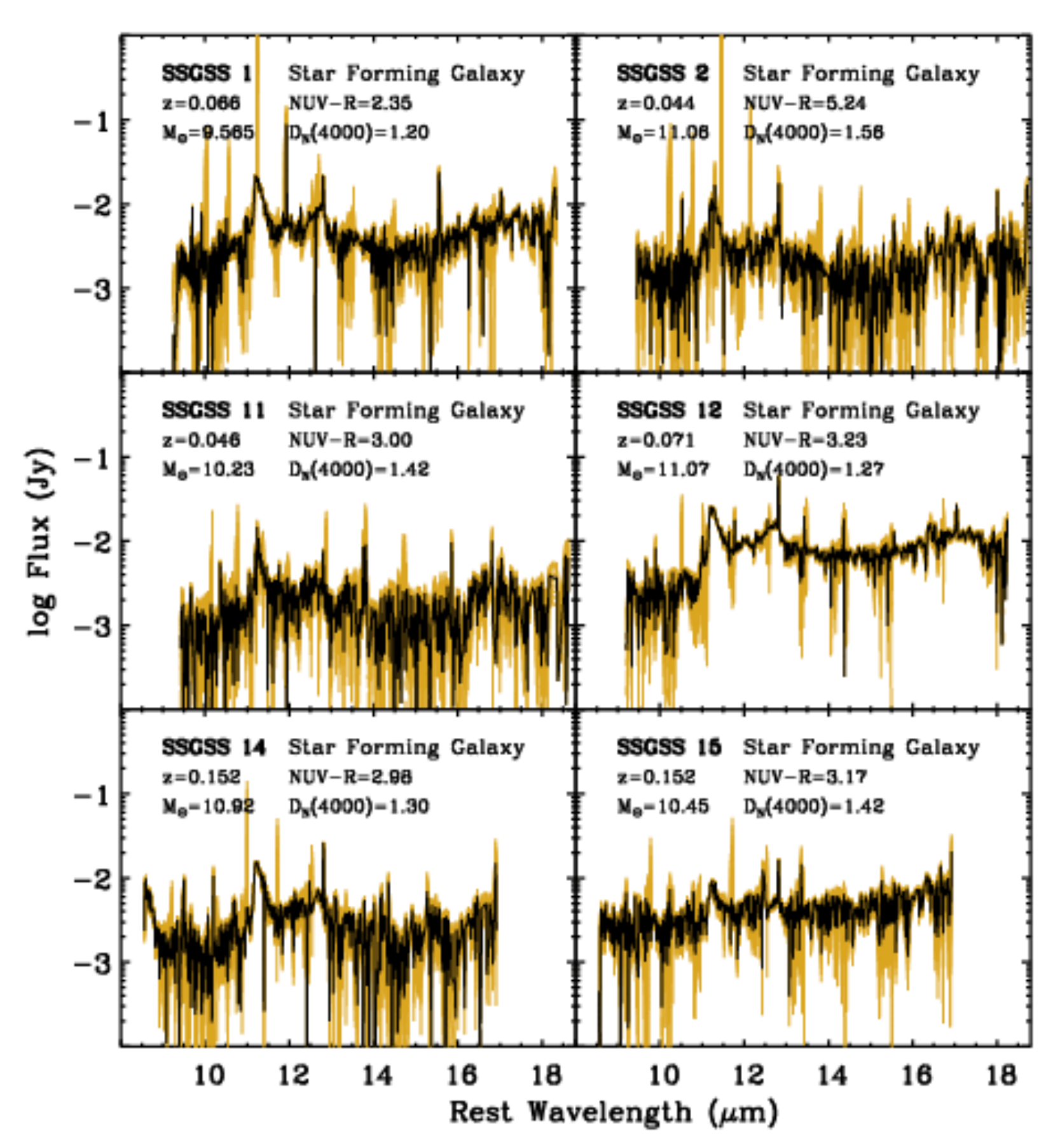}
\caption{Short-high IRS spectra, reduced and stitched as
  described in Section~\ref{hiresreduction}. The yellow shaded region shows
  the $\pm 1$-sigma errors. Also given are the SSGSS catalog
  number, BPT type \citep{BPT}, redshift, stellar mass, 
 NUV-R color, and $D_n(4000)$ age diagnostic.
\label{hiresspectra}}
\end{figure*}
\begin{figure*}
\centering
\includegraphics*[width=14cm]{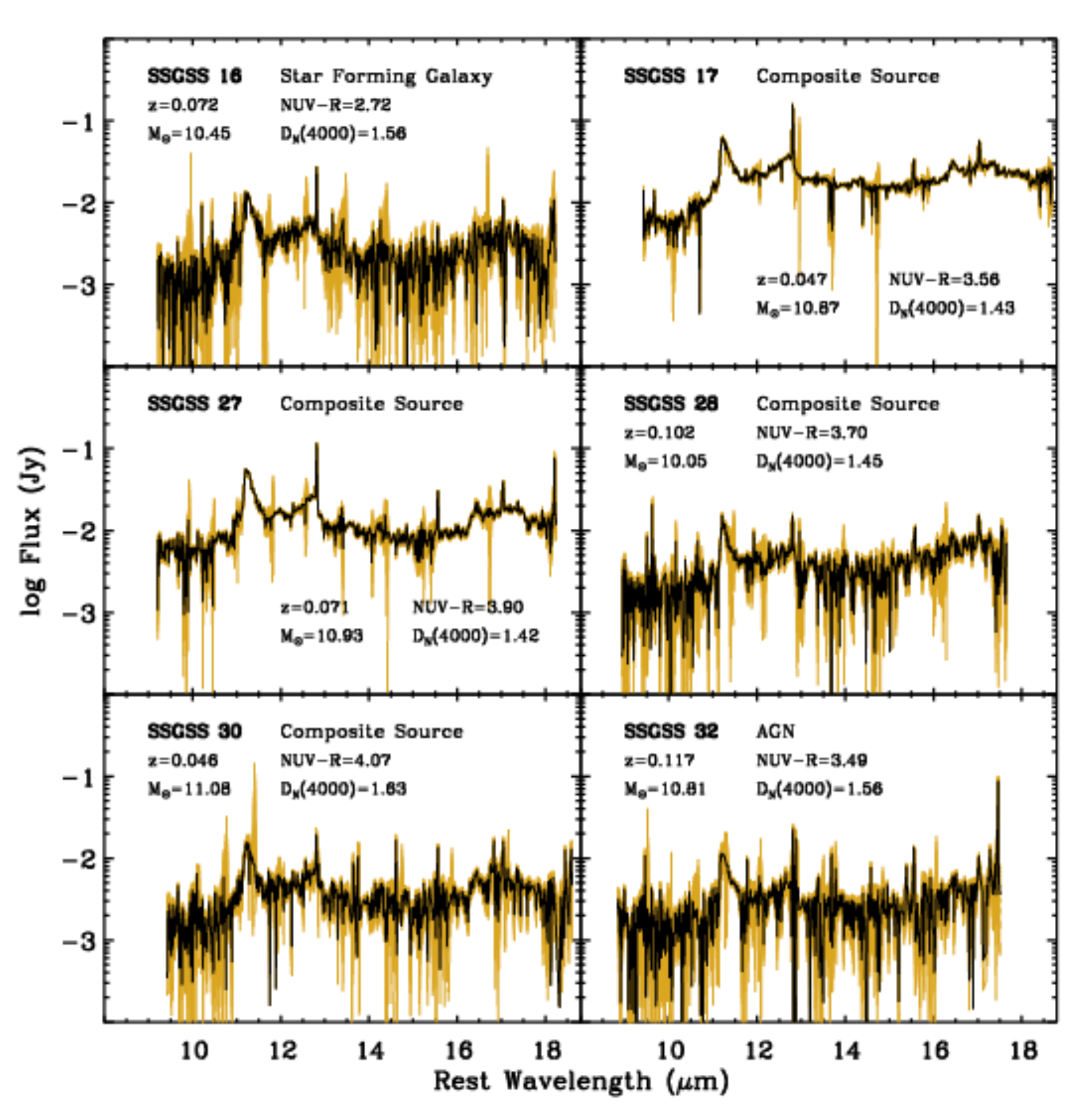}\\
Figure~\ref{hiresspectra} (continued)
\end{figure*}
\begin{figure*}
\centering
\includegraphics*[width=14cm]{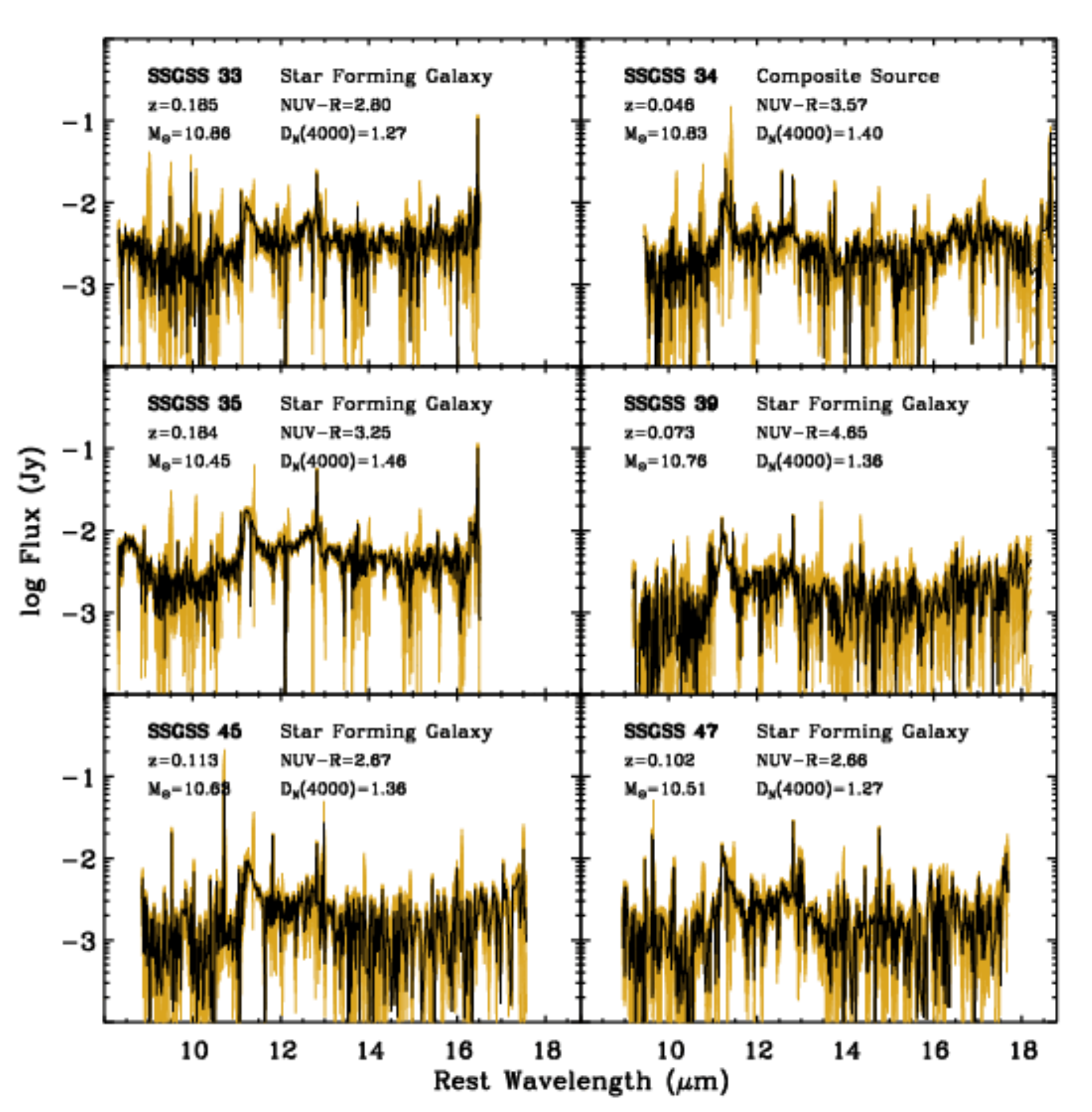}\\
Figure~\ref{hiresspectra} (continued)
\end{figure*}
\begin{figure*}
\centering
\includegraphics*[width=14cm]{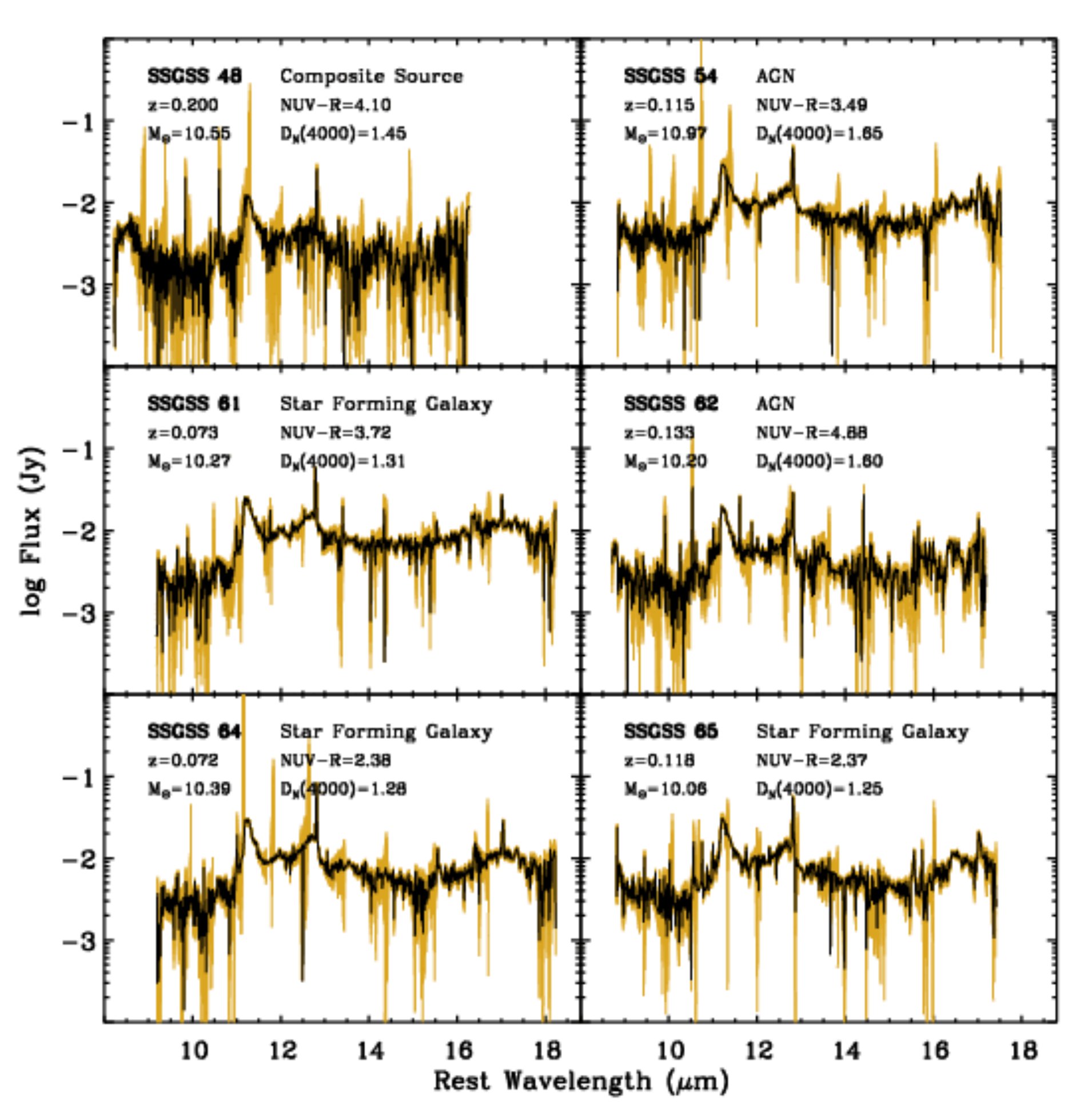}\\
Figure~\ref{hiresspectra} (continued)
\end{figure*}
\begin{figure*}
\centering
\includegraphics*[width=14cm]{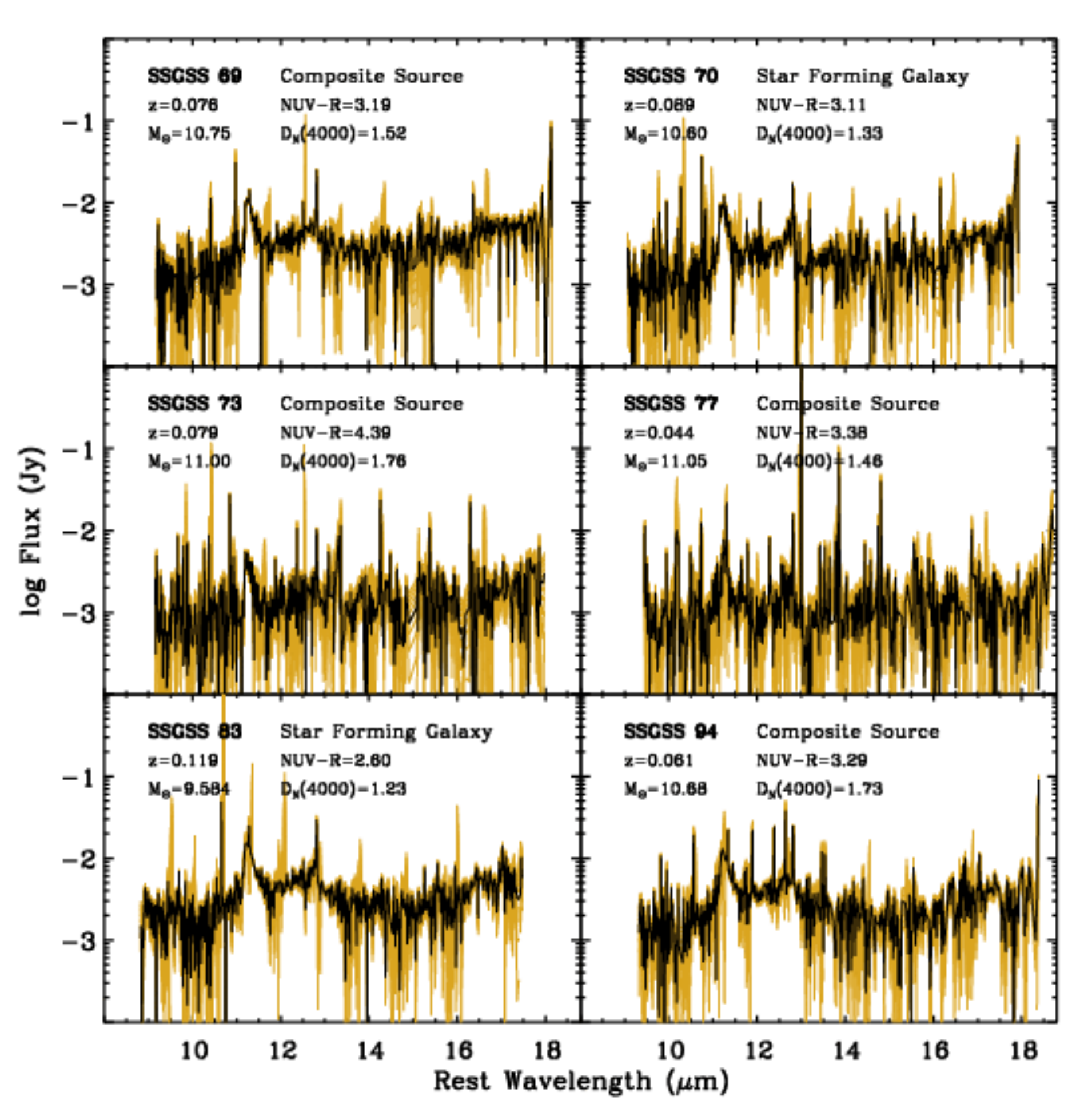}\\
Figure~\ref{hiresspectra} (continued)
\end{figure*}
\begin{figure*}
\centering
\includegraphics*[width=7cm]{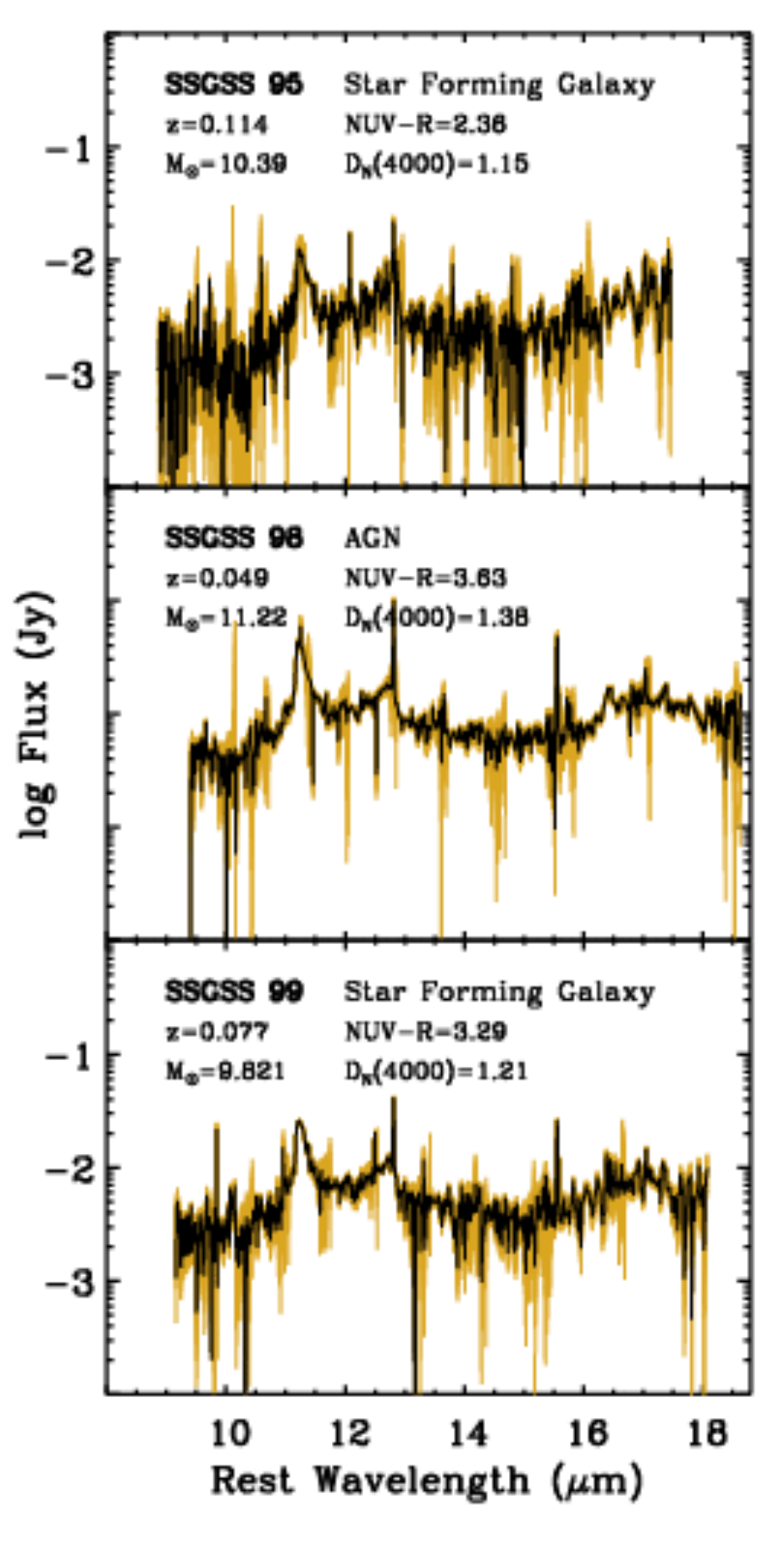}\\
Figure~\ref{hiresspectra} (continued)
\end{figure*}

\end{document}